\documentclass[apj, letterpaper]{emulateapj}  

\pdfoutput=1

\usepackage{amsmath}
\usepackage{slantsc}
\usepackage[usenames,dvipsnames]{xcolor}
\usepackage{verbatim}

\usepackage{xcolor}

\begin{document}

\slugcomment{\bf}
\slugcomment{Submitted to the Astrophysical Journal}

\title{Migration and Growth of Protoplanetary Embryos III:
Mass and Metallicity Dependence for FGKM main-sequence stars}



\author{ Beibei Liu  \altaffilmark{1,2}$^*$, Xiaojia Zhang\altaffilmark{3}, Douglas N. C. Lin \altaffilmark{2, 3, 4, 5}}
\altaffiltext{1}{Anton Pannekoek institute for Astronomy, University of Amsterdam, the Netherlands}
\altaffiltext{2}{Kavli Institute for Astronomy \& Astrophysics, Peking University, Beijing, 100871, China}
\altaffiltext{*}{E-mail: \href{mailto:bbliu1208@gmail.com}{bbliu1208@gmail.com}}
\altaffiltext{3}{Department of Astronomy and Astrophysics, University of California, Santa Cruz, CA 95064, USA}
\altaffiltext{4}{Institute for Advanced Studies, Tsinghua University, Beijing, 100086, China}
\altaffiltext{5}{National Astronomical Observatory of China, Beijing, 100012, China}
\begin{abstract}
\label{abstract}

Radial velocity and transit surveys have found that the fraction of FGKM stars 
with close-in super-Earth(s) ($\eta_\oplus$) is around $30 \%- 50\%$, 
independent of the  stellar mass $M_\ast$ and metallicity $Z_\ast$.
 In contrast,  the fraction of solar-type stars harboring one or more 
gas giants ($\eta_J $) with masses $M_{\rm p} > 100 \ M_\oplus $ is nearly $ 10\%-15\%$,  and it appears 
to increase with both $M_\ast$ and $Z_\ast$. Regardless of the properties 
of their host stars, the total mass of some multiple super-Earth systems 
exceeds the core mass of Jupiter and Saturn.  We suggest that  both super-Earths 
 and supercritical cores of gas giants were assembled from a 
population of embryos that underwent convergent type I migration 
from their birthplaces to a transition location between 
viscously heated and irradiation heated disk regions. 
We attribute the cause for the $\eta_\oplus$-$\eta_{\rm J}$ dichotomy to conditions 
required for embryos to merge and to acquire supercritical core mass 
($M_c \sim 10  \ M_\oplus$) for the onset of efficient gaseous envelope 
accretion. We translate this condition into a critical disk accretion rate, and 
our analysis and simulation results show that it weakly depends on  $M_\ast$ 
and decreases with metallicity of disk gas $Z_{\rm d}$. 
We find that embryos are more likely to  merge into supercritical cores 
around relatively massive and metal-rich stars.  
This dependence accounts for the observed $\eta_{\rm J}$-$M_\ast$. 
We also consider  the $Z_{\rm d}$-$Z_\ast$ dispersed relationship 
and reproduce the observed $\eta_J$-$Z_\ast$ correlation.

\end{abstract}



\section{Introduction}
\label{Introduction}

The rapidly accumulating data on exoplanets' mass ($M_{\rm p}$), radius
($R_{\rm p}$), semimajor axis ($a$), period ($P$), multiplicity, and their 
host stars' mass ($M_\ast$) and metallicity ($Z_\ast$) 
\citep{Winn-2014} provide valuable clues and meaningful 
constraints on the theory of planet formation.
The widely adopted sequential accretion scenario 
is based on the assumption that the formation of gas giants is 
preceded by the assemblage of cores with masses in access of a 
critical value ($M_c \simeq 10 \  M_\oplus$) \citep{Pollack-1996}. 
The building-blocks of these cores are evolved from protoplanetary embryos with
dynamical isolation masses \citep{Ida-2004a}.
The observed kinematic and structural diversity of exoplanets
has provided solid evidence to support a major paradigm shift
from the traditional {\it in situ} formation hypothesis based on 
the minimum mass nebula model \citep{Hayashi-1981} to the concept
that the proto-gas-giant planets'  cores and  super-Earth progenitors
may have migrated extensively in their natal disks. 

During their formation and growth, these embryos and cores
tidally interact with their natal disks \citep{Goldreich-1980, Kley-2012, Baruteau-2014}. 
With masses up to  a few $M_\oplus$, embedded embryos 
and cores do not exert sufficiently strong perturbations to 
modify the disk structure \citep{Lin-1993}.  Nevertheless, the
disk torque induces them to undergo type I 
migration \citep{Ward-1997}. Massive proto-gas giants do open 
gaps in the vicinity of their orbits, and they undergo type II 
migration \citep{Lin-1986}. Models based 
on these effects have been invoked to account for the origin of 
hot Jupiters \citep{Lin-1996}, resonant gas giants
\citep{Lee-2002}, and super-Earths 
\citep{Papaloizou-2005}.

Early linear torque analysis \citep{Tanaka-2002} indicates that  planets 
with $M_{\rm p} \simeq M_{\rm c}$ and $a=1$ AU in disks with gas surface 
density ($\Sigma_{\rm g}$) and temperature ($T_{\rm g}$) comparable to those 
of the minimum mass solar nebula (MMSN)
\citep{Hayashi-1981}, undergo  inward type I migration on 
a time scale $\tau_a (= a/{\dot a})$ of $\lesssim 0.1$ Myr.  This 
time scale is smaller than the time scales of disk depletion  
($\tau_{\rm dep}$) and cores' growth through gas accretion 
($\tau_{\rm c,acc}$).  This issue has the embarrassment that 
type I migration may lead to a substantial loss of cores before 
they evolve into gas giants \citep{Ida-2008}.

In order to suppress the migration speed and solve this time scale
challenge for the sequential accretion scenario, two approaches have 
been investigated. \cite{Masset-2006} proposed that planets may be trapped
near disk radii where the surface density $\Sigma_g$ and temperature $T_g$ undergo a transition.  
Such a transition may occur near the  inner disk edge \citep{Terquem-2003}, boundary 
between dead and active zones \citep{Hasegawa-2010}, the outer 
edge of the gap opened by giant planets \citep{Pierens-2008} 
or  the snow-line region \citep{Cuzzi-2004}.  Near these transitional
locations, planets' net tidal interaction with the disks may be 
suppressed. Alternatively, turbulent fluctuations in MHD disks
produce additional stochastic torques, and the survival of the planets can 
be statistically determined by this turbulent amplitude \citep{Laughlin-2004b,
Nelson-2005}.

Based on a series of numerical simulations for non-isothermal disks, 
\cite{Paardekooper-2010,
Paardekooper-2011} (hereafter PBK10, PBK11) systematically analyzed
the tidal interaction between isolated embryos and disks with continuous 
$\Sigma_g$ and $T_g$ distributions.  They showed that (1) embryos' 
migration rate and direction are determined by the sum of differential 
Lindblad ($\Gamma_{L}$) and corotation ($\Gamma_c$) torque, (2) $\Gamma_{L}$
generally leads to orbital decay, (3) for some $\Sigma_g$ and $T_g$ 
distributions (characterized by their logarithmic gradient $s \equiv 
\partial {\rm ln} \Sigma_g / \partial {\rm ln} r$ and $\beta \equiv 
\partial {\rm ln} T_g / \partial {\rm ln} r$), $\Gamma_c$
may induce a positive angular momentum transfer to the embryos, and 
(4) the strength of $\Gamma_c$ may be saturated unless the unperturbed 
values of $s$ and $\beta$ can be retained in the corotation region 
by the combined effects of viscous and thermal diffusion.
Several authors \citep{Lyra-2010, Bitsch-2013, Baillie-2015}
have suggested that embryos may migrate to and accumulate near some 
trapping radius ($r_{\rm trap}$) where $\Sigma_g$ and $T_g$ distributions
undergo such transitions.

PBK10, 11 summarized their numerical
results into a set of a very useful analytic approximations 
for the torque formula.  Several investigators have 
applied this prescription to various disk models and 
constructed modified N-body schemes to simulate 
the outcome of embryos migration.
 \cite{Pierens-2013} showed that multiple 
embryos may converge into resonant convoys and these merger 
barriers may be bypassed by a large number of embryos or 
stochastic force due to disk turbulence. \citet{Cossou-2014}
proposed that $r_{\rm trap}$ is located near the opacity 
transition region, where the disk temperature gradient is steep.
 They showed that convergent embryos
may merge during the early phase of disk evolution. 
\citet{Coleman-2014} used a different disk model to simulate both 
the formation and migration of super-Earths and gas giants. They
found that unless gas giants were formed at large radii during the
advanced stages of disk evolution, a large fraction of them would
migrate to the proximity of their host stars. 

In this series papers (\citealp{Zhang-2014, Liu-2015}, hereafter Papers I 
and II respectively), we also constructed a Hermite-Embryo code, based 
on the application of PBK10's torque formula to a self-consistent 
disk model (\citealp{Garaud-2007}, hereafter GL07).
With this approach, we systematically determine the condition 
for core formation around solar-type G stars.  We show the following (1) 
In the viscously heated inner disk, the corotation torque
leads to a net transfer of angular momentum from the protostellar
disks to the embryo, whereas the direction of angular momentum 
transport is reversed in the irradiated outer region.  (2) 
The corotation torque is saturated (weakened) for embryos with 
$M_p$ outside a factor of $(2h)^{-2/3}$ ($h$ is the aspect ratio of the disk) 
from an optimum value ($M_{\rm opt}$); thus, embryos with 
$ (2h)^{2/3} M_{\rm opt} < M_p < (2h)^{-2/3}M_{\rm opt}$ 
undergo convergent migration toward a 
transition radius ($r_{\rm trans}$) separating these regions
\citep{Kretke-2012}. The magnitudes of $M_{\rm opt}$
and $r_{\rm trans}$ are $4 \ M_\oplus$ and $1.4$ AU in disks
with ${\dot M}_g = 10^{-8} \ M_\odot$ yr$^{-1}$; $ 11 \ M_\oplus$ 
and $7.1$ AU for ${\dot M}_g = 10^{-7} \ M_\odot$ yr$^{-1}$.
(3) We also demonstrated that in disks with ${\dot M}_g \lesssim 10^{-8} \  
M_\odot$ yr$^{-1}$, embryos converge relatively slowly and 
they capture each other into their mutual mean motion resonances 
(MMRs).  Thereafter, they migrate together as a convoy of 
embryos with $M_p < M_c$ and are trapped at $r_{\rm trap} 
= r_{\rm trans}$.  (4) However, in disks with ${\dot M}_g \sim
10^{-7} \ M_\odot$ yr$^{-1}$, migrating embryos converge 
rapidly toward $r_{\rm trans}$, bypass the MMR barrier, and cross 
each other's orbits. (5) After many close encounters, embryos
collide and merge into large cores, retain outside and evolve into 
gas giants.  These results are qualitatively in agreement with 
previous contributions by others. We provided  the quantitative
criteria to indicate that the ubiquitous presence of super-
Earths and  the limited frequency of gas giants around solar-type 
stars are the manifestation of a threshold condition 
for core formation that 
depends on the magnitude of the disk accretion rate (${\dot M}_g$).  

In this paper, we generalize these results to determine 
the dependence of $\eta_J$ and $\eta_\oplus$ around stars 
with different $M_\ast$ and $Z_\ast$. 
Some basic formulae used here are already presented in Paper II.
In \S2, we first obtain observational clues from multiple
transiting planet systems in the latest Kepler database.  
We highlight that  $\eta_\oplus$ appears to be independent of $M_\ast$
and $Z_\ast$, but  $\eta_J$ generally increases with $M_\ast$
and $Z_\ast$.  In order to account for these
observational properties with our threshold core-formation
scenario, we briefly recapitulate the dependence of $M_{\rm opt}$
and $r_{\rm trans}$ on ${\dot M}_g$.  Based on the observational
data, we introduce an approximate prescription for ${\dot M}_g$'s
dependence on $M_\ast$ and $Z_\ast$.  With much more limited observational
constraints, we assume  that (1) The metallicity of the disk gas $Z_d$ 
increases with $Z_\ast$ with a dispersion ($\Delta_Z$) and 
(2) ${\dot M}_g$ is independent of $Z_\ast$.  The isolation mass 
of embryos ($M_{\rm iso}$) is determined by the surface density 
of heavy elements ($\Sigma_d \propto 10^{-Z_d}$).
We show that the magnitudes of $\tau_a$, $r_{\rm trans}$, viscosity,
and $M_{\rm opt}$ are also functions of $Z_d$ through the opacity 
dependence in the disk structure.  

We recapitulate the necessary disk condition for the formation of 
critical-mass cores in \S3. We generalize the prescription and
analytic approximations in Paper II to a range of $M_\ast$ and 
$Z_d$.  We present several simulation models for embryos' migration
around different stellar masses in \S4.  These
results are generated with the Hermite-Embryo code. We 
show that regardless of $Z_d$, a population of super-Earths may 
accumulate near $r_{\rm trans}$.  But they would not be able to 
bypass the MMR barrier and merge into cores unless  the gas accretion rate (${\dot M}_g$) in their
natal disks exceeds some critical value (${\dot M}_{res}$).  Based on 
the observed ${\dot M}_g$ dependence on $M_\ast$, we find (1) only a 
small fraction of solar-type stars ($\eta_{{\dot M}}$) satisfy this 
core-formation criterion and (2) $\eta_{{\dot M}}$ increases
with $M_\ast$.  Under the assumption that $\eta_{{\dot M}}$ corresponds
to $\eta_J$, we reproduce the observed $\eta_J$-$M_\ast$ correlation
among FGKM main-sequence stars.


In \S5, we focus on the $\eta_J$-$Z_\ast$ correlation around solar-type
stars.  We show that although $M_{\rm iso}$ and $r_{\rm trans}$
are increasing functions of $Z_d$, the formation probability of cores does not 
depend sensitively on the initial total mass of embryos. 
However, in a layer accretion disk model,
the effective viscosity $\nu$ may decrease and $\Sigma_g$ increase 
with $Z_d$ such that migrating embryos are likely to bypass the 
MMR barrier and converge near $r_{\rm trans}$ in metal-rich disks.  
The close packing of embryos enhances their merger probability 
and promotes the emergence of cores and gas giants. We reproduce
the observe $\eta_J-Z_\ast$ correlation by taking into account
the correlation and dispersion between $Z_d$ and $Z_\ast$.  Finally,
in \S6, we summarize our results and discuss their implications.

\section{Observational properties}
\label{sec:observationaldata}

\subsection{Frequency of planets in different stellar environment}
\label{sec:keplerdata}


Radial velocity (RV) and transit surveys indicate that while nearly $10\%-15\%$ 
of solar-type stars harbor one or more gas giant planets \citep{Marcy-2008, 
Cumming-2008}, they are rarely found around late dwarfs 
\citep{Endl-2006,Bonfils-2013}.  This fraction ($\eta_J$) appears to 
increase with $M_\ast$ among subgiant and giant stars more massive than 
the Sun \citep{Johnson-2007, Johnson-2010,Jones-2016}. However, the fraction of stars 
that contain  super-Earths ($\eta_\oplus$) is almost $ 30\%-50\%$ and appears to be at least as abundant 
in M stars as in FGK stars \citep{Bonfils-2013, Howard-2012, Fressin-2013, 
Dressing-2013,Mulders-2015}.  Here super-Earths categorically
refer to the 
low-mass ($M_p <30 \ M_{\oplus}$) and modest-size ($R_p<4 \ R{\oplus}$)  planets 
with period $P<100$ days, in contrast to the more massive gas giants with 
period up to a few years. The magnitude of $\eta_J$ is also an increasing 
function of $Z_\ast$\citep{Santos-2004, Fischer-2005, Sousa-2011,
Mortier-2013}, whereas $\eta_\oplus$ appears to be independent of $Z_\ast$
\citep{Sousa-2008,Schlaufman-2011,Buchhave-2012, Wang-2015, Buchhave-2014,Schlaufman-2015}.
Combined the RV measurement and transit light curve, some inferred super-Earths appear to  have
substantial gaseous atmospheres. 
In contrast to the formation history of terrestrial planets in our solar system,  these super-Earths may have acquired
most of their masses prior to the depletion of the disk gas \citep{Lopez-2014}.

There have been several attempts to account for both the 
$\eta_J$-$M_\ast$ and $\eta_J$-$Z_\ast$ correlations
\citep{Laughlin-2004a, Ida-2004b, Ida-2005, Mordasini-2012}. 
These models generally assume that the total mass of building-block
 planetesimals and embryos in the disk is a 
fixed fraction of the heavy elements in the central stars.

In Paper II, we proposed an alternative minimum planetary building-block 
scenario based on the confirmed planets in the multiple-planet systems.  
Kepler data showed that even though a large number of  individual super-Earths have mass ($M_p$)
smaller than the critical mass ($M_c$), the total mass $M_s$ in most multi-planetary systems around 
individual host stars exceeds $M_c$ (Figure 1 in Paper II). The 
common existence of such multiple super-Earth systems suggests that there is no lack of heavy elements in their natal disks.  Nevertheless, the 
lack of gas giants around most solar-type stars may be due to the inability  of 
 a sufficient fraction of all available building-block materials to be 
collected into a few supercritical cores with $M_p \geq M_c$. 

\begin{figure*}[htbp]
\includegraphics[width=0.5\linewidth,clip=true]{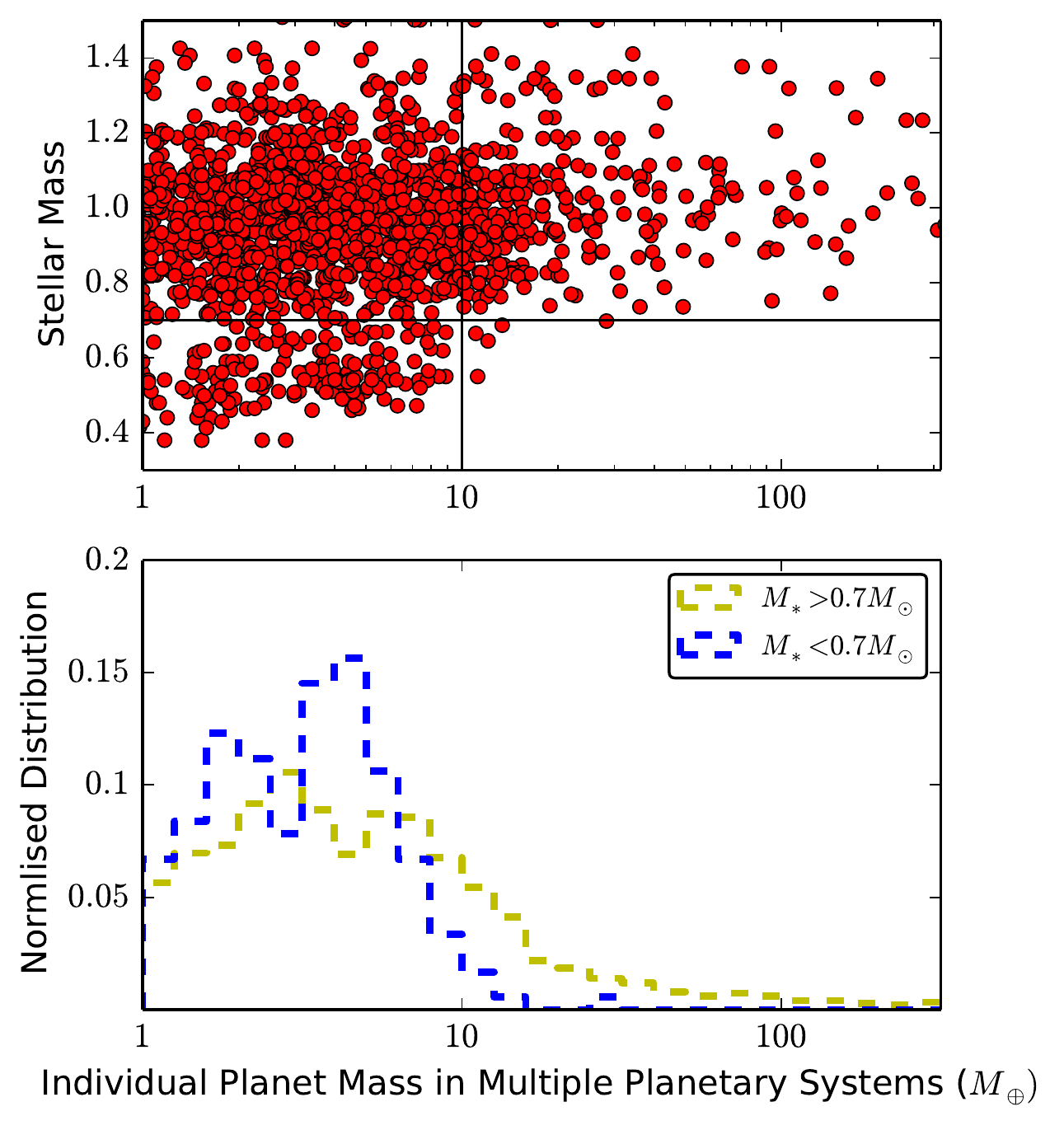}
\includegraphics[width=0.5\linewidth,clip=true]{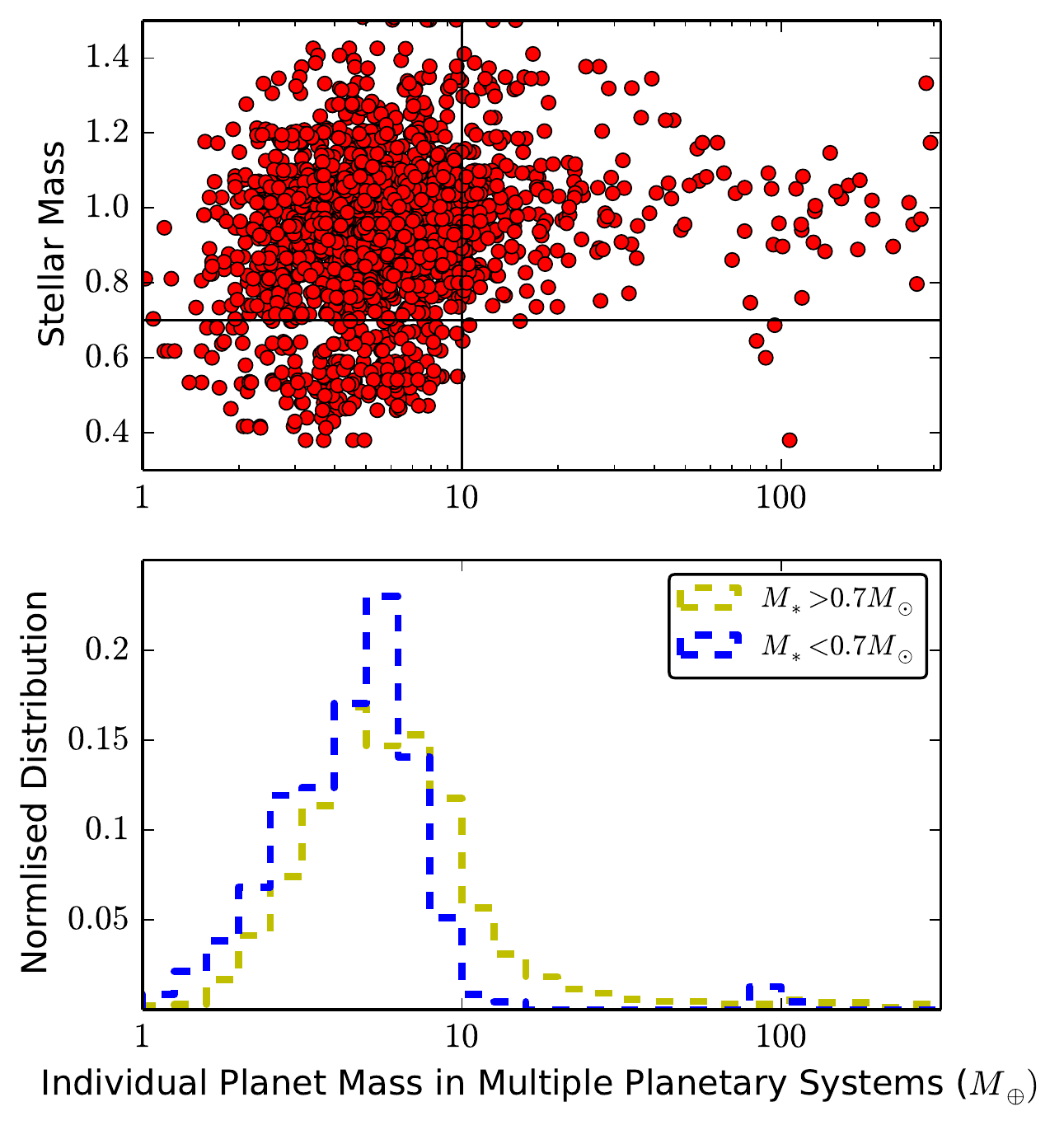}
\caption{
single planet mass $M_p$ (red dots)  versus the mass of their host stars ($M_\ast$). 
The horizontal black line indicates the critical stellar mass ($M_{\ast}$) 
and vertical black line marks the critical planet mass  $M_c$
for efficient gas accretion. The left panel adopts $\eta_M =2.06$ and $M_1=1$
and the right panel uses $\eta_M =1$ and $M_1=3$.  The planets' data are obtained from the Kepler's database in NASA Exoplanet Archive, and the stellar parameters are adopted from the \cite{Huber-2014} catalog.\\
}
\label{figX}
\end{figure*}

In this paper, we consider the formation efficiency of cores around stars 
with different $M_\ast$ and $Z_\ast$.  We first analyze the implication of
observational data.  The ground-based observations are contaminated by 
heterogenous observational selection effects. But the Kepler target stars 
provide a controlled sample with well-stated observational selection 
criteria \citep{Dong-2013}. The latest  Kepler data release
(http://exoplanetarchive.ipac.caltech.edu/index.html) categorizes 4664 
planetary candidates. Most of these transiting objects are potentially
super-Earths 
with $R_p \sim 1-4 \ R_\oplus$. A fraction of them have been confirmed by 
 follow-up RV or transit timing observations \citep{Marcy-2014}.

In order to infer $M_p$ from the Kepler data, we adopted
an empirical $M_p-R_p$ relationship extrapolated from solar system planets  in Paper II 
\citep{Lissauer-2011} that
\begin{equation}
M_{\rm p} \simeq  M_{1} (R_{\rm p}/ R_{\oplus})^{\eta_M},
\label{eq:radiusmass0}
\end{equation}
where $\eta_M=2.06$ and the normalized mass $M_1 = 
1 \ M_\oplus$.  Based on the follow-up RV 
determination, \cite{Dressing-2015} find that the solar system 
extrapolation  
 from \cite{Zeng-2013} matches well with the $M_p-R_p$ 
correlation among super-Earths with $R_p < 1.6 \ R_\oplus$. 
The latest  two-layer metal/rock planet model  
\citep{Zeng-2016} suggests   $\eta_M=3.7$, $M_1 = 
(1.07-0.21 \times \rm cmf)^{0.27}$ for planets with  $R_p<1.75 \ R_{\oplus}$, where $\rm cmf$ refers to a core mass fraction between 0 and 0.4.  There is a 
large dispersion among more massive super-Earths.     
\citet{Wu-2013} used  measured mass  from transit timing variations 
(TTVs)  to get $\eta_M =1$ and $M_1 = 3 \ M_\oplus$.
\citet{Weiss-2014} 
fitted both  RV and TTV data with similar $\eta_M=0.93$ and $M_1 =2.69  \ M_\oplus$ for $1.5 \ R_{\oplus}< R_p<4 \ R_{\oplus}$.

Combining the data obtained from both RV and TTV methods, \cite{Rogers-2015} and 
\cite{Wolfgang-2015} found that $M_p$ generaly increases with $R_p$
 but with a wide dispersion.  
Quantitatively, for each observationally measured $R_p$, the 
inferred $M_p^\prime$ may have a Gaussian distribution, 
\begin{equation}
{d N (M_{\rm p}^\prime, R_p ) \over d M_{\rm p}^\prime} 
= {N \over M_\oplus} {\rm exp}   \left[- \frac{
 \rm log M_{\rm p} ^\prime (R_{\rm p}) - 
{\rm log} M_{\rm p} (R_{\rm p}) } 
{ \Delta_{M_{\rm p}}}
 \right]^2
\label{eq:radiusmass2}
\end{equation}
where $\Delta_{M_{\rm p}} \simeq 0.3$ is a factor of 2 mass dispersion in  logarithm.
Instead of a piecewise function, we choose observed masses and 
radii relation with a single power law plus extra intrinsic scatter shown above 
where $\eta_M = {1.8}$ and $M_1 = 1.6 \ M_\oplus$ for
  planets with $R_p <8 \ R_{\oplus}$  \citep{Wolfgang-2015}.


\begin{figure*}[htbp]
\includegraphics[width=0.5\linewidth,clip=true]{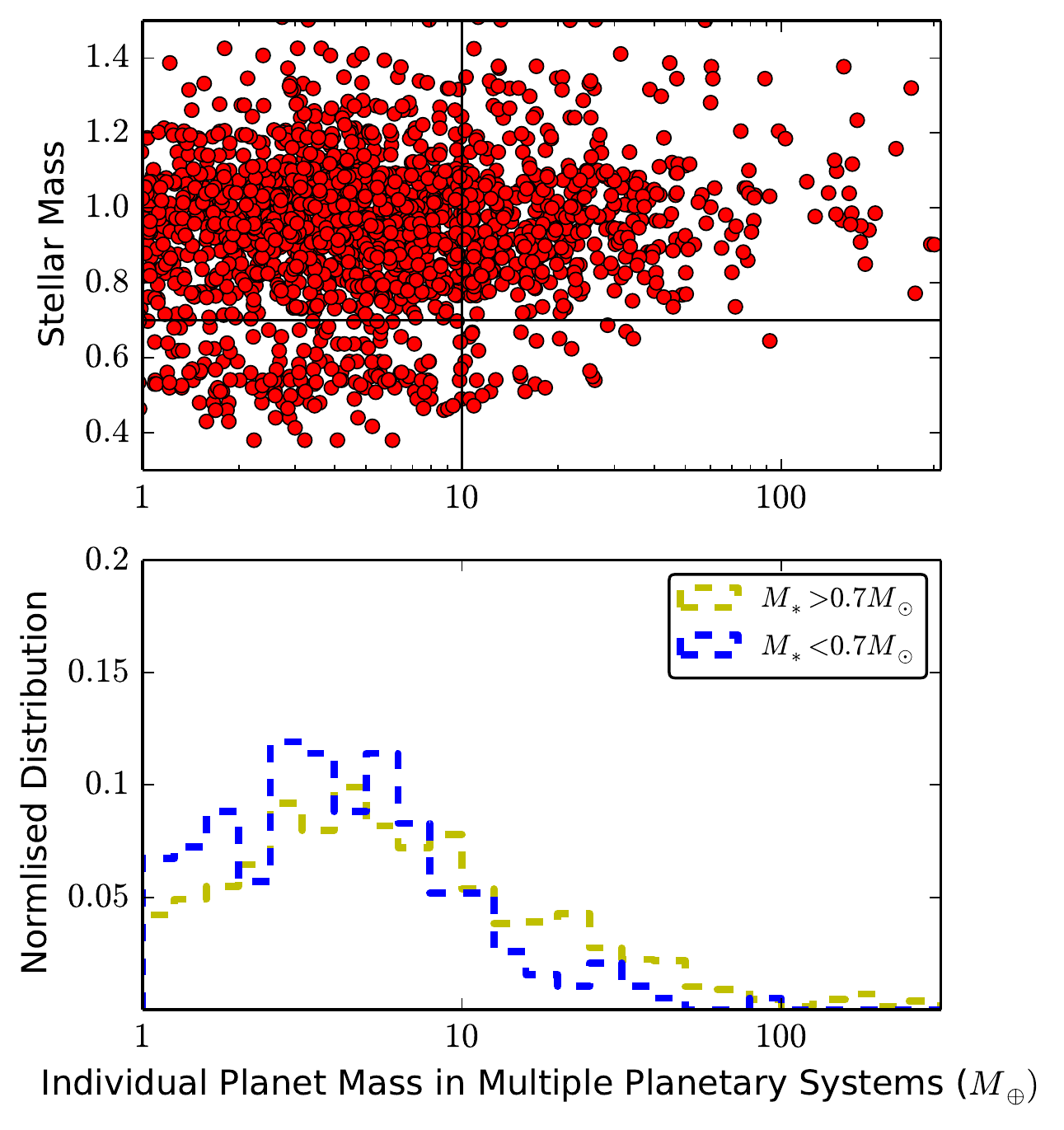}
\includegraphics[width=0.48\linewidth,clip=true]{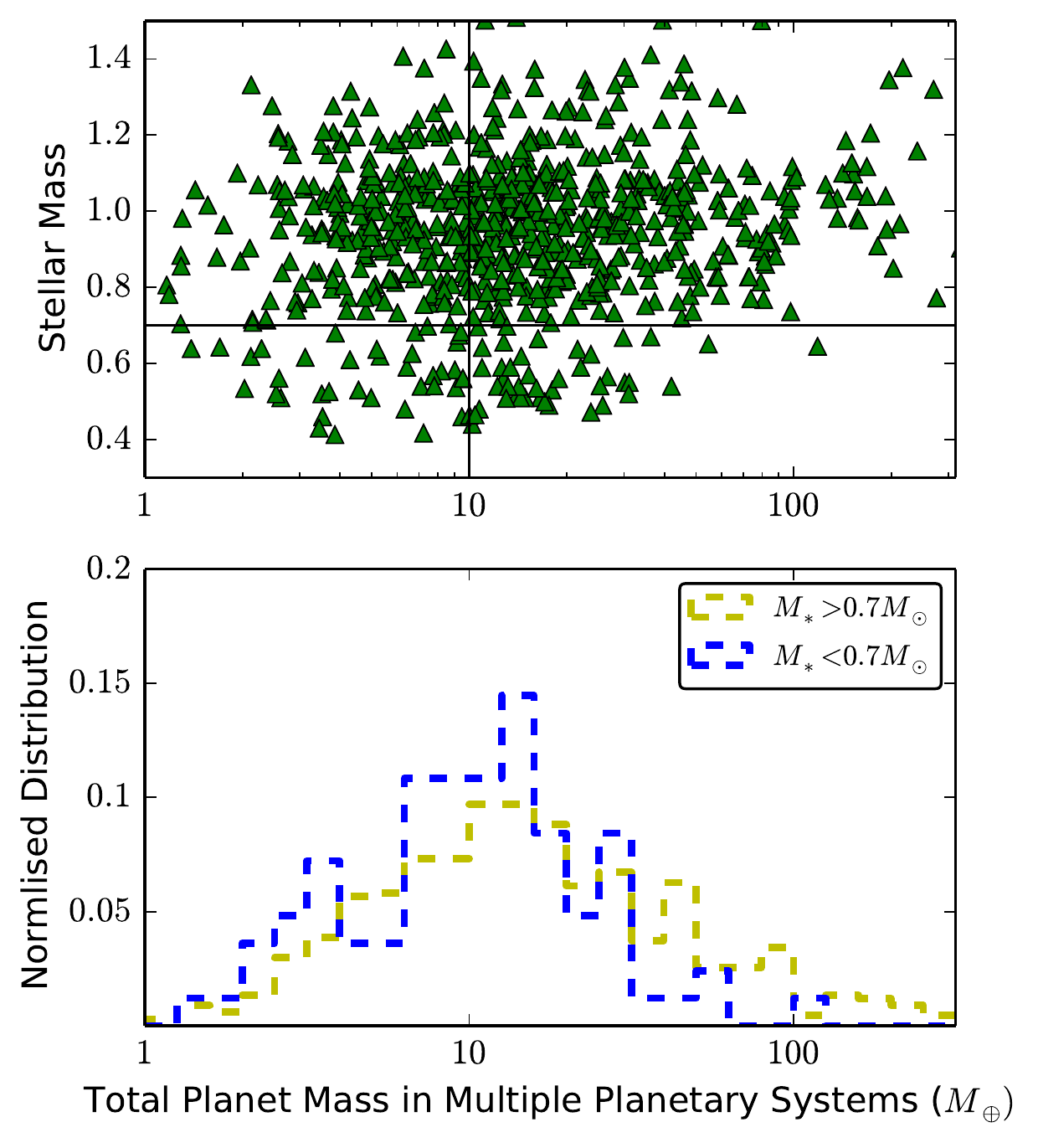}
\caption{
Top panels: single planet mass $M_p$ (red dots in left panel) and 
total (detected) planet mass $M_s$ in individual systems (green 
triangles in right panel) versus the mass of their host stars ($M_\ast$). 
The values of $M_p$ and $M_s$ are obtained from the statistical prescription in Eq \ref{eq:radiusmass2}.  
The horizontal black line indicates the critical stellar mass ($M_{\ast}$) 
and the vertical black line separates the critical planet mass  
($M_{c} =10 \ M_{\oplus}$) for efficient gas accretion. \\
Bottom panels: normalized distribution of $M_p$ (left) and $M_s$ (right)
as functions of $M_\ast$. The yellow dashed histogram represents the 
samples with $M_{\ast}>0.7 \ M_{\odot}$, while the blue dashed histogram represents 
samples with $M_{\ast}<0.7 \ M_{\odot}$. 
All the planets' data are obtained from the Kepler's database in NASA Exoplanet Archive, and the stellar information is from the \cite{Huber-2014}'s catalog.
}
\label{figa0}
\end{figure*}

\begin{figure*}[htbp]
\includegraphics[width=0.5\linewidth,clip=true]{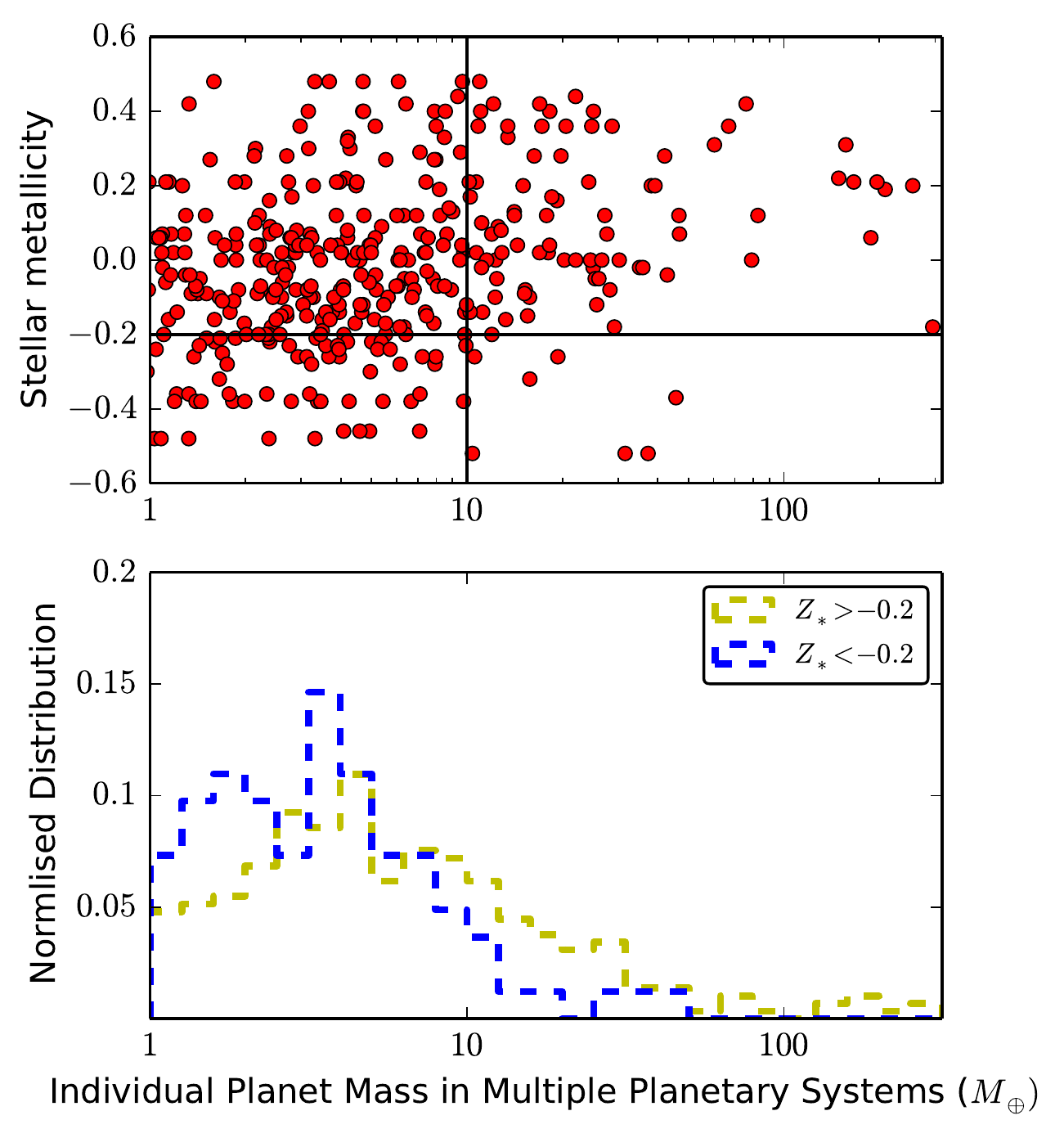}
\includegraphics[width=0.48\linewidth,clip=true]{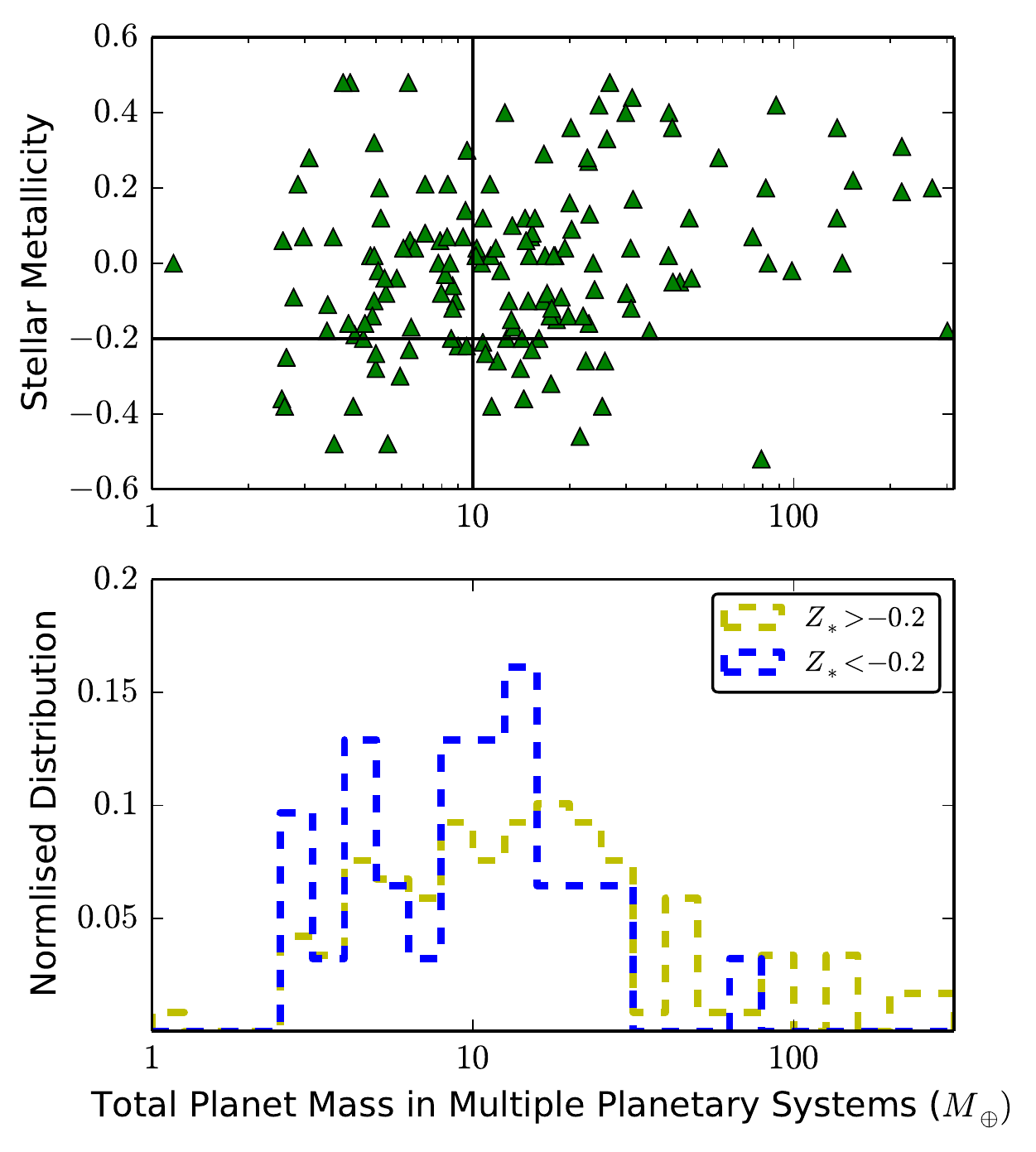}
\caption{
Top panels: single planet mass $M_p$ (red dots on the left) and $M_s$
(green triangles on the right) versus their host stars' metallicity 
($Z_\ast$). The values of $M_p$ and $M_s$ are obtained from the statistical prescription in Eq \ref{eq:radiusmass2}.  
The horizontal black line indicates $Z_{\ast}=-0.2$ which appears
to be the threshold metallicity for stars with gas giant planets,  
and the vertical black line corresponds to $M_p=M_{c}(=10 \ M_{\oplus})$, 
which is the conventional condition for efficient gas accretion. \\
Bottom panels: normalized distribution $M_p$ (left) and $M_s$ (right) distributions. 
The yellow dashed histograms represent the samples with $Z_{\ast}>-0.2$, while 
the blue dashed histogram represents samples with $Z_{\ast}<-0.2$.
The planets' data are obtained from the Kepler's database in NASA Exoplanet Archive, and the stellar parameters are adopted from the \cite{Huber-2014} catalog. Only the SPE samples (with sufficiently accurate spectroscopic  $Z_\ast$) 
are included in this figure. 
}
\label{figb0}
\end{figure*}

 We plot the individual $M_p$ 
for the 2007 confirmed/candidate planets from the dataset 
as a function of their host stellar mass.   We obtain $M_\ast$ 
from the  \cite{Huber-2014}'s  stellar catalog and the 
Dartmouth Stellar Evolution Program (DSEP). We adopt the values of $\eta_M$ 
and $M_1$ for two deterministic prescriptions from equation 
(\ref{eq:radiusmass0}) ( Figure \ref{figX}) and one probabilistic $M_p-R_p$ 
prescription from equation (\ref{eq:radiusmass2}) (Figure \ref{figa0} 
and  \ref{figb0}). Comparison of the results obtained with different $M_p-R_p$ 
prescriptions shows no noticeable differences in the inferred 
$M_p$ and $M_s$ distribution around stars with different
$M_\ast$ and $Z_\ast$ values.

Based on the statistical prescription of equation (\ref{eq:radiusmass2}),
we plot in Figure \ref{figa0}, the individual planetary mass ($M_p$) 
and the total (detected) planetary mass ($M_s$) distribution
for the 820 confirmed/candidate multiple-planet systems.
Using the same approach, we also plot in Figure 
\ref{figb0} the $M_p$ distribution for the 421 confirmed/candidate 
planets and the total mass $M_s$ distribution for the 159 multiple-planet 
systems around metal-rich and metal-deficient stars respectively.
The number of data entries in Figure \ref{figb0} is smaller
than that in Figure \ref{figa0} because it includes only the high-resolution spectroscopy  
``SPE'' sample, which excludes host stars with average
metallicity error bars larger than 0.15 dex.

In Figures \ref{figa0} and \ref{figb0}, red dots and green triangles 
represent the single planetary mass $M_p$ and the total mass $M_s$ 
of multiple-planet systems around individual host stars. The top panels 
of Figure \ref{figa0} show $M_p$ (left) and $M_s$ (right) as a function of 
$M_\ast$. In Figure \ref{figa0}, black lines correspond to $10 \ M_{\oplus}$ 
for the critical core mass and $0.7 \  M_{\odot}$ for stellar mass of K and M main-sequence
 dwarfs.  Bottom panels show the histogram of their normalized $M_p$ 
(left) and $M_s$ (right) distributions with host stars  $M_\ast > 0.7 \ 
M_\odot$ (yellow) and $M_\ast < 0.7 \ M_\odot$ (blue).

The same symbols are used in Figure \ref{figb0} for the $M_p-Z_\ast$ and 
$M_s$-$Z_\ast$ distribution.  The horizontal black lines in the top panels 
present the critical value  of $Z_{\ast}=-0.2$. Precise measurement of 
$M_\ast$ and $Z_\ast$ of all Kepler samples including some excluded planets 
may be significantly improved by follow-up observations 
with spectroscopic survey telescopes such as SDSS or LAMOST 
(private discussion with  Yang Huang and Subo Dong).

Figures \ref{figa0} and \ref{figb0} show that very few individual
planets have $M_p$ in excess of the critical core mass 
($ M_c \simeq 10 \ M_{\oplus}$) typically around late-type K and M dwarfs 
(with $M_{\ast}<0.7 \ M_{\odot}$) and metal-deficient stars (with $Z_{\ast}<-0.2$).  
But the total mass of planets in multiple systems $M_s$ is substantially 
larger than $ 10 \ M_\oplus$ regardless of their host stars's $M_\ast$ and 
$Z_\ast$. Around some metal-deficient and K and M dwarf stars, $M_s$ can extend 
to $30-40 \ M_{\oplus}$.  While the $M_p$ distribution above $M_c$ (left bottom panel of Figure \ref{figa0} and \ref{figb0})
for relative massive and metal-rich stars differs intrinsically from  those
with smaller $M_\ast$ and $Z_\ast$, 
the discrepancy of their total mass distribution is relatively small 
for all stars with $M_s<40 \ M_{\oplus}$. 

These figures pose challenges to the conventional planet formation
scenarios based on the following assumptions: (1) a nearly constant fraction 
of the heavy elements in their host stars is retained to form 
protoplanetary building blocks, and (2) the formation probability of 
gas giants is determined by the total mass of protoplanetary embryos 
in their natal disks.

These figures also indicate that even around low-mass and metal-deficient
stars, there is still an adequate supply of building-block materials to  form a  critical mass
core and  initiate runaway gas accretion if  these low-mass super-Earths were able to merge with each other.
How these low-mass super-Earths are organized and assembled together in protoplanetary disks is a more crucial  issue that modulates the gas giant formation efficiency. 
Based on that consideration, we search for decisive factors that affect the efficiency of gas giant formation. In a straightforward approach we fix one condition ( the total mass of embryos) and vary the other variables ( \textit{i.e.}, $M_{\ast}$ and $\dot M_g$ in section 4) in our simulations.

Another interesting issue is the critical condition for the onset of 
efficient gas accretion. The critical mass $M_c$ is determined by the
rate of heat transport from the core ($R_c$) to the Roche radius ($R_R$).  
This process is dominated by convection close to $R_c$ and by radiative 
diffusion in the tenuous region near $R_R$.  Since the radiative flux
is inversely proportional to the opacity, the magnitude of $M_c$ determined
from 1D quasi-hydrostatic calculations increases with the local $Z_d$
\citep{Ikoma-2000, Hubickyj-2005, Movshovitz-2010}.  The implication
of these models is that around metal-deficient stars, a larger fraction 
of cores with $M_p \sim 10 \ M_\oplus$ may evolve into gas giants rather than 
become super-Earths.  The results in Figures \ref{figb0} show that 
(1) the fraction of super Earths with $M_p \sim 5-10 \ M_\oplus$ remains 
modest and gas giants are extremely rare around metal-deficient stars, 
and (2) the drop-off of the normalized $M_p$ distribution increases with 
$Z_\ast$.  These features suggest that the transitional threshold from 
super-Earths to gas giants may be weakly but not principally determined 
by opacity of the accreted gas.

\subsection{Link Type I torque with observational properties of protostellar disks}
\label{sec:diskdata}

The common existence of super-Earths and rarity of gas giants, 
especially around low $M_\ast$ and $Z_\ast$ stars, suggests that
it is inappropriate to simply attribute the observed $\eta_J$-$M_\ast$ 
and $\eta_J$-$Z_\ast$ correlation to a dependence on the availability 
of building-block material.  Instead, we suggest that they may be
due to the marginal conditions required for protoplanetary embryos to 
migrate, converge, cross each other's orbit, and merge into proto-gas-giant 
cores in their natal disks. In this scenario, the condition for 
super-Earth embryos to evolve into supercritical cores is determined 
by the rate of their migration, $\dot a$ (Paper II). Analytic derivation 
and numerical simulations \citep{Goldreich-1980, Baruteau-2014} 
show that several Earth-mass planet exerts a net torque 
\begin{equation}
\Gamma=f_\Gamma (s, \beta, p_\nu, p_\xi) \Gamma_0 = 
f_\Gamma (s, \beta, p_\nu, p_\xi)  (q/h)^2 \Sigma_p r_p^4 \Omega_p^2
\label{eq:torque}
\end{equation}
through their Lindblad and corotation resonances. 
In the above equation,
$q=M_p/M_{\ast}$, $\Sigma_p$, $h$, and $\Omega_p$ are the disk surface 
density, aspect ratio, and Keplerian frequency at the planet's location 
$r_p$.  The coefficient $f_\Gamma$ is a function of $s$, $\beta$, dimensionless parameters
$p_\nu = (2/3) (R_e x_s^3)^{1/2}$ and $p_\xi =(R_e x_s^3/P_t)^{1/2}$
where $R_e = \Omega_p r_p^2 / \nu$ and $P_t= \nu/\psi$ are the 
Reynolds and Prandtl numbers, respectively, $\nu$ and $\psi$ are the 
viscosity and radiative diffusion coefficients, and $x_s \simeq 
(q/h)^{1/2}$ is the dimensionless width of the horseshoe region.

The total torque leads to a change in the  planet's semimajor axis at a rate
\begin{equation}
{\dot a} = {2 f_a (s, \beta, p_\nu, p_\xi) 
q \over h^2} {\Sigma_p r_p^2 \over M_\ast} r_p \Omega_p
\label{eq:adot}
\end{equation}
where $f_a (s, \beta, p_\nu, p_\xi) = \sum f_{\Gamma, i} 
(s, \beta, p_\nu, p_\xi) $ and the index $i$ refers to components
including the differential Lindblad and corotation torque.  In the 
above expression, the migration rate is an increasing function of
the disk surface density $\Sigma_p$ at $r_p$.  

Direct observational determination of disks' $\Sigma_g$ and $T_g$ is 
challenging since the inner region of the disk is optically thick. 
Although the disk becomes optically thin at relatively large radii, the 
bulk disk mass derived from (sub)millimeter dust observation still 
depends on poorly known dust opacity and gas-to-dust ratio.  The 
uncertainties of CO measurement are due to the condensation of 
gas onto grains and photodissociation from external UV radiation 
by nearby OB stars. However, both the $\Sigma_g$ and $T_g$ distributions 
(i.e. $s$ and $\beta$) are expected to be correlated with ${\dot M}_g$. 
Based on the observed values of ${\dot M}_g$ from disks around classical 
T Tauri and  the well adopted $\alpha$ prescription for viscosity
\citep{Shakura-1973, Ruden-1986, Hartmann-1998BOOK},
we find that the condition for the formation of supercritical-mass 
cores is marginally satisfied in protostellar disks around solar-type
 stars (Paper II).  

\subsection{Disk properties around different-mass host stars}
\label{sec:dependence}

Here we extend this embryo migration scenario for stars with 
different $M_\ast$ and $Z_\ast$ through the dependence of 
$\Sigma_g$ and $T_g$ distribution in their circumstellar disks.
Statistical observational data of protostellar disks indicate that (1)  
the average accretion rate (${\dot M}_a$) is an increasing function 
of $M_\ast$ with a dispersion ($\Delta {\dot M}_g$), 
(2) the magnitude of ${\dot M}_a$ decreases with time, 
and (3) there is no direct evidence that ${\dot M}_a$
correlates with $Z_\ast$ or $Z_d$.
  
Quantitative measurements 
\citep{Hartmann-1998, Natta-2006, Manara-2012, 
Ercolano-2014, DaRio-2014} have been approximated by  
\begin{equation}
{\dot M}_a = {\dot M}_{a \odot} { m_\ast^{\eta_b} 
(t/\tau_{\rm dep})^{-\eta_a }} 
\label{eq:mdota}
\end{equation}
where $m_\ast= M_\ast/M_\odot$ and ${\dot M}_{a \odot} 
\sim 5 \times10^{-8} M_\odot$ yr$^{-1}$
corresponds to the average value of ${\dot M}_g$ for 
solar-mass T Tauri stars. The time and mass dependences are
fitted power-law functions with indices $\eta_a \sim 1.0-1.4$ and
$\eta_b \sim 1.3-2.0$. 

 For  individual stars, we also introduce a Gaussian distribution function 
\begin{equation} 
\begin{split}
\mathrm{dN} / \mathrm{d {\dot M}}_g   = & A_0 \   
{\rm exp}\left[ - \left( {{\rm log}{\dot M}_g - {\rm log}{\dot M}_a \over
\Delta_{{\dot M}_a}} \right)^2  \right] \\
& = A_0 \
{\rm exp}\left[ - \left( {{\rm log}({\dot M}_g/{\dot M}_a) \over 
\Delta_{{\dot M}_a}} \right)^2   \right] 
\label{eq:gaussian}
\end{split}
\end{equation}
where $ \Delta_{{\dot M}_a} = {\rm log}
(\Delta {\dot M}_a/{\dot M}_a)$ and 
$A_0$ is a normalized factor.  Based on this expression,
we can determine  that the fraction of stars ($\eta_{\dot M}$) with a 
given age and mass has ${\dot M}_g$ larger than some fiducial 
value ${\dot M}_f (M_\ast, Z_\ast)$ such that 
\begin{equation}
\eta_{\dot M} ({\dot M}_f, M_{\ast})= \frac{N({\dot M}_g> {\dot M}_f )}{N_{\rm tot}}= 
 \frac{1}{2} \mathrm {erfc} \left(   \frac
 {    {\rm log} {\dot M}_f / 
{\dot M}_a (M_{\ast})   }  
{  \Delta_{{\dot M}_a}}
 \right).
\label{eq:etadotm}
\end{equation}

\begin{figure*}[htbp]
\includegraphics[width=0.48\linewidth,clip=true]{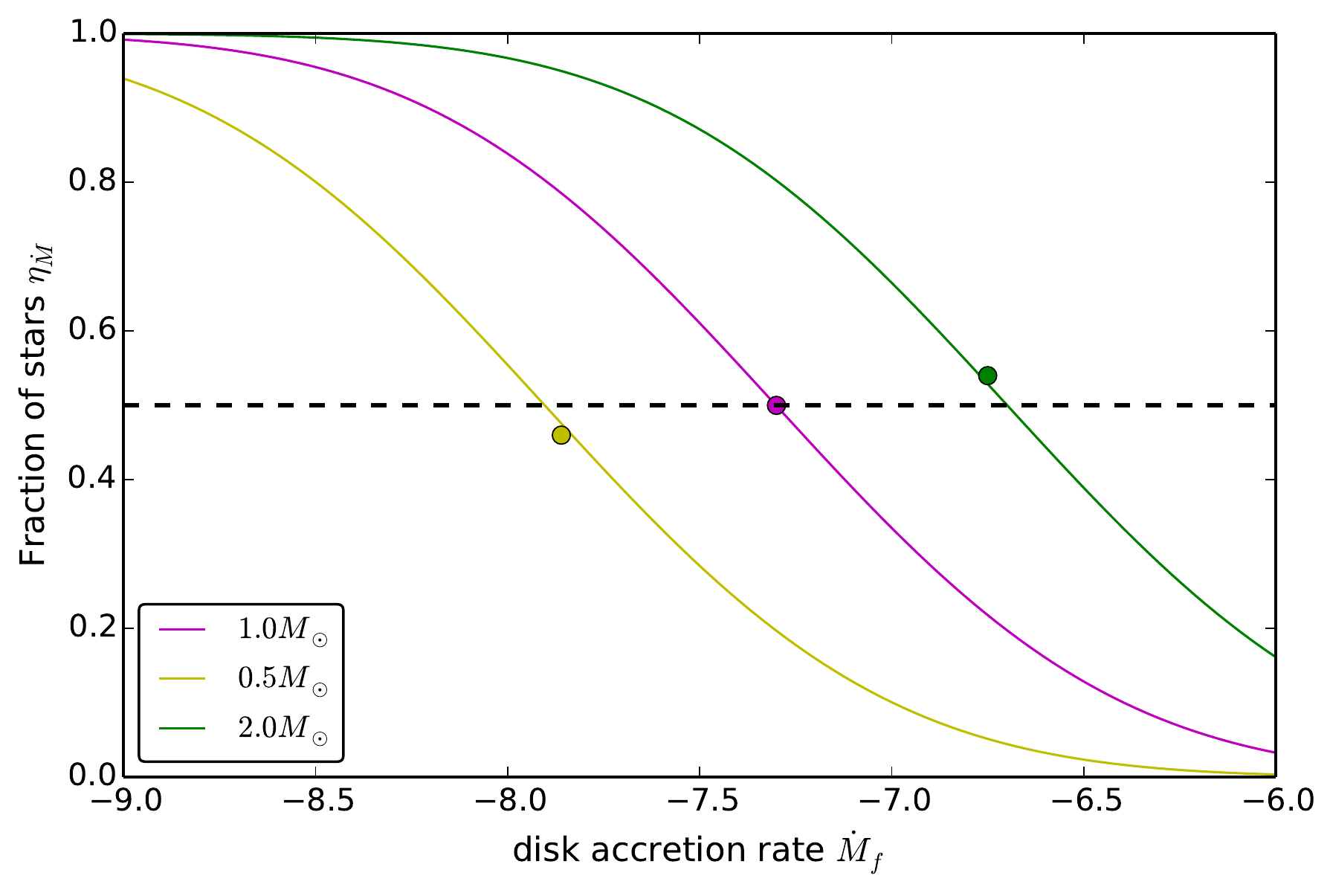}
\includegraphics[width=0.5\linewidth,clip=true]{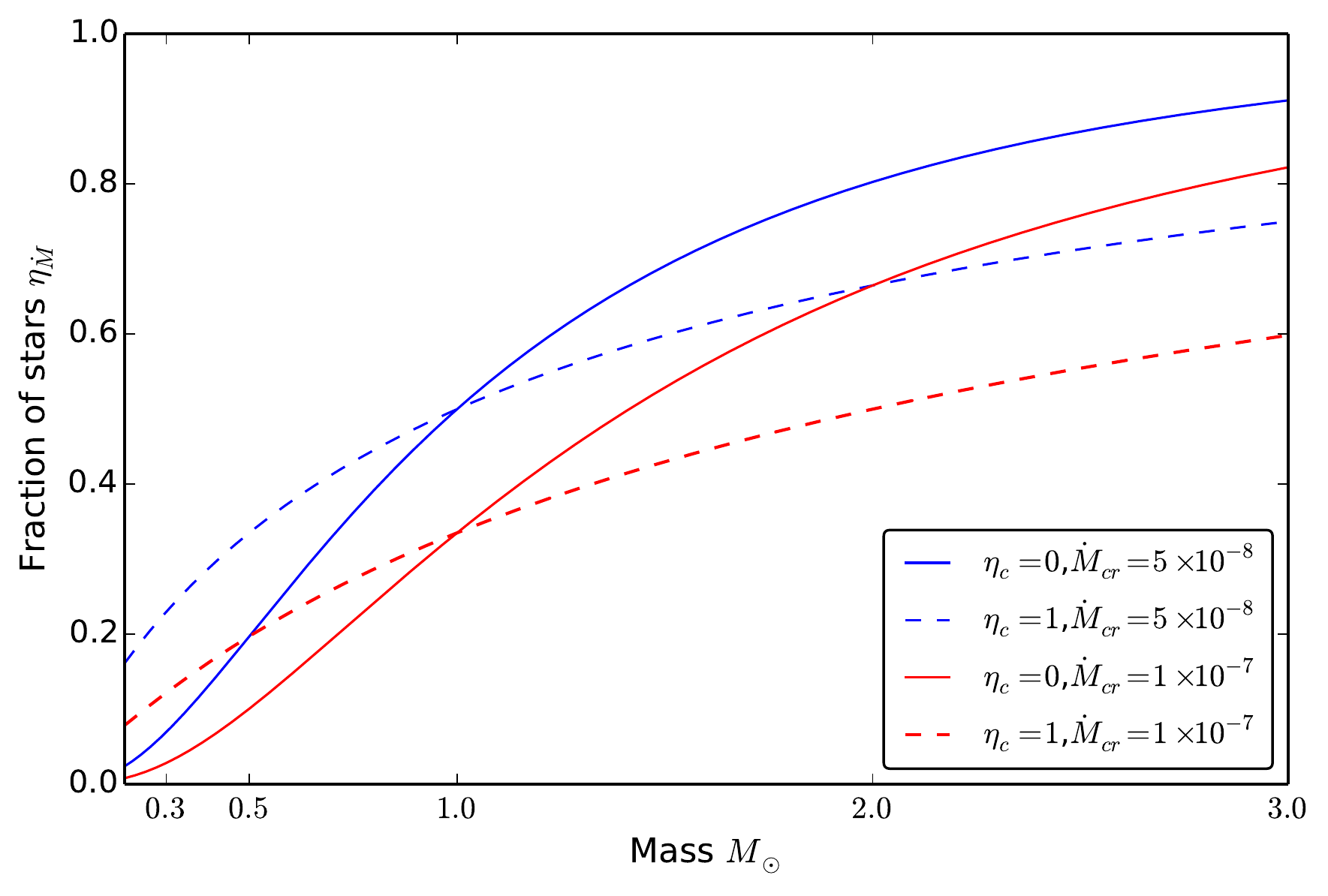}
\caption{
{\bf Left}: The $\eta_{\dot M}$-${\dot M}_g$ correlation for three 
different stellar masses (yellow, purple, and green lines for $0.5$, $1$, and $2 M_\odot$, respectively). The black 
dashed line represents $\eta_{\dot M}=0.5$,  and three colored dots represent the average value of the  observed accretion rate for different stellar masses ($M_{\ast} = 0.5, 1$ and $2 \ M_{\odot}$) in \cite{GarciaLopez-2006}'s work.  
\ \
{\bf Right}: $\eta_{\dot M}-M_\ast$ correlation for different 
$\eta_c$ (solid and dashed lines for $\eta_c=0$ and $\eta_c=1 $
respectively). The red and blue colors correspond to  ${\dot M}_{cr}  
= 1 \times 10^{-7} M_\odot$ yr$^{-1}$
and $5 \times 10^{-8} M_\odot$ yr$^{-1}$, respectively. For solar-type stars, $\eta_{\dot M} \sim 30\%-50\%$ as an upper limit  for $\eta_J$.  
}
\label{fig1}
\end{figure*}

The  expression of equation (\ref{eq:mdota}) is for stars with age comparable to or longer than 
the disk depletion time scale $\tau_{\rm dep} (\sim 3-5$ Myr).
We are mostly interested in the active phase II of
disk evolution around classical T Tauri stars when
${\dot M}_a$ is relatively large.  Within $\sim 
\tau_{\rm dep}/2$, we can neglect the time dependence
in ${\dot M}_a$ so that equation (\ref{eq:mdota}) 
reduces to ${\dot M}_a = {\dot M}_{a \odot} m^{\eta_b}$. 
For illustration, we choose ${\dot M}_{a \odot}
= 5 \times 10^{-8} M_\odot$ yr$^{-1}$ 
$\Delta_{{\dot M}_a} = 1$  (from Figure 3 of \citealp{GarciaLopez-2006}), 
and  $\eta_b=2$. The $\eta_{\dot M}$-${\dot M}_g$ 
correlations are shown 
in the left panel of Figure \ref{fig1} for three different $M_\ast (=0.5, 1, 2M_\odot)$.  

In the next three sections, we determine the threshold disk accretion
rate ${\dot M}_{f} (M_\ast, Z_d)$ above which embryos merge into retainable
cores.  For illustrative purposes, we introduce an $\eta_{\dot M}$-$M_\ast$ 
correlation with an idealized power-law $m_\ast$ dependence such that
\begin{equation}
{\dot M}_{f} (M_\ast) = {\dot M}_{cr} (M_\odot) m_\ast ^{\eta_c}
\label{eq:mdotcf}
\end{equation} 
where the normalization factor ${\dot M}_{cr} 
(M_\odot) \sim 10^{-8}-10^{-7} \ M_\odot$ yr$^{-1}$ for 
$\alpha_\nu=10^{-3}$ (Paper II). 
Although equation (\ref{eq:mdotcf}) takes into account
 the intrinsic dispersion in ${\dot M}_g$ for different 
values of $M_\ast$, we have neglected the $Z_d$ dependence
due to the lack of any direct measurement of $Z_d$ and 
$\Sigma_d$.  It is customary to assume $Z_d = Z_\ast$
because all of their contents were accreted onto the central
stars through protostellar disks. In \S\ref{sec:metaldep}, we 
list some physical effects, which may introduce dispersions to 
the $Z_d-Z_\ast$ correlation.  This dispersion is incorporated
with a general expression for ${\dot M}_{f} (M_\ast, Z_d)$
(as ${\dot m}_{\rm 9 \ res}$ in Eq. \ref{eq:criticalmdot2}) 
in the evaluation of $\eta_J (M_\ast, Z_\ast)$.
 
In the right panel of Figure \ref{fig1}, we plot the $\eta_{\dot M}$-$M_\ast$
diagram with different $\eta_c$   (solid line for $\eta_c=0$ and dashed 
line for $\eta_c=1 $). The red and blue colors correspond to ${\dot M}_{c0}  
= 1\times 10^{-7} M_\odot$ yr$^{-1}$ and $5\times 10^{-8} M_\odot$ 
yr$^{-1}$, respectively.  When considering the range of ${\dot M}_{cr}$ 
for $M_\ast= 1M_\odot$,  we obtain  an $\eta_{\dot M}$ of 
$ \sim 0.3-0.5$. 

The quantity $\eta_{\dot M}$ indicates the fraction of stars
around which cores may form with $M_p > M_c$.  Some of the 
cores may have $M_p > M_{\rm opt} + \Delta M$    such that their 
corotation torque would be saturated (weakened). Unless they
can significantly modify the $\Sigma_g$ distribution near 
their orbit (i.e., open up gaps), such massive 
cores would migrate toward and be consumed by their host 
stars. In this regard, the magnitude of $\eta_{\dot M}$ 
should be considered as an upper limit for $\eta_J$.  

\section{Migration and Growth from embryos to gas giants }
\label{sec:migrategrowth}

In this section, we construct an analytic expression for the 
critical accretion rate ${\dot M}_f (M_\ast, Z_\ast)$ above which 
embryos can bypass the MMR barriers and merge into cores.  
We normalize the disk parameters (see Paper II) ${\dot m}_9 
\equiv {\dot M}_g/10^{-9} M_\odot$ yr$^{-1}$, 
$m_\ast \equiv M_\ast/M_\odot$, $l_\ast \equiv L_\ast/L_\odot$,  
$r_{\rm AU} \equiv r/{\rm AU}$, $\alpha_3 \equiv \alpha_\nu/10^{-3}$
where $L_\ast$ and $L_\odot$ are the stellar and solar luminosity,
respectively, the opacity is $\kappa= 0.02 {\kappa_0} T_g$ representing
grain abundance similar to that of protostellar disks, 
$P = m_\ast^{-1/2} r_{\rm AU}^{3/2}$ yr is the orbital 
period,  $Z_d$ is the disk-metallicity enhancement factor 
relative to the Sun and  $\eta_{\rm ice}$ (1 inside the ice line
and 4 outside it) represents the enhancement factor for the 
condensed-ice contribution to the planetesimal building-block material.

\subsection{Embryos' Isolation Mass and Type I Migration Timescale}

With a feeding zone width of $\sim 10 R_R$, 
the embryos attain their isolation mass  
\begin{equation}
M_{\rm iso} \simeq 0.16   
\left( \Sigma_d \over 10 {\rm \ g cm^{-2}}\right)^{3/2} 
m_\ast^{-1/2} r_{AU}^3 M_{\oplus}
\label{eq:iso}
\end{equation}
on a growth time scale
\begin{equation}
\begin{split}
\tau_{\rm c, acc}  
\simeq 
\left( {r \over R_p} \right) \left({ M_p \over \Sigma_d r^2} 
\right) \left( {M_p \over M_\ast} \right)^{-1/3} P
\label{eq:taucin}
\end{split}
\end{equation}
(see Eq (5) and (20)  in \citealp{Ida-2004a}).

In the inner viscously heated region of the disk 
$(r < r_{\rm trans})$, the isolation mass is 
\begin{equation}
\begin{split}
M_{\rm iso <}  \simeq 5 \times 10^{3 Z_d/2-3} \eta_{\rm ice} ^{3/2} 
\alpha_3 ^{-9/8} m_\ast^{-5/16} {\dot m}_{9}^{3/4} 
\kappa_0^{-3/8} r_{\rm AU}^{39/16} \ M_\oplus 
\label{eq:misoin}
\end{split}
\end{equation}
(this expression corrects a normalization error in Equation 25, Paper II).
The outer irradiated region of the disk $(r > r_{\rm trans})$
is generally outside the snow line (GL07) such that
$\eta_{\rm ice}=4$ and 
\begin{equation}
\begin{split}
M_{\rm iso >} 
= 10^{3 Z_d/2-2} m_\ast^{13/28} {\dot m}_{9} ^{3/2} l_\ast^{-3/7}
\alpha_3^{-3/2} r_{\rm AU}^{39/28} \ M_\oplus 
\label{eq:misoout}
\end{split}
\end{equation}
is relatively large due to the condensation of the volatile ices.

If the embryos' size $R_p \simeq 
(M_p/M_\oplus)^{1/2.06} \ R_\oplus$ \citep{Lissauer-2011}, 
$\tau_{\rm c, acc}$ would be a weakly increasing function of 
$M_p$ but strongly correlated with $r$.  
In the inner disk region 
$\tau_{\rm c, acc} (M_{\rm iso}, r < r_{\rm trans}) 
\propto r^{7/8} < <  \tau_{\rm dep} $,
and beyond the transition radius 
$ \tau_{\rm c, acc} (M_{\rm iso}, 
r > r_{\rm trans}) \propto  r^{11/7} \sim  \tau_{\rm dep}$.
In these regions, embryos acquire their isolation mass.  But at
very large disk radii ({\it e.g.}, $r > 10$ AU), $\tau_{\rm c, acc} (M_{\rm iso}) 
> \tau_{\rm dep}$, so that their growth to dynamical isolation may 
not be completed before disk gas is depleted \citep{Ida-2004a}. 

Recent models also suggest that the embryos could form inside-out at either the inner edge of the dead zone \citep{Chatterjee-2014} or the magnetic cavity boundary \citep{Li-2016}. At these locations, gas pressure in the disk attains local maxima and its azimuthal speed reaches the local Keplerian value.  Consequently, pebbles are stalled there as their orbital decay due to aerodynamic drag from upstream (\textit{i.e.}, at large radii).
Although disruptive impacts provide growth barriers for meter-size pebbles, 
collisional fragments remain in the proximity of these specific  radii.  The accumulation of these building blocks eventually leads to gravitational instability and formation of planetesimals.  Subsequent cohesive mergers lead to the emergence of embryos.  When their masses increase to several $M_{\oplus}$, they undergo rapid 
outward type I migration.  Through this mechanism, we anticipate the rapid formation of a population of super-Earth embryos.  
We still adopt the isolation mass formula from \cite{Ida-2004a} in the following analytical approach; the  $M_{\rm iso}-m_{\ast}$ correlation is robust, but the $M_{\rm iso} - \Sigma_d$ relation may not be  quantitatively valid any more. Nevertheless, $M_{\rm iso} $ approximately equals to  $ M_{\rm opt}$ within a reasonable range of parameters.

While the formation of embryos is discussed elsewhere, we focus here on their tidal interaction with the disk during their growth.
From Equation (\ref{eq:adot}), the type I migration time scale is given by
\begin{equation}
\tau_{I}\equiv {r_p \over {\dot a}} = \left(h^2\over 2 f_a \right)
\left(M_{\ast} \over \Sigma_g r^2 \right)  \left(M_{\ast} \over 
M_p\right) \left(P\over 2 \pi \right) yr
\label{eq:taui}
\end{equation} 
where $h=H/r$ is the aspect ratio and $H$ is the thickness of the disk. In 
the viscously heated inner region,  embryos undergo outward type I migrate on a time scale
\begin{equation}
\begin{split}
& \tau_{I<} (M_p)  \simeq ({0.62 \ M_\oplus /  f_a M_p}) m_\ast^{3/4}. 
\alpha_3^{1/2} \kappa_0^{1/2} r_{\rm AU}^{-1/4} {\rm Myr} 
\\
& \simeq \left({123 M_{\rm iso} \over f_a M_p} \right) 
10^{-3 Z_d/2} \eta_{\rm ice} ^{-3/2} \alpha_3 ^{13/8} 
m_\ast^{17/16} \kappa_0 ^{7/8} {\dot m}_{9} ^{-3/4} 
r_{\rm AU} ^{-43/16} {\rm Myr}.\\
\label{eq:tauia}
\end{split}
\end{equation}

Embryos grow and reach their local isolation mass provided that  $\tau_{c, acc} (M_{\rm iso}) < \tau_{I<} (M_{\rm iso})$.  But the isolation mass increases with $\Sigma_d$ as more pebbles accumulate.  When embryos attain masses $M_p$ with 
$\tau_{c, acc} (M_{\rm iso})> \tau_{I>} (M_{\rm iso})$,  they would migrate outward.  

Embryos may also form in special disk locations far from their central stars such as the condensation/evaporation boundary \citep{Cuzzi-2004, Kretke-2007, Ros-2013}.  In the outer irradiated region embryos undergo inward type I migration on a time scale
\begin{equation}
\begin{split}
\tau_{I>} (M_p) \simeq 
({3.97 /  f_a} ) ( {M_\oplus / M_p} ) 
m_\ast^{-2/7} {\dot m}_{9}^{-1} 
l_\ast^{4/7} \alpha_3 r_{\rm AU}^{8/7} {\rm Myr} 
\\
\simeq ({397 / f_a}) ({M_{\rm iso} / M_p} ) 
10^{-3 Z_d/2} \alpha_3 ^{5/2} m_\ast^{-3/4} l_\ast {\dot m}_{9}^{-5/2}
r_{\rm AU} ^{-1/4} {\rm Myr}.
\label{eq:tauib}
\end{split}
\end{equation}
In this paper, we consider both possibilities by placing an initial convoy of super-Earth embryos in the inner viscously heated and outer irradiated regions. 
They migrate extensively and may  get trapped at the trapping radius separated by different heating sources. Regardless of their initial spacing, the convergent migration  eventually  gathers them toward to relatively small separation nearly the trapping radius. We could assume a range of values and start our simulations at  that typical time (this  is not the time embryos formed but the time embryos approach close to $r_{trap}$).    We will demonstrate the effect of different  separations on the  final outcome in  section 4.2.

\subsection{From Embryos to Cores}
\label{sec:2cores}

The migration time scales (Eq. \ref{eq:tauia} and  \ref{eq:tauib}) 
are determined by the sum of differential Lindblad and corotation 
torque through the magnitude of the efficiency factor $f_a$ (PBK10, 11).  
In the irradiated outer region, both torques lead to inward migration.  
In the viscously heated inner region, embryos with 
\begin{equation}
M_{\rm opt} \simeq m_\ast^{13/48} {\dot M}_{9}^{7/12} 
\alpha_3^{3/8} \kappa_0^{7/24} r_{\rm AU}^{-7/48} \ M_\oplus
\label{opt}
\end{equation}
experience a maximum-strength, unsaturated corotation torque 
that is stronger than the differential Lindblad torque and it
induces to an outward migration (see Eq [16] in Paper II). 
The corotation torque for embryos with $M_p < < M_{\rm opt}$ 
or $M_p > > M_{\rm opt}$ is saturated (weakened) and falls 
below the differential Lindblad torque.  The ratio between 
the upper and lower mass range for outward migration is 
$\sim (2h)^{-2/3}$ which is a few.  This mass range would be 
more extended if the disk had a more complicated layer 
structure \citep{Kretke-2012}.  

Embryos with $M_p \sim M_{\rm opt}$ migrate and converge toward 
the trapping radius located at
\begin{equation}
r_{\rm trans} \simeq 0.26 m_\ast^{0.74} 
l_\ast^{-0.41} {\dot m}_{9}^{0.72}
\alpha_3^{-0.36} \kappa_0^{0.36} {\rm AU}
\label{eq:trans1}
\end{equation}
(see Eq. 14 in Paper II) on time scales $\tau_{I<}(M_{\rm opt})$ 
(from the inner region) or $\tau_{I>} (M_{\rm opt})$ (from the 
outer region) (see Eq. \ref{eq:tauia} and  \ref{eq:tauib}).

The local isolation mass around the transition radius may 
be substantially enhanced (with a relatively large $Z_d$) 
by a local concentration of trapped building-block embryos 
\citep{Liu-2015}.  We also showed (in Eq. 32, 
Paper II) that convergent embryos in disks with 
${\dot m}_{9}<{\dot m}_{9 \ {\rm res}}$ 
where the critical magnitude is
\begin{equation}
{\dot m}_{9 \ {\rm res}} \simeq 6 f_{\rm res}^{0.95} 
m_\ast^{-1.33} \alpha_3^{0.97} \kappa_0^{-0.026} l_\ast^{0.70},
\label{eq:criticalmdot}
\end{equation}
tend to trap each other on their mutual MMRs 
and form a convoy of super-Earths.  The magnitude of a dimensionless 
coefficient $f_{\rm res}$ is of order a few, and it is smaller 
for 3:2 than 2:1 MMR (Papers I \& II).     
In disks with ${\dot m}_{9} > {\dot m}_{9 \ {\rm res}}$,
embryos can bypass their MMR barrier, enter into each other's 
feeding zone, undergo close encounters and coalesce.  

Convergent migration to $r_{\rm trans}$ enhances the local concentration
of building-block material.  Provided that $\tau_{\rm c, acc} (M_{\rm opt}) $ is substantially smaller 
than $\tau_{\rm dep}$, the trapped embryos grow to the optimum mass 
at $r_{\rm trans}$ with 
\begin{equation}
M_{\rm opt} (r_{\rm trans}) \simeq  1.2 m_\ast^{0.16} 
{\dot m}_{9} ^{0.48} \alpha_3^{0.43} \kappa_0^{0.24} 
l_\ast^{0.06} \ M_\oplus.
\label{eq:mopttrans}
\end{equation}
Embryos' growth beyond a few times $M_{\rm opt} (r_{\rm trans})$ 
would lead to corotation torque saturation and orbit decay. 
If $M_{\rm opt} (r_{\rm trans})  > M_c$, large cores that  emerged 
out from convergent embryos can also retain and accrete gas 
around the trapping radius.

Close encounters between optimum-mass and residual 
embryos trapped at $r_{\rm trans}$ excite their eccentricity and 
widen their semimajor-axis separation. The extent of orbital change 
increases with the ratio between embryos' surface escape 
($V_e$) and the Keplerian ($V_k$) speed at
$r_{\rm trans}$,
\begin{equation}
f_V \equiv V_e (M_{\rm opt}) / V_k (r_{\rm trans})
\label{eq:fvdef}
\end{equation}
From Equation (\ref{eq:trans1}) and (\ref{eq:mopttrans}) we find
\begin{equation}
\begin{split}
f_V \simeq 0.2 m_\ast^{-0.09} {\dot m}_{9}^{0.48} 
\alpha_3^{-0.07} l_\ast^{-0.19}\kappa_0^{0.24}.
\label{eq:fvscat}
\end{split}
\end{equation}

Encounters with a small or moderate value of $f_V (\ll 1)$ do not 
strongly modify the orbits of interacting embryos.  Their 
eccentricity is rapidly damped as they resume their convergent 
type I migration---from a scattered location $r_{\rm scat}$ 
inside and outside $r_{\rm trans}$ to its proximity--- on time scales of
$\tau_{I<}  (M_p, r_{\rm scat} < r_{\rm trans})$ and
$\tau_{I>}  (M_p, r_{\rm scat} > r_{\rm trans})$ where 
\begin{equation}
\begin{aligned}
& \tau_{I<} \simeq
\left( {0.62 m_\ast^{23/48} \alpha_3^{1/8} \kappa_0^{5/24}
\over f_a {\dot m}_{9}^{7/12}} \right)
\left({ M_{\rm opt} \over M_p} \right)
r_{\rm scat, AU}^{-5/48}  {\rm \  Myr}.
\label{eq:tauiin}
\end{aligned}
\end{equation}
\begin{equation}
\begin{aligned}
& \tau_{I>} \simeq
\left( { 4 m_\ast^{-187/336} \alpha_3^{5/8}  l_\ast^{4/7} \over
f_a {\dot m}_{9}^{19/12}} \right)
\left({ M_{\rm opt} \over M_p} \right)
r_{\rm scat, AU}^{1.3}  {\rm \ Myr}
\end{aligned}
\end{equation}
where $r_{\rm scat, AU} =r_{\rm scat}/ 1 \rm AU $.

\subsection{Onset of Gas Accretion}
\label{sec:gasaccret}

In disks with $M_{\rm opt} (r_{\rm trans}) > M_c (\sim 10 \  M_\oplus)$,
efficient gas accretion proceeds on a Kelvin-Helmholtz time scale 
\citep{Pollack-1996, Ikoma-2000, Ida-2004a} 
\begin{equation}
\tau_{\rm KH} 
\simeq \kappa_{\rm acc} (M_{\oplus}/M_p)^3 {\rm Gyr}.
\label{eq:tau_KH} 
 \end{equation}
 
Under the assumption that the opacity of the accreted gas 
$\kappa_{\rm acc} \sim \kappa_0$, 
\begin{equation}
\begin{split}
&\tau_{\rm KH} (M_{\rm 
opt}, r_{\rm trans} ) \simeq 0.58 \kappa_0^{0.28} m_\ast^{-0.48}  {\dot m}_{9} ^{-1.44} \\
& \alpha_3^{-1.29}l_\ast^{-0.18} 
 \left( {M_{\rm opt} \over M_p}
\right)^3 {\rm Gyr}
\label{eq:taukh}
\end{split}
\end{equation}
for $M_p \simeq M_{\rm opt}$ at $r_{\rm trans}$.  
In the limit of ${\dot M}_g =   10^{-7} M_\odot \rm yr^{-1}$,  $M_{\rm 
opt} (r_{\rm trans}) \sim M_c$ and  $ \tau_{\rm KH} \simeq 0.6 {\rm  Myr}$,   
growth beyond $M_{\rm opt}$
would further reduce $\tau_{\rm KH}$ and increase the gas accretion 
rate onto the cores.

\subsection{Core-formation threshold around stars with 
different $M_\ast$ and $Z_\ast$}

In order to determine the $\eta_J$-$M_\ast$  and $\eta_J$-$Z_\ast$  correlations,  we use Equation (\ref{eq:mdota}) to replace ${\dot M}_g$ with a
fiducial average value ${\dot M}_a$ and assume 
that (1) $l_\ast \propto m_\ast^2$; (2) $\eta_b=2$, i.e., ${\dot m} \propto m_\ast^2$;
and (3) $Z_d$ is independent of $M_\ast$.  
Based on  Eq. (\ref{eq:trans1}, \ref{eq:tauiin}, \ref{eq:taukh}, 
\ref{eq:mopttrans}, \ref{eq:criticalmdot} and \ref{eq:fvscat}), 
we summarize here: 
\begin{equation}
\begin{aligned}
M_{\rm iso <}  \simeq  5   {\dot m}_{a\odot9}^{3/4}  
10^{3Z_d/2-3} \eta_{\rm ice}^{3/2} 
m_\ast^{19/16} \kappa_0^{-3/8} r_{\rm AU}^{39/16}  \ M_\oplus,
\end{aligned}
\end{equation}
\begin{equation}
\begin{aligned}
\tau_{I<}(M_{\rm opt}) \simeq 0.62  {\dot m}_{a\odot9}^{-7/12}  
m_\ast^{-33/48} \alpha_3^{1/8} \kappa_0^{5/24} r_{\rm AU}^{-5/48}  {\rm Myr},
\end{aligned}
\end{equation}
\begin{equation}
\begin{aligned}
r_{\rm trans} \simeq 0.26  {\dot m}_{a\odot9}^{0.72}  m_\ast^{1.36} \alpha_3^{-0.36} \kappa_0^{0.36}    {\rm AU}
\label{eq:rtrans2}
\end{aligned}
\end{equation}
\begin{equation}
M_{\rm opt} (r_{\rm trans}) \simeq  1.2  {\dot m}_{a\odot9}^{0.48}  m_\ast^{1.24} 
\alpha_3^{0.43} \kappa_0^{0.24} \ M_\oplus.
\label{eq:moptdiffm2}
\end{equation}
\begin{equation}
{\dot m}_{9 \ {\rm res}} \simeq 6 f_{\rm res}^{0.95} 
m_\ast^{0.07} \alpha_3^{0.97} \kappa_0^{-0.026} 
\label{eq:criticalmdot2}
\end{equation}
\begin{equation}
\tau_{\rm KH} (M_{\rm opt}, r_{\rm trans}) \simeq 0.58  
{\dot m}_{a\odot9}^{-1.44} 
m_\ast^{-3.72} \alpha_3^{-1.29} \kappa_0^{0.28} {\rm Myr}.
\label{eq:taukh2}
\end{equation}
\begin{equation}
f_V \simeq 0.2 m_\ast^{0.49}
\alpha_3^{-0.07} \kappa_0^{0.24} 
{\dot m}_{a\odot9}^{0.48} .
\label{eq:fvscat2}
\end{equation}
where  ${\dot m}_{a\odot9} = {\dot M}_{a \odot} /  
10^{-9} M_\odot$ yr$^{-1}$ and  ${\dot M}_{a \odot}$ 
(average accretion rate for solar-mass protostars) is 
in the range of $\sim 1-5  \times 10^{-8} M_\odot$ 
yr$^{-1}$ during  the T Tauri phase. 

It is often assumed that the amount of heavy elements
in the disk (or $\Sigma_d$) is proportional to $M_\ast$
such that embryos' growth timescale decreases and 
isolation mass increases with $M_\ast$.  The supercritical cores 
are easier to form around relatively massive stars, which is 
in agreement with the $\eta_J$-$M_{\ast}$ correlation. 

However, $M_{\rm iso}$ also increases with $Z_d$ which may be enhanced
by the convergent migration. 
The luminosity of T Tauri stars $L_\ast
\propto M_\ast^2$ {\citep{DAntona-1994}. 
Taking into account of this 
dependence,  these results imply a threshold increase 
in $\eta_J$ for relatively massive stars because

\noindent
1. $r_{\rm trans} \propto m_\ast^{1.36}$ and is beyond the snow line
where volatile ices contribute to both $\eta_{\rm ice}$ and $\kappa_0$
of the building-block material; 

\noindent
2. $M_{\rm iso} \propto m_\ast^{19/16}$ and is larger than $M_c$
so that efficient gas accretion may be initiated;  

\noindent
3. ${\dot m}_{9 \ {\rm res}} \propto m_\ast^{0.07}$ and $  {\dot m}_a \propto m_\ast^2$ so that
the MMR barriers may be bypassed;

\noindent
4. $M_{\rm opt} \propto m_\ast^{1.25}$ and is larger than $M_c$; 

\noindent
5. $\tau_{\rm KH} \propto m_\ast^{-3.72}$ and is comparable to or less than 
$\tau_{\rm dep}$.  

Note that 
 ${\dot m}_{9 \ {\rm res}}$ 
is essentially independent of the stellar mass while ${\dot M}_a 
\propto M_\ast^2$ and  $L_\ast \propto M_\ast^2$, and a larger fraction of embryos around relatively 
massive ($M_\ast  \sim 2 \ M_\odot$) stars are able to bypass the MMR 
barrier and converge onto crossing orbits near $r_{\rm trans}$.  
These embryos undergo close encounters with each other, including large-angle deflection and physical collisions.  
Scattered embryos repeatedly
return to $r_{\rm trans}$ until eventually they collide and merge 
into supercritical cores. 
Since $M_{\rm opt} (r_{\rm trans}) > M_c (\sim 
10 \ M_\odot)$ and $\tau_{\rm KH}  (M_{\rm opt}, r_{\rm trans}) $ decreases rapidly with stellar mass, 
rapid gas accretion onto relatively massive retained cores promotes
the formation of gas giants around intermediate-mass and massive stars.

In disks with ${\dot M}_g \sim {\dot M}_a$, changes in the orbital 
properties of the scattered embryos due to close encounters 
at $r_{\rm trans}$ increase with the mass of the host stars.
Equation (\ref{eq:fvscat}) implies that $f_V \propto m_\ast$.  
In the large $f_V ( \geq 1)$ limit, most embryos are either ejected 
or scattered to distances far well beyond $r_{\rm trans}$ where 
$\Sigma_g$ is low and $\tau_{I>}$ becomes longer than 
$\tau_{\rm dep}$ (Eq. \ref{eq:tauib}).  This consideration 
introduces the possibility of transporting cores to large 
distances and the formation of long-period gas giants.
We defer further discussion on multiple-planet formation in 
evolving disks to subsequent papers in this series.


\begin{figure*}[htbp]
\includegraphics[width=0.35\linewidth,clip=true]{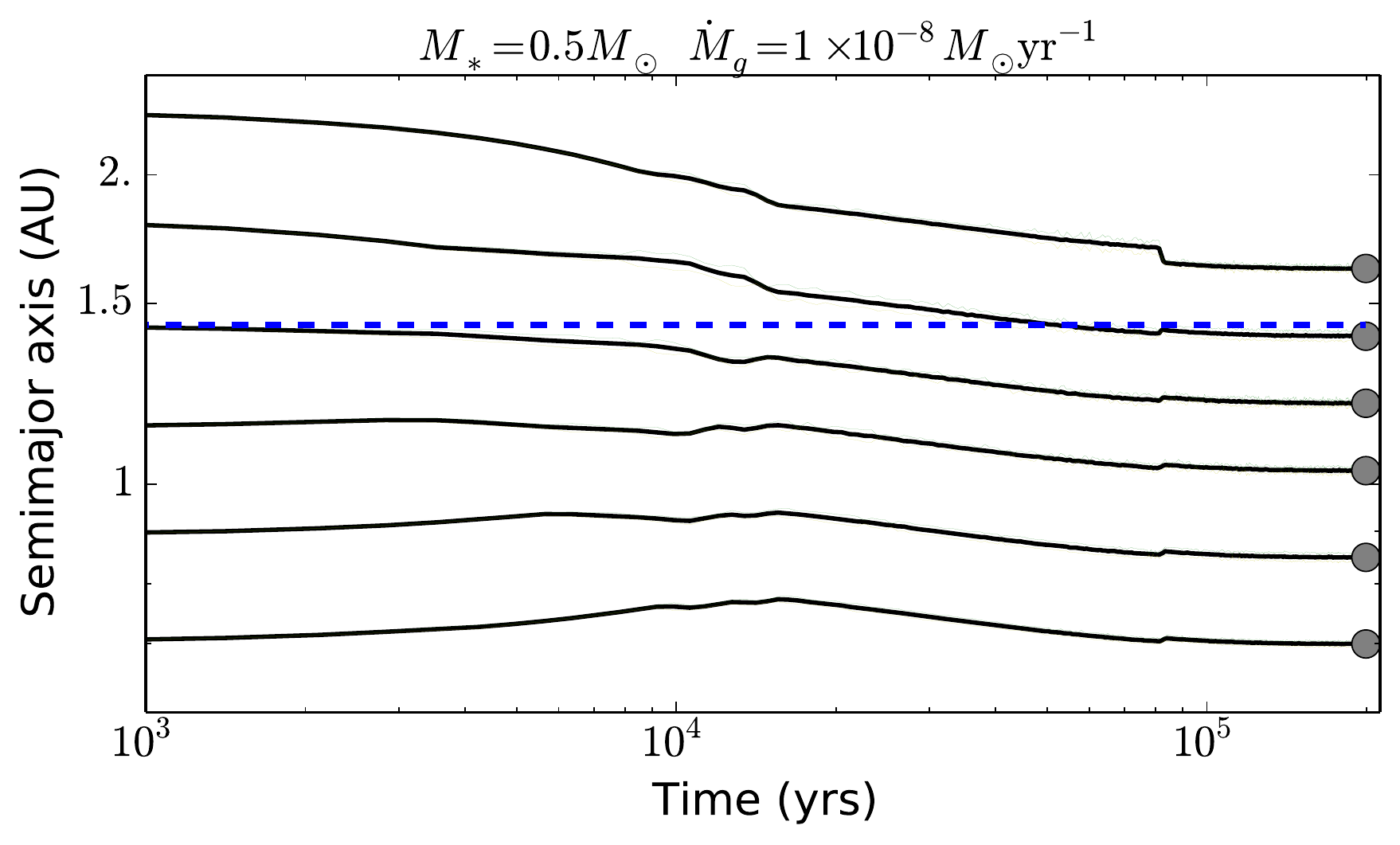}
\includegraphics[width=0.35\linewidth,clip=true]{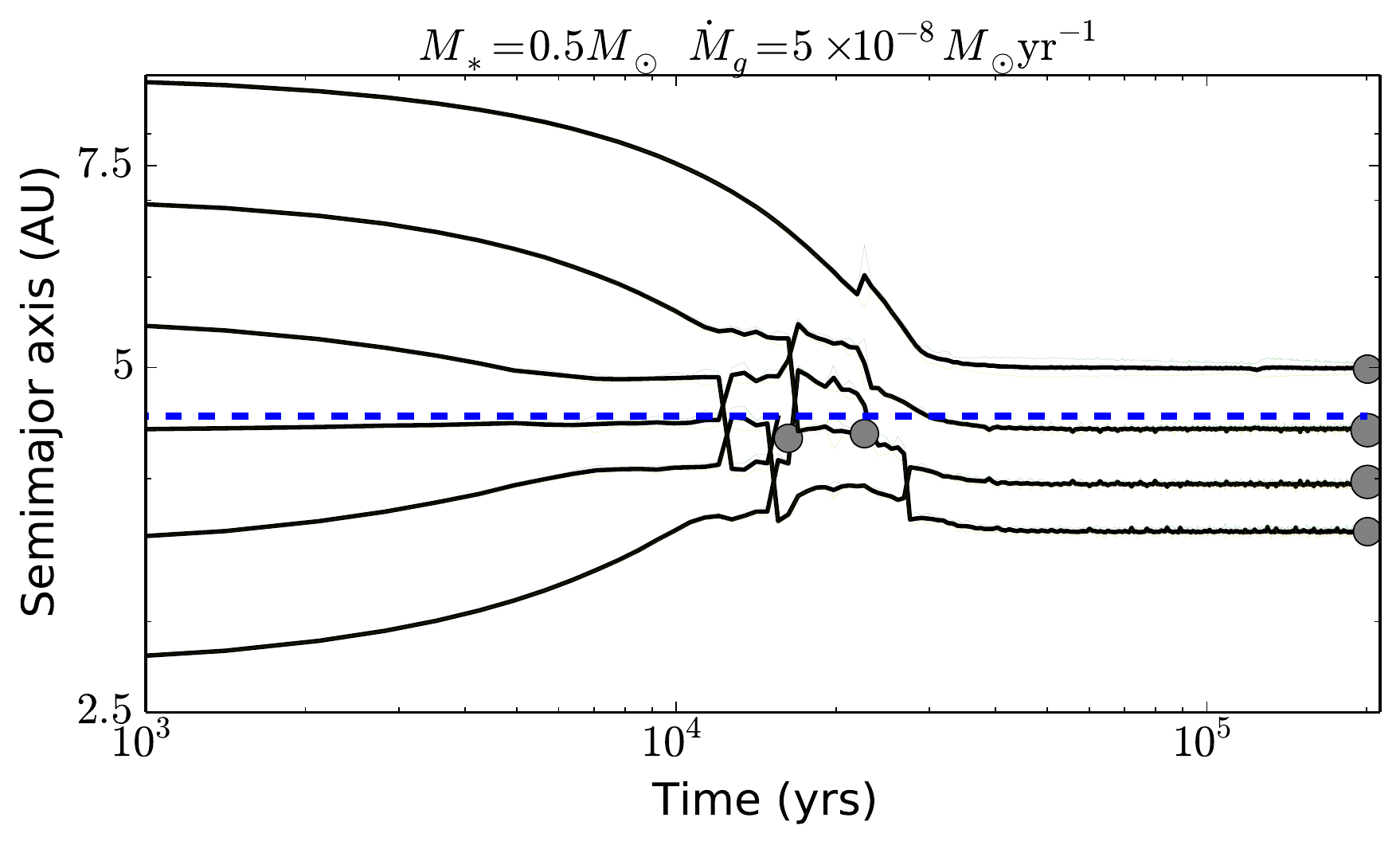}
\includegraphics[width=0.35\linewidth,clip=true]{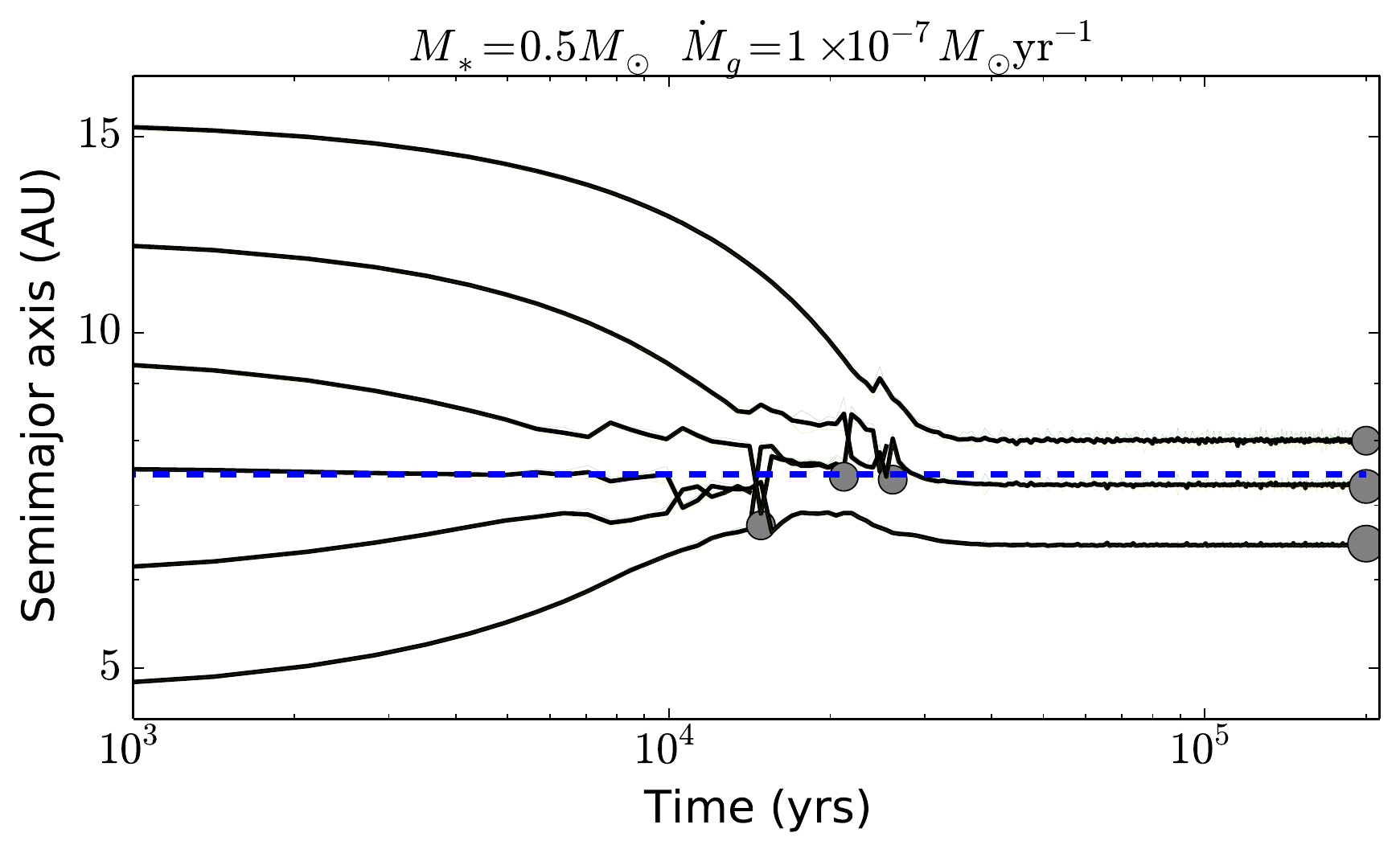}
\includegraphics[width=0.35\linewidth,clip=true]{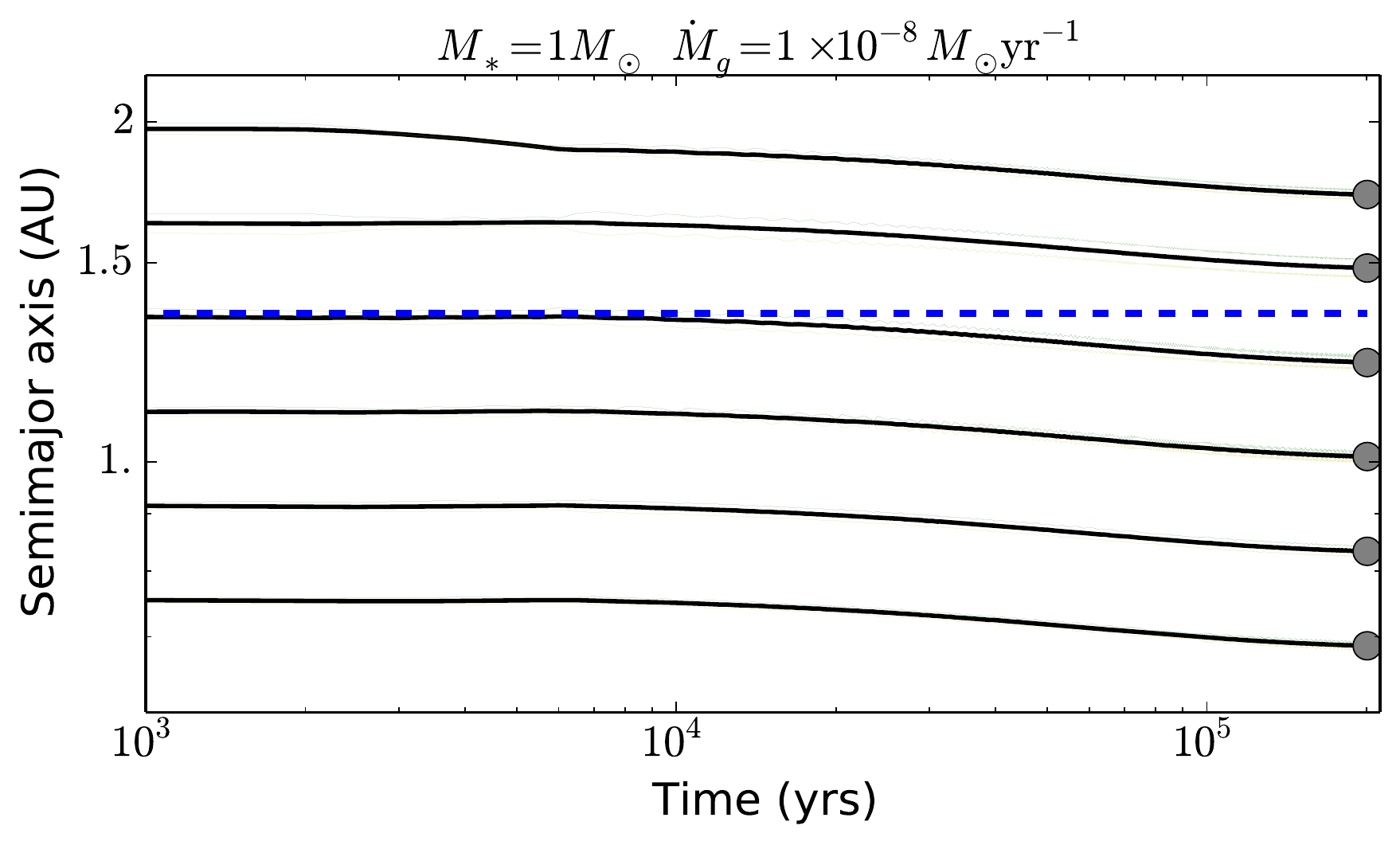}
\includegraphics[width=0.35\linewidth,clip=true]{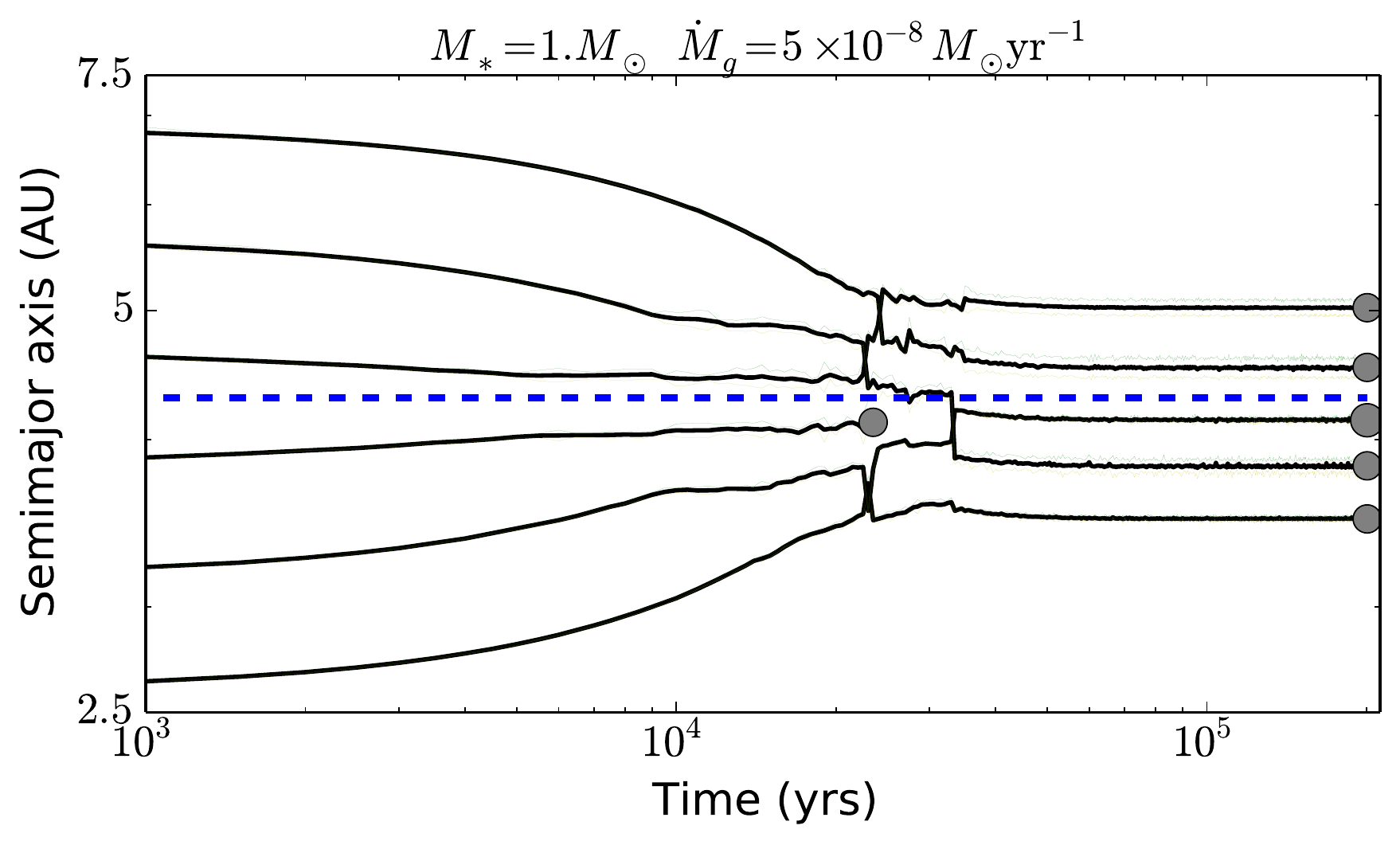}
\includegraphics[width=0.35\linewidth,clip=true]{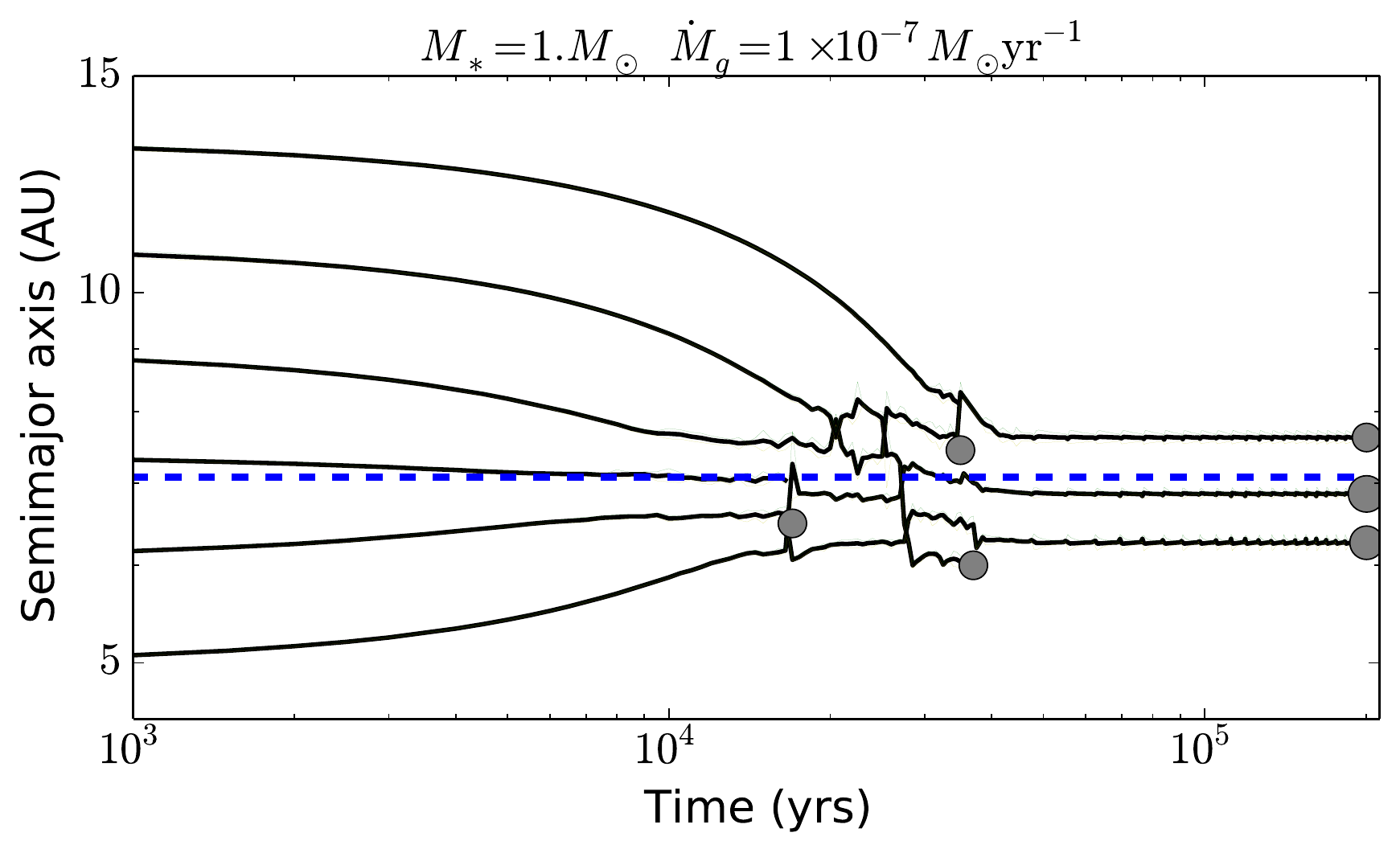}
\includegraphics[width=0.35\linewidth,clip=true]{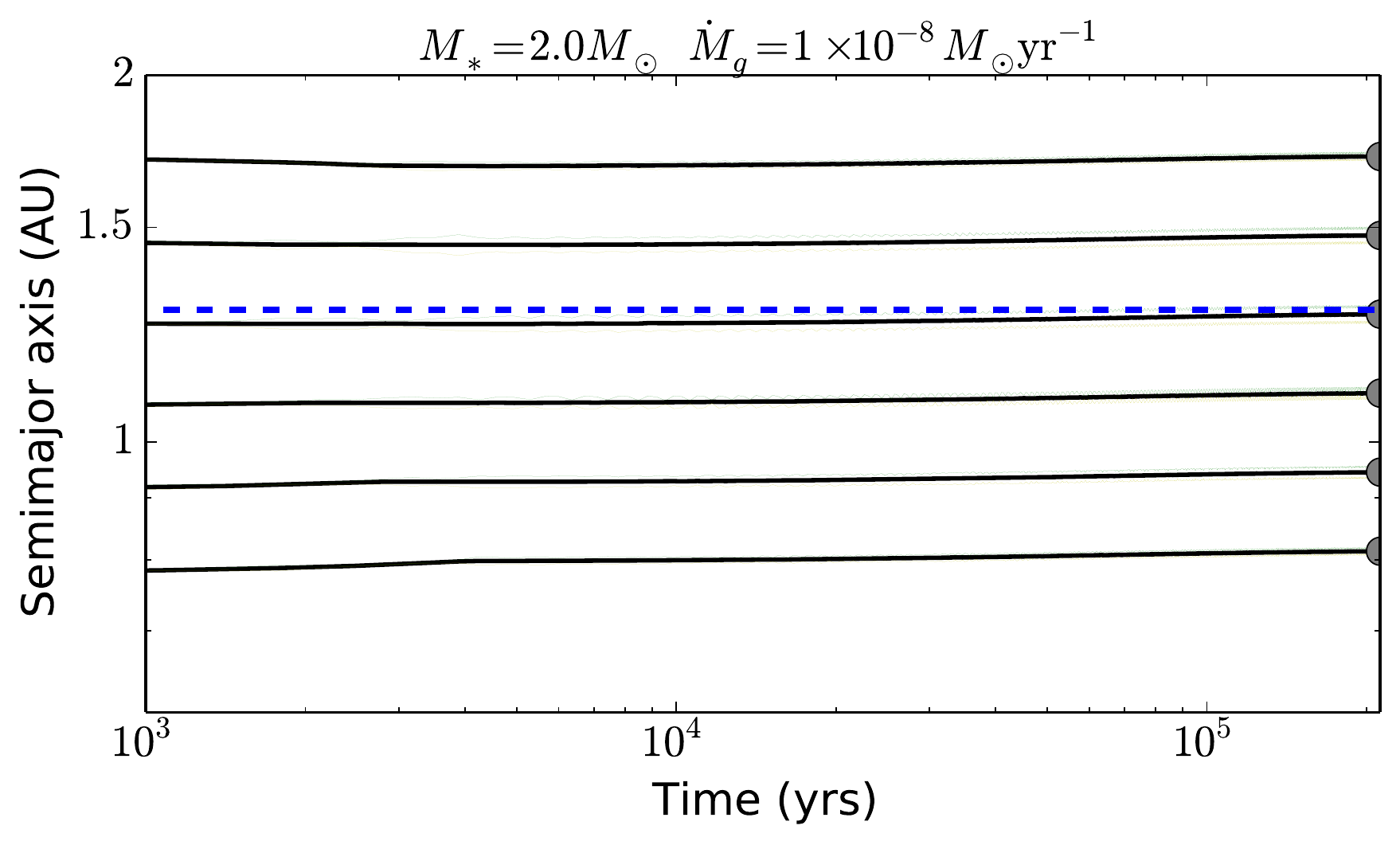}
\includegraphics[width=0.35\linewidth,clip=true]{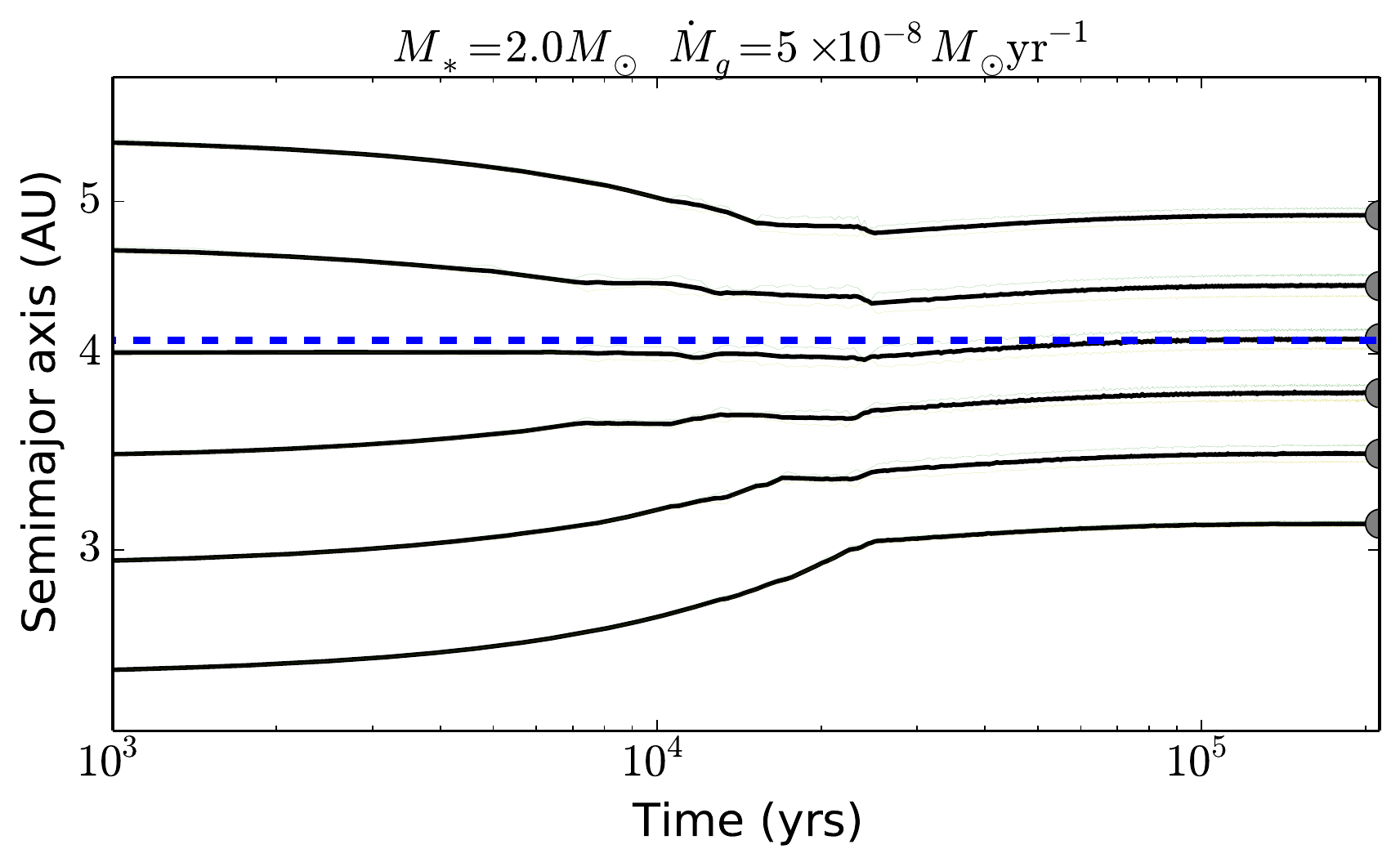}
\includegraphics[width=0.35\linewidth,clip=true]{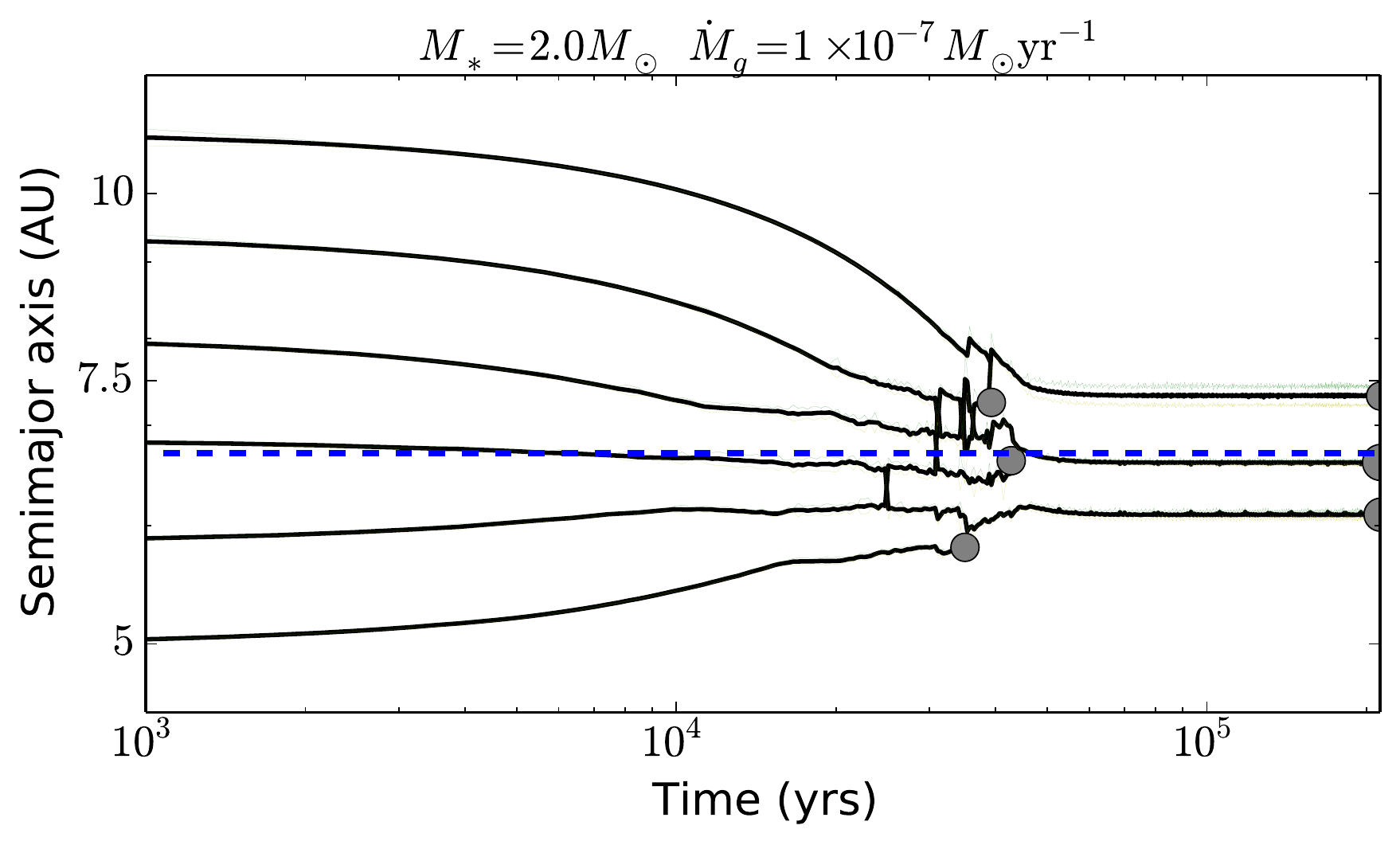}
\caption{
 Mutual interaction between embryos and their natal disks for 
different stellar masses $M_\ast$ and disk accretion rates ${\dot M}_g$. 
The black lines trace the evolution of embryos' semimajor axes 
and the blue dashed line indicates the location of $r_{\rm trap}$. 
The green and yellow lines are embryos' apocenter and pericenter 
distance. Disk parameters are chosen to be those of model A. 
All systems contain six $5 \ M_{\oplus}$ embryos that  are initially 
distributed  on either side of the trapping radius with $10R_R$ separation.  
The accretion rate (${\dot M}_g =   10^{-8} M_\odot \rm yr^{-1}$, 
$ 5 \times 10^{-8} M_\odot \rm yr^{-1}$ and $ 10^{-7} M_\odot \rm yr^{-1}$) 
increases from left to right panels. The stellar mass  ($M_{\ast} 
= 0.5, 1 $ and $2 \ M_{\odot}$) increases from top to 
bottom panels. 
}
\label{xxx}
\end{figure*}

\begin{table*}[!ht]
\centering
\caption{Models with a range of stellar mass and disk accretion rate}
\begin{tabular}{|c|c|c|c|c|}
\hline
\hline
Models & Mass of the Planets  ($M_\oplus$) &  Number of Planets  & 
Initial Separation  $k_0$ ($R_{R}$) & Opacity $\kappa_0$  \\
\hline
Series 1    &   5  & 6  & 7 &1 \\
Series 2    &   5  & 6  & 10 &1 \\
Series 3    &   5  & 6  & 12 &1 \\
Series 4    &   5  & 6  & 15 &1 \\
Series 5    &   3  & 10  & 10 & 1\\
\hline
\end{tabular}

\begin{tabular}{|c|c|c|c|}
\hline
 Model & Stellar Mass  ($M_{\odot}$)  & 
Accretion Rate ${\dot M}$ ($M_\odot$ yr$^{-1}$)   \\
\hline
Model A &  2.0    &     $10^{-8}$    \\
Model B &  2.0    &     $5 \times 10^{-8}$    \\
Model C &  2.0    &     $10^{-7}$   \\
Model D &  1.0    &     $10^{-8}$    \\
Model E &  1.0    &     $5 \times 10^{-8}$     \\
Model F &  1.0    &     $10^{-7}$   \\
Model G &  0.5    &     $10^{-8}$  \\
Model H &  0.5    &     $5 \times 10^{-8}$    \\
Model J &  0.5    &     $ 10^{-7}$   \\
\hline
\end{tabular} 
\label{tab1}
\end {table*}

\section{Formation of critical cores among different stellar mass systems}
\subsection{standard case}
\label{standard case}

In this section, we present numerical models to verify the analytic results
presented in the previous section.  We investigate the role of disk 
accretion rate (${\dot M}_g$) and stellar mass ($M_\ast$) on assembling 
of supercritical cores with our Hermite-Embryo code.

In principle, the disk accretion rates may be extraordinarily high  and $M_{\rm opt} \gg 10 \ M_\oplus$ during the brief ($\sim 10^5$ yr) embedded phase when their central stars acquire most of their masses.  However, if disks have retained a large fraction 
of their initial angular momentum content  and  the disk wind is 
not the dominant mechanism for mass and angular momentum removal, 
the Class I  phase disks would probably be much more compact and hotter than 
typical protoplanetary disks around T Tauri stars. Although grains may condense 
in the outer region of rapidly evolving  disks with very high ${\dot M}_g$,
 it is not clear whether there is adequate time for them to 
coagulate and grow into embryos with isolation masses of  a few $ M_\oplus$. 
 
 The magnitude of ${\dot M}_g$
decreases with time, and that of the corresponding  
optimum mass ($M_{\rm opt}$) for the trapped embryos (Eq. \ref{eq:moptdiffm2}) is well 
below $M_c$ in the transitional or debris disks.  Therefore, the necessary 
condition for core formation ({\it i.e.}, the magnitude of $\eta_J$) is 
during the early Class II stage of disk evolution. 
Following  Paper II, we start our simulations when the average accretion rate ${\dot M}_a$ is comparable to that of classical T Tauri stars.

We limit model parameters to those in Table \ref{tab1}. 
For simplicity, equal-mass embryos are treated here with $M_p$ in the
range that is consistent with the analytical estimation of  
Equation (\ref{eq:misoin}) and (\ref{eq:misoout}). 
We assume perfect coalescence
with the conservation of total mass and angular momentum  when either (1) the 
separation of two embryos is smaller than the sum of their physical 
radii estimated by \cite{Lissauer-2011}  or (2) they form a close-in binary 
with a common orbital period less than $10$ days.
The realistic  super-Earth-mass embryos contain a  modest 
envelope, and the drag of accreted gas would increase the capture cross 
section during close encounters (\citet{Ormel-2012}). And separation of such 
pairs shrinks, so they eventually merge due to their tidal interaction with 
the disk and common envelope \citep{Baruteau-2011}.  
We specify criteria (2) for both  physical consideration and  computational 
idealization. Our simulation results  show that almost $90\%$ of mergers are 
due to  criteria (1).  We examine the embryos' evolution under 
different  disk accretion (${\dot M}_g$)  and stellar mass ($M_\ast$) 
in  Figure \ref{xxx}. Nine panels are shown with ${\dot M}_g =10^{-8}$,
$5 \times 10^{-8}$, and $10^{-7} M_\odot$ yr$^{-1}$ (left to right panels) 
and $M_\ast=0.5, 1.0,$ and $2.0 M_\odot$ (top to bottom panels). Six $5 \ M_\oplus$  embryos are embedded on  each side 
of the trapping radius with $10R_{R}$ initial separation.  We also verify  the non-gap-opening criterion (Equation [19] in Paper II) for parameter space ($M_{\ast}$, ${\dot M}_g$) adopted here and ensure the type \uppercase\expandafter{\romannumeral1} migration assumption in our simulations is justified.   
In a layered  disk, midplane viscosity becomes much smaller than that of the surface layer due to the limited penetration depth of stellar  ionizing photons.  As demonstrated in \cite{Kretke-2012} and  Paper II,   the lower mass range for outward migration is roughly $ (2h)^{2/3} M_{opt}$, which increases  with viscosity through $M_{opt}$'s dependence on $\alpha$ (see Eq \ref{opt}).  In this section, we adopt the prescription and  fiducial  values of model C in  Paper II, in which $ \alpha_\nu = \alpha_{ M} $  when $ \ R_{R} <  R_{\rm dz}$ and 
 $ \alpha_\nu=\alpha_{\rm M}+(\alpha_{H} -\alpha_{M}) \left( \frac{(R_{\rm R}/R_{\rm dz})^{2}-1 } {(R_{\rm R}/R_{\rm dz})^{2}+1}\right)$  
when $R_{R} > R_{\rm dz}$. $R_{\rm R}$ is the planetary Roche radius, 
  and the size of  the dead zone is  $R_{\rm dz}= H(\rm r) \Sigma_{g}(\rm r)/\Sigma_{\eta} $, 
  where  $\alpha_{ H}=10^{-3}$, $\alpha_{ M}=1.4 \times 10^{-4}$  and   $\Sigma_{\eta} =10^4  \rm g cm^{-2}$.  This condition ensures that  $3-5 \ M_{\oplus}$  are within the  optimal mass range for outward migration in disks around stars with $ M_{\ast} = 0.5 \ M_{\oplus}-2 \ M_{\oplus}$.

The orbital  evolution of embryos  around an  $M_\ast = 0.5M_\odot$ central 
star is displayed in the first row of Figure \ref{xxx}.  The group of embryos undergo convergent 
migration with a  slow pace in a low-accretion disk and 
with  a fast pace in high-accretion disk.  In the low accretion rate  model (${\dot M}_g =  
10^{-8} M_\odot \rm yr^{-1}$ for top left panel), embryos' migration 
is stalled as they capture each other into their mutual MMR  and evolve into chains of low-mass planet systems.  This result is consistent with the  Kepler's data, which indicate the common existence of  multiple super-Earth systems with $M_p < M_c$
and $M_s > M_c$ (\S \ref{sec:keplerdata}).  

In the intermediate and high accretion rate models (${\dot M}_g = 5 \times 10^{-8} M_\odot \rm yr^{-1}$ for top middle panel and  $ {\dot M}_g =10^{-7} 
M_\odot \rm yr^{-1}$ for  top right panel), orbital crossing between embryos 
happens repeatedly after $10^{4}$ yrs and some  eventually  lead to 
a merger.  The asymptotic mean separations between neighboring residual 
embryos in above three runs are around $ 6.2R_R$, $3.5R_R $ and $3R_R$ respectively.
These results indicate that the lack of gas giants around M stars may be associated 
with their relatively low ${\dot M}_a$ in gaseous disks rather than 
their smaller $M_s$, the total supply of building-block material. 

Similar results (middle and bottom rows) are also found for the solar-type
 and intermediate-mass stars (with $M_\ast = 1$ and $2M_\odot$). For the same ${\dot M}_g$,  the location of the trapping 
radius is essentially independent of $m_\ast$ (see Eq \ref{eq:trans1})
(from top to bottom rows). However, $r_{\rm trans}$ increases with 
${\dot M}_g$ (from left  to right  panels)  for the same $M_{\ast}$.    These numerical results of type I migration timescales are also consistent with the estimation  of Equation (\ref{eq:tauia}) and 
(\ref{eq:tauib}).  

At a similar ${\dot M}_g (\sim  5 \times 10^{-8}  \ M_\odot$ yr$^{-1})$, 
the disk torque is sufficiently strong for  embryos to cross each 
orbit and merge together to form critical cores. The formation of supercritical cores 
appears to be significantly dependent on gas accretion rate (from left 
to right panel), but the correlation for stellar mass is shown to  be relatively  weak (from top 
to bottom panels).  These results may imply that the value of $\eta_c$  
in ${\dot M}_{f}$ (Eq.  [\ref{eq:mdotcf}]) is close to zero, which will be further discussed in the next subsection. 

\cite{Bitsch-2010} showed that corotation torque would also be saturated due to  non-negligible eccentricity. This effect is modest as eccentricity is smaller  than $ 0.03$ and the embryo can still undergo outward migration (their Figure 3). In our simulations, the eccentricities maintain a relatively low equilibrium value of $\sim 0.02$ except for  close encounters.   Corotation torque would  be weakly suppressed due to nonzero eccentricity during the convergent migration.  However, the other neglected work here is additional stochastic torques due to the fluctuation nature of disk turbulence.   \cite{Pierens-2013} proved that  these random stochastic torques can disrupt the resonant configurations and enhance the growth of massive cores. We expect that our outcomes will be compromised when  both physical processes are taken into account, and  the main conclusion of the paper is still  convincing.

\subsection{Dispersion in the critical accretion rate}
\label{sec:dispersion}

In addition to the disk accretion rate and stellar mass, embryos' migration rate is 
determined by the saturation of their corotation torque from $f_a$
(Eq.  \ref{eq:taui}),  Consequently, ${\dot M}_{f}$ is a function
of $M_p$ through $f_a$ (see \S\ref{sec:2cores}).  We now assess
the dispersion in the threshold accretion rate for a range of $M_p$ values.

Using the Hermite-Embryo code, we carry out three series of simulations
(see Table \ref{tab1}) with either six $5 \ M_\oplus$ or ten $3 \ M_\oplus$ 
embryos and various initial separations (7 ,10,12 and 15 $R_R$).  Total 
planet mass is identical ($30 \  M_\oplus$) for all series.  Embryos' eccentricity is
chosen from Rayleigh distribution with a mean value $e_0=0.02$, and they 
are all in coplanar orbits with random orientations and phase  
angles, and with initial locations interior 
and exterior to the trapping radius.  In each series, we perform $9$ 
different models, with three different $M_\ast (=0.5, 1, 2 \ M_\odot$)
and ${\dot M}_g$ ($10^{-8},  5 \times 10^{-8}, 10^{-7} \ M_\odot$ yr$^{-1}$).

Resonant capture, orbit crossing, and physical collisions are stochastic
processes.  In order to boost the statistical significance of our results,
we simulate 5 independent runs with slightly different initial semimajor 
axes and orbital phases in each model.  For series 1, 2, and 5, we show, in the top, 
middle, and bottom panels of Figure \ref{fig2}, the mean asymptotic 
separation $k_0$ (normalized by $R_R$). Models with $2 \  M_\odot$, $1 \ M_\odot$ and $0.5 \ M_\odot$ are respectively 
represented by red circles, green pentagons and yellow rectangles. 

For a given set of model parameters (such as $M_\ast$ and ${\dot M}_g$), 
variations in the initial conditions generate a  limited range of 
$k_0$ (indicated by the error bar).  
We find that embryos’ final mean separation is dependent on the  initial spacing only for the  low-accretion cases (e.g., $\dot {M_g} < 2-3 \times 10^{-8} \ M_{\odot} \rm yr^{-1}$ ). In disks with relatively high accretion rates, they undergo fast migration pace, bypass their mutual MMR barriers  and attain similar  final spacing with modest variations.
Similarities between series $1-4$ 
results for $\dot {M_g} >3 \times 10^{-8} \ M_{\odot} \rm yr^{-1}$   indicate that convergent migration gathers embryos toward $r_{\rm
trans}$ along similar paths regardless of their initial configurations. 
Comparison between series 1 and 5 indicates that  provided $M_p$ is 
comparable to the optimum mass $M_{\rm opt}$ calculated in Eq. [\ref{opt}],
the asymptotic $k_0$ and embryos' collision probability are insensitive to
the embryos' initial isolated mass.  In all cases, these asymptotic values
of $k_0$ are reached well within the gas depletion time scale $\tau_{\rm dep}$.

In the limit of small ${\dot M}_g (\leqslant 10^{-8} \ M_\odot$ yr$^{-1}$), 
embryos capture each other onto their MMRs.  But $k_0$ is a decreasing 
function of ${\dot M}_g$ for all 3 values of $M_\ast$. The black solid 
lines in Figure \ref{fig2} denote the critical separation ($k_0=5$) 
less than which neighboring embryos perturb and cross each 
other's orbits within a few times  $10^{4}$ yr \citep{Zhou-2007}. Subsequently, they 
undergo frequent close encounters, collide and merge into supercritical 
cores.  Since the orbit crossing time is a rapidly increasing function 
of $k_0$ (see \S \ref{sec:2cores}), we approximate ${\dot M}_{f}$ by the
magnitude of ${\dot M}_g$ which leads to $k_0 =5$.  Similarities between the  
three panels indicate that our results are insensitive to  the initial separation 
and mass of embryos around the optimum value  for the parameter ranges explored here. The approximate values ({\it i.e.}, ${\dot M}_g$ which leads 
to $k_0=5$) of ${\dot M}_{f}$ are $2.7, \ 4.0, \ 6.3 \times 10^{-8}$ for
$M_\ast = 0.5, \ 1.0, \ 2.0 \ M_\odot$ respectively ( see  top panels of Fig. \ref{fig2}).  These results can be fitted
to Equation (\ref{eq:mdotcf}) with $\eta_c \sim 0.6$.  But the analytic 
approximation in Equation (\ref{eq:criticalmdot2}) indicates $\eta_c \sim 0.07$.
 The minor $\eta_c$ difference is caused by the condition we used 
for the analytic approximation  and numerical computation of the threshold
orbit crossing condition.

The  critical accretion rate (Eq (\ref{eq:criticalmdot2})) is  derived  from  \cite{Murray-1999},  which requires that the time scale for the differential migration ($\tau_{\Delta a}$) between planet pairs through their characteristic liberation width is longer than the resonant angle liberation time scale ($\tau_{lib}$).  For the liberation time scale, it relies on a coefficient factor $f_{res}$ which is a function of the semi-axis ratio between the two planets, e.g $f_{res}$ is smaller for $4:3$ than $3:2$ resonance.  \cite{Ogihara-2013}  indicates that this coefficient factor also depends on the outer-to-inner planet mass ratio when the value is larger than 0.1.
 The difference in our cases is that $f_{res}$ is also $M_p$ dependent due to the saturation of corotation torque, so we are only interested in the  limited mass range close to $M_{\rm opt}$. 
 They simulated two planets' migration and found that for equal-mass planets the critical migration time scale to pass through $2:1$ resonance is one order of magnitude shorter than that for a  very small mass ratio.  In our work, the initial separation between embryos is much smaller than that in $2:1$ resonance.  According to their simulation results (see the Table 3 in \cite{Ogihara-2013}), the critical migration time scale is only reduced by a factor of $2$ for $4:3$ and $5:4$  MMRs when the planets have equal mass.  The situation is suspected  to be more complicated, and resonance configurations would also be disrupted when the number of embryos is large \citep{Pierens-2013}. The exploration of $M_p$ dependence on resonance capture (Eq \ref{eq:criticalmdot2}) is not a proper and prior task for this study.  At least for the numerical experiments implemented here, we find the planet mass is insensitive  to embryos' collision probability  and final asymptotic $k_0$.


\begin{figure}[htbp]
\includegraphics[width=0.99\linewidth,clip=true]{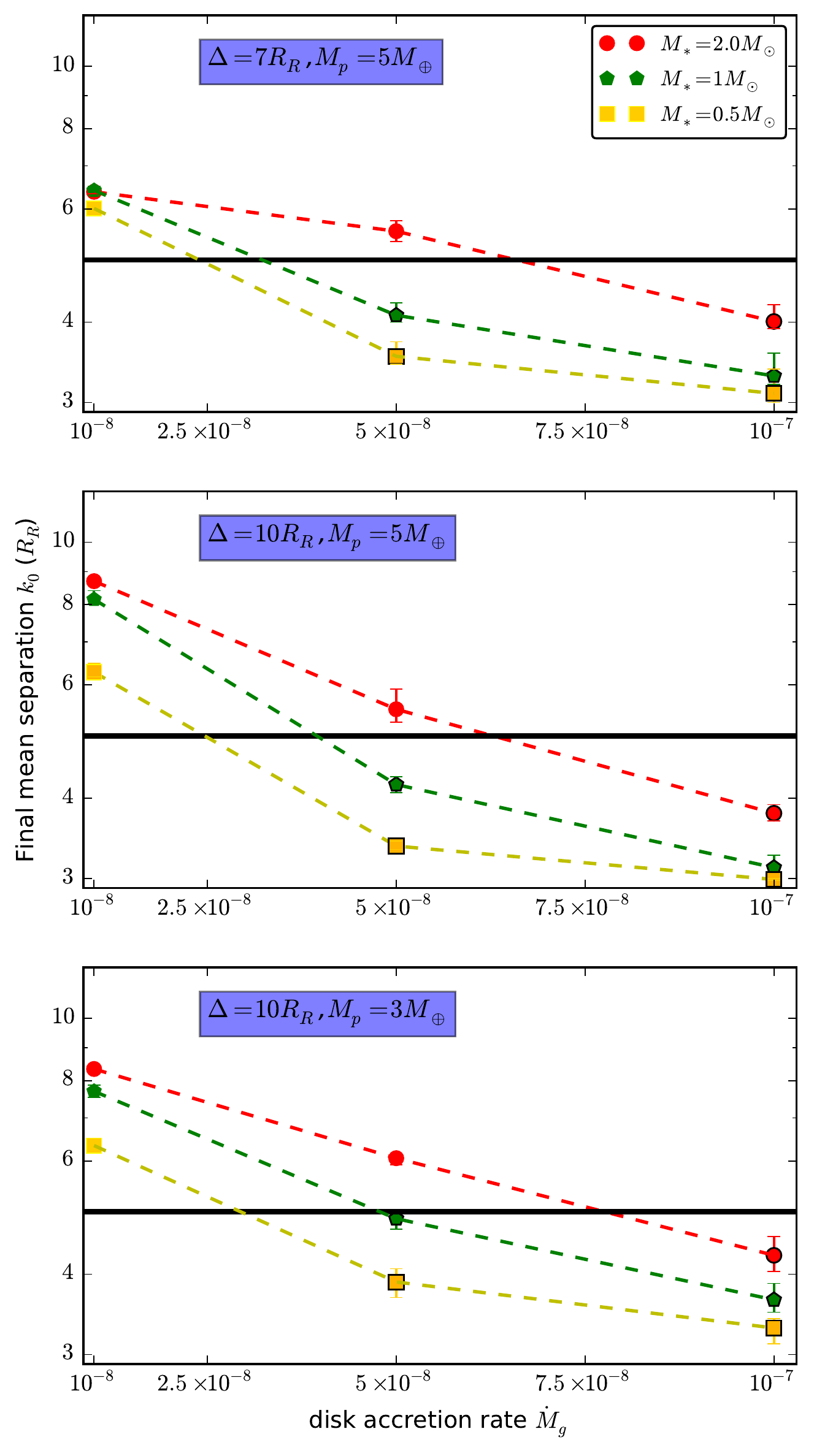}
\caption{
Asymptotic mean separation $k_0$ vs. disk accretion rate ${\dot M}_g$
for series 1, 2, and 5 (from the top to the bottom panels).
Red circles,  green pentagons and yellow rectangles represent the models with
$M_\ast=2 \ M_\odot$,$M_\ast=1 \ M_\odot$ and $M_\ast=0.5 \ M_\odot$, respectively. 
Symbols with black borders represent models in which merger occurred
during the evolution.  Embryos which captured each other in MMR without
coalescence are denoted by symbols without borders. The thick horizontal 
black lines mark the critical final mean separation, {\it i.e.} $5R_{ R}$. 
If the planets enter into a more compact configuration, dynamical instability
would lead to orbit crossing and collisions within $10^5$ yr.
}
\label{fig2}
\end{figure} 

\begin{figure}[htbp]
\includegraphics[width=0.99\linewidth,clip=true]{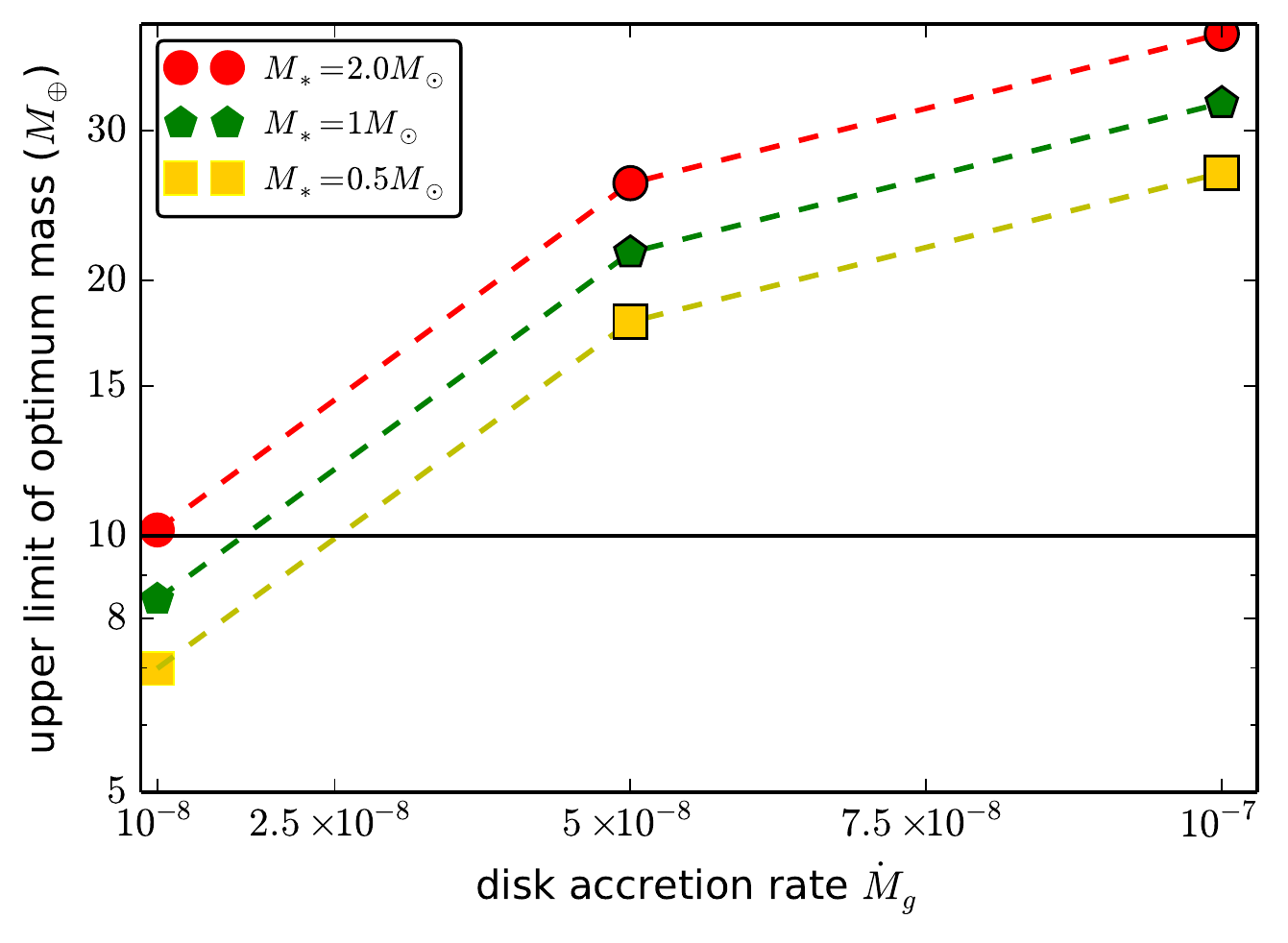}
\caption{
Maximum value of optimum planet mass as a function of $M_\ast$ and 
${\dot M}_g$. The critical core mass $M_c (=10 \ M_\oplus$) is marked by
the black line.  The red circles, green pentagons, and yellow rectangles 
represent the maximum mass of the trapped embryos for three stellar masses
($M_\ast=2, 1, 0.5 \ M_\odot$,  respectively). 
}
\label{fig2b}
\end{figure}


In the numerical fit to the observed
${\dot M}_a$ (Eq \ref{eq:mdota}), $\eta_b \sim 1.3-2.0$.  Provided
$\eta_b > \eta_c$, the actual accretion rate increases with $M_\ast$
faster than the threshold accretion rate for inducing embryos to
bypass their MMR barriers and cross each other's orbits.  
Cohesive collisions increase $M_p$.  The migration of mergers
continues to converge around $r_{\rm tran}$, provided their
$M_p$ reach  $M_c$ within a  relatively short timescale.
(\S\ref{sec:2cores}). 

The corotation torque is saturated (weakened) for embryos with $M_p > M_{\rm retain} \simeq (2h)^{-2/3} M_{\rm opt}$ (see \S\ref{sec:2cores}). Values of $M_{\rm retain}$ are shown in Figure \ref{fig2b}.  Torque on embryos with $M_p > M_{\rm retain}$ is dominated by the contributions from differential Lindblad resonances, which generally induce them to migrate inward.
 The magnitude of $M_{\rm retain}$
is typically $\sim 2.2 M_{\rm opt}$ (Paper II), which increases 
with both $M_\ast$ and $ {\dot M}_g$ (see Eq \ref{eq:mopttrans}). 
It also decreases with the distance to the central stars (see Eq 
\ref{opt}). Around $M_\ast=0.5 \  M_\odot$, $M_{\rm retain} > M_c$ in the
limit ${\dot M}_g > 3 \times 10^{-8} \ M_{\odot}$ yr$^{-1}$
(yellow rectangle).  Around $M_\ast=2 \ M_\odot$, $M_{\rm retain} > M_c$ 
in the limit ${\dot M}_g > 10^{-8} \ M_{\odot}$ yr$^{-1}$
(red circle). Together with the condition of threshold accretion rate, these results imply that if 
${\dot M}_g$ is sufficiently large to form critical-mass cores,
they would be retained near the trapping radius.

Supercritical cores accrete gas on the time scale $\tau_{KH}(M_{\rm retain}) 
(\propto m_\ast^{-3.72} )$ (see Eq \ref{eq:taukh} in \S\ref{sec:gasaccret}). 
Although their $M_p$ may reach and exceed $M_{\rm retain}$ 
during their subsequent gas accretion, they would accrete gas and 
evolve into gas giants {\it in situ} if the migration timescale 
$\tau_I (M_{\rm retain}) > \tau_{KH}(M_{\rm retain})$. As they gain mass, 
proto-gas-giant planets strongly perturb their natal disks by opening up gaps and transit from type I to slow type II migration \citep{Lin-1986}.  

In general, critical-mass cores form more readily, are more likely
to be retained, and can grow into gas giants more rapidly around 
relatively massive stars.  Quantitatively, the transition from convoys 
of resonant super-Earths to gas giants occurs in the $\eta_{\dot M}$-$M_\ast$ 
domain between the red and  blue lines (represent $M_{cr}=1 \times 10^{-7} 
\ M_\odot$ yr$^{-1}$ and  $M_{cr}=5 \times 10^{-8} \ M_\odot$ yr$^{-1}$    
respectively) in the right panel of Figure \ref{fig1}.  
This 
correlation  generally agrees with the observed $\eta_J$-$M_\ast$ correlation, 
though the magnitude of the simulated $\eta_{\dot M}$ is somewhat higher than 
the observed $\eta_J$ (see further discussions in \S\ref{sec:etajzast}).

\section{Formation of critical cores around stars with different metallicities}
\label{sec:metaldep}

In this section, we first examine the correlation between $Z_d$ and 
$Z_\ast$ prescription.  Following similar methods from the previous 
two sections, we determine the dependence of critical ${\dot M}_f$ 
on the $Z_d$ of the disk gas.  We then combine these two sets of 
results to determine the main causes of the observed $\eta_J$-$Z_\ast$ 
correlations.

Equation (\ref{eq:etadotm})  indicates 
that $\eta_J$ is determined by both the critical accretion ${\dot M}_f$ 
for core formation and the actual observed value of the disk accretion 
rate ${\dot M}_a$. We suggest that the average values of opacity 
$\kappa_0$, isolation mass $M_{\rm iso}$ and trapping radius $r_{\rm trans}$ 
are increasing functions of $\Sigma_d$.  These factors affect the 
magnitude of ${\dot M}_f$.

\subsection{Metallicity of grains and embryos in protostellar disks versus stellar 
metallicity}
\label{sec:grainopacity}

In the sequential accretion scenario, heavy elements are not only 
the building blocks of embryos and cores, but also the main opacity 
source such that $\Sigma_d$ determines not only the dynamical evolution 
of the embryos and cores but also the disk structure.  However, $\Sigma_d$ 
cannot be directly determined from observations of optically thick
protostellar disks.  The metallicity of T Tauri stars is also poorly
determined such that there are no well-established  $Z_d$-$Z_\ast$ 
correlation or  any ${\dot M}_a$-$Z_\ast$ dependence (in contrast to
the observed ${\dot M}_a$-$M_\ast$ correlation in Eq. \ref{eq:mdota}).
In typical accretion disk models \citep{Hartmann-1998BOOK}, it is customary to 
construct a $\Sigma_g$ distribution  and 
to determine $\Sigma_d$ based on the assumption that $Z_d \propto 
Z_\ast$ throughout the disk \citep{Laughlin-2004a, Ida-2005, 
Mordasini-2012}.   

Based on the following four circumstantial evidences, we revisit the validity of this 
assumption.  

\noindent
A1) Perhaps the strongest evidence for $Z_d \propto Z_\ast$ 
is the correlation of abundance distribution between the solar atmosphere
and chondritic meteorites for nearly all refractory elements
\citep{Anders-1989}.  There are clear indications of volatile-element 
depletion either due to evaporation or inability to condensate.  

\noindent
A2) During 
their formation, all of the stellar mass content, including H, He, and 
heavy elements, passes through protostellar disks.  Efficient retention
of heavy elements in the disk would not only enhance $Z_d$ but also
reduce $Z_\ast$ from the protostellar clouds.  Upper limits on the
iron abundance variation ($\Delta$ [Fe/H]$_\ast \leq 0.02$) among 
solar-type stars in young stellar clusters (such as Pleiades and IC 4665) 
\citep{Wilden-2002, Shen-2005} suggest that the total amount of 
heavy elements and its dispersion in protostellar disks is limited
to less than twice that in the solar nebula. They are generally
smaller than the range of $Z_\ast$ between different open clusters.  
Metallicity dispersion in planet-hosting binary stars (such as 16 
Cyg and XO2) \citep{Ramirez-2011, Ramirez-2015} can also be interpreted
to support the assumption that protostellar disks are formed with $Z_d 
\sim Z_\ast$.  

\noindent
A3) The typical accretion rate in protostellar disks
around solar-type stars is a few times  $10^{-8} \ M_\odot$ yr$^{-1}$.
If their viscous diffusion time scale is comparable to the age
(a few Myr) of their central stars, the mass of gas $M_g$ in these disks
would be comparable to or a few times larger than that of the MMSN
(i.e., a few times $10^{-2} \ M_\odot$).  
Total mass ($M_z$) in millimeter-size (both volatile and refractory) grains in 
these disks \citep{Beckwith-1990} is comparable to or a few times 
that of heavy elements in the MMSN model.  
The ratio $M_z/M_g (\sim 
10^{-2}$) of T Tauri stars is comparable 
to that in the Sun. 

\noindent
A4) In \S\ref{sec:dependence},
we indicate that ${\dot M}_d \propto M_\ast^{\eta_b}$ and $\eta_b\sim 
1.3-2.0$ (Eq.~\ref{eq:mdota}).  Most of the gas is distributed in the 
outer irradiated region where $\Sigma_g \propto {\dot M}_d$ (GL07).  
Recent millimeter observation \citep{Andrews-2013} suggests that $\Sigma_d
\propto M_\ast$.  These inferred ($\Sigma_g$-$M_\ast$ and $\Sigma_d$-$M_\ast$)
correlations suggest that the ratio $\Sigma_d/\Sigma_g$ may vary
significantly with $M_\ast$.

Bear in mind that the millimeter continuum observations of $\Sigma_d$ only determine
the heavy-element content in the millimeter-size range. We do not 
have any reliable information on the abundant small-size grains or 
large-size planetesimals. Due to the lack of constraints on the gas/dust
depletion factors, it is also difficult to extract quantitative
information on both $\Sigma_d$ and $\Sigma_g$ from spectroscopic (CO) 
observations.  Although $M_\ast$ can be directly determined from photometric 
observations, we also do not have any reliable data on the metallicity
of the infant host stars $Z_{\ast}$.  Nevertheless, from the above circumstantial 
evidences, we infer, within an order of magnitude, the ratio 
$\Sigma_d/\Sigma_g$ in protostellar disks is comparable to heavy elemental 
fraction in their central stars.  We now assess the dispersion in the 
$Z_d$-$Z_\ast$ correlation in terms of  theoretical considerations.     

\noindent
B1) During the formation and early phase of their evolution when 
disks are more compact, hotter, and vigorously turbulent,  the 
gas accretion rate is high ($>10^{-6} \ M_\odot$ yr$^{-1}$) and 
dust grains are well coupled to the dense disk gas, it is likely 
that $Z_d \simeq Z_\ast$.  At that epoch,  the dust destruction 
radius $r_{\rm ref}$ extends to a large fraction of the disk so 
that grains may only be available in the outer disk regions 
(beyond $1$ AU). The rapid evolution of the disk may also limit the 
available time window (a few times  $10^5$ yr) for embryos to grow to a 
few $M_\oplus$.

\noindent
B2) During the main evolution stage of protostellar disks, several 
physical processes may induce the decoupling of gas and solid, and  lead to  a dispersion in $Z_d$.  
They include a) grain growth and embryo formation that may deplete 
micron-size grains and reduce $\kappa_0 $\citep{DAlessio-2001}, 
b) extensive orbital evolution of  pebbles ($\sim$ cm-size grains) 
due to the hydrodynamic drag and turbulent diffusion \citep{Whipple-1972,
Supulver-2000, Ormel-2010, Lambrechts-2012} to modify 
and bypass $M_{\rm iso}$, and c) a local 
concentration of grains and enhancement of $\Sigma_d$ and $Z_d$ 
near the refractory dust and volatile ice condensation/sublimation 
fronts $r_{\rm ref}$ and $r_{\rm ice}$ (see \S\ref{sec:graingrowth})
\citep{Cuzzi-2004, Ciesla-2006,Chiang-2010}.

\noindent 
B3) The modest $\Delta_Z$ assumption 
is likely to break down with $Z_d > > Z_\ast$
during the advanced evolutionary phase of transitional and debris disks
when the residual gas is depleted through viscous accretion and photoevaporation. However, 
the diminishing ${\dot M}_a$ (see Eq. \ref{eq:mdota}) is generally too 
small to enable residual embryos to evolve into cores unless they 
have already formed at earlier phases.  Therefore, in the determination 
of $\eta_J$, we can generally neglect advanced stages of disk evolution.  

The aim of our study is to determine the average probability 
of a supercritical core and an $\eta_J$-$Z_\ast$ correlation.
We have already indicated that there is no information on 
any $Z_d$ or $Z_\ast$ dependence in ${\dot M}_a$. 
All of  disks' structural and evolutionary (B1-B3) effects 
contribute to a dispersion $\Delta_Z$.  We represent $\Delta_Z$ 
to be the logarithm of the disk-metallicity range relative to 
the stellar metallicity.  Here  we assume $\Delta_Z \sim 0.5-1$ 
which corresponds to a factor of $3-10$ dispersion in metallicity
between disks and their central stars. Based on the discussions 
on the $Z_d$-$Z_\ast$ correlation (A1-A4) and its dispersion (B1-B3), 
we assume that the amount of heavy elements contained in both the 
opacity sources ( micron-size grains) and building-block planetesimals 
($\Sigma_d$) have the following metallicity dispersion:
\begin{equation} 
\begin{split} 
{\mathrm{d^2N} \over \mathrm{d {\dot M }_g \ d {Z_d }}  } =
&  A_0 {\rm exp} {\left[ - \left( \frac{{\rm log} (\dot M_g/\dot M_a) }{ \Delta_{{\dot M}_a}} \right)^2 \right]}\\
& \times {\rm exp} {\left[ - \left(  \frac {(Z_d - Z_\ast) }{ \Delta_Z} \right)^2 \right]}
\end{split} 
\end{equation}
which is a more general form of Equation (\ref{eq:gaussian}).

Analogous to Equation (\ref{eq:etadotm}),  the fraction of stars ($\eta_Z$) 
that have ${\dot M}_g$  larger than some fiducial accretion rate ${\dot M}_f$
with a given set of age, mass $M_\ast$, metallicity $Z_\ast$  and $Z_d$ dispersion is
\begin{equation}
\begin{split} 
\eta_{Z} ({\dot M}_f, M_\ast, Z_{\ast}) 
= & \frac{1}{2} \int \rm erfc \left(\frac{{\rm log}
 [{\dot M}_f (M_\ast, Z_d)/{\dot M}_a (M_\ast)]}
{  \Delta_{{\dot M}_a}} \right)\\
& {\rm exp} {\left[ - \left( \frac{(Z_d - Z_\ast)}{\Delta_Z } \right)^2 \right]}
d Z_d.
\label{eq:etaz}
 \end{split}
\end{equation}
In the above expression, we have taken into account the
$M_\ast$ dependence but neglected any $Z_d$ and $Z_\ast$ 
dependence in ${\dot M}_a$.   
The $\eta_J$-$Z_\ast$ correlation can be obtained from 
$\eta_Z$ as a function of $M_\ast$ and $Z_\ast$ by substituting 
${\dot M}_f$ with the critical accretion rate (${\dot M}_{\rm res} = 
{\dot m}_{9 \ {\rm res}} 10^{-9} \ M_\odot$ yr$^{-1}$) for embryos
to bypass resonant trapping locations
(see \S\ref{sec:etajzast}).

\subsection{Grain condensation, growth, opacity, and viscosity}
\label{sec:graingrowth}

Equation (\ref{eq:criticalmdot2}) indicates that the  metallicity
of the disk gas contributes to ${\dot m}_{9 \ {\rm res}}$ through
both the opacity $\kappa_0$ and viscosity $\alpha_\nu$.  
We first consider how the disk opacity depends on $Z_d$. 
We adopt the customary assumption that $\kappa_0 \propto 10^{Z_d}$
but take into account its difference across condensation fronts.
In the protostellar disks where very close 
to the central star, heavy elements are in the gas phase.
Refractory grains condense outside the dust destruction 
front ($r_{\rm ref}$) where $T_g \sim 2 \times 10^3$ K, while 
volatile grains sublimate interior to the snow line ($r_{\rm ice}$) where $T_g 
\sim 170$ K.  We parametrize the metallicity and opacity enhancement factors 
$\eta_{\rm ice}$ to be $0$, $1,$ and 4 at $r <r_{\rm ref}$, 
$r_{\rm ref} < r < r_{\rm ice}$, and $r > r_{\rm ice}$ respectively. 

The midplane temperature in the viscously heated inner disk (GL07) is 
\begin{equation}
T_g = 120 m_\ast^{3/8} {\dot m}_9 ^{1/2} \alpha_3^{-1/4} \eta_{\rm ice}^{1/4}
\kappa_0^{1/4} r_{\rm AU}^{-9/8} \rm K,
\label{eq:tvis}
\end{equation}
so that
\begin{equation}
r_{\rm ref, ice}= r_\nu m_\ast ^{1/3} {\dot m}_9 ^{4/9} \alpha_3 ^{-2/9}  \rm AU
\eta_{\rm ice}^{2/9} \kappa_0 ^{2/9}  
\label{eq:r_ice}
\end{equation}
where $r_\nu = 0.08$, $\eta_{\rm ice} = 1$ for $r_{\rm ref}$ and 
$r_\nu = 0.73$, $\eta_{\rm ice} = 4$ for $r_{\rm ice}$ respectively. 
$r_{\rm ref}$ and $r_{\rm ice}$ also depend on $\alpha_\nu$ which might be a function of $\eta_{\rm ice}$.
Besides, compared with $r_{\rm trans}$ in Eq. (\ref{eq:trans1}), 
the snow line would be within the viscously heated region (i.e. $r_{\rm ice}
< r_{\rm trans}$) if ${\dot M}_g$ is greater than
\begin{equation}
{\dot M}_{\rm ice} = 4.3 \times 10^{-8} m_\ast^{3/2} \alpha_3 ^{1/2} 
\kappa_0^{-1/2} \eta_{\rm ice}^{-1/2}  \ M_{\odot} \rm yr^{-1}.
\label{mdotice}
\end{equation}

At $r > r_{\rm trans}$, stellar irradiation leads to
\begin{equation}
T_g = 300 l_\ast ^{2/7} m_\ast^{-1/7} r_{\rm AU}^{-3/7} \rm K,
\label{eq:tirr}
\end{equation}
so that $r_{\rm ref, ice}= r_{\rm irr}$ where 
$r_{\rm irr}= 0.012$ AU, $\eta_{\rm ice} =1$ for $r_{\rm ref}$ 
and $r_{\rm irr}= 3.76$ AU, $\eta_{\rm ice}=4$ for $r_{\rm ice}$ 
respectively.  For ${\dot M}_g < {\dot M}_{\rm ice}$,
$r_{\rm trans} < r_{\rm ice}$ which can significantly modify
the $T_g$ distribution from those in Equation (\ref{eq:tvis})
and (\ref{eq:tirr}), and modify the migration rate ${\dot a}$ in 
Equation (\ref{eq:adot}) as well (see \S\ref{sec:snowline}).  

In addition to condensation/sublimation processes, coagulation/fragmentation processes 
may also modify the size distribution of grains and 
the value of $\kappa_0$. If these processes lead to an unique equilibrium 
Mathis, Rumpl, and Nordsieck (MRN, 1977) size ($s$) distribution (as
in the interstellar medium), 
in which $dN/ds \propto s^{-3.5}$ \citep{Kim-1994}, 
most of the mass would be contained in the large 
grains, whereas the disk opacity is mainly contributed by grains with 
size comparable to the wavelength ($\lambda$) of 
the reprocessed or emitted photons in micron or submillimeter scale. 
If the MRN size distribution 
is maintained as grains coagulate into planetesimals and embryos, 
the magnitude of $\kappa_0$ would decrease with their growth. 
Fragmentation, especially collisional cascade, can also replenish small 
grains. 


These uncertainties are taken into account by the dispersion 
in \S\ref{sec:grainopacity} (see Eq.~\ref{eq:etaz}).  
However, a wide (a factor 
of ten) range of $\kappa_0$ alone does not significantly 
modify ${\dot M}_{9 \ {\rm res}}$ due to its weak dependence
on $\kappa_0$ (Eq. \ref{eq:criticalmdot2}). We now consider 
how $Z_d$ may modify the viscosity, {\it i.e.} the 
effective magnitude of $\alpha_3$. 

It is widely assumed that the dominant angular momentum 
transport mechanism in accretion disks, including protostellar
disks, is MHD turbulence. The disk midplane between $r_{\rm ref}$ 
and $r_{\rm ice}$ is often thought to be inert because it is
cold and neutral \citep{Sano-2000}.  But the disk surface is 
partially photoionized by the stellar UV flux. Charged particles
recombine on grains and the disk gas establishes an ionization 
equilibrium.  The ionization fraction is an increasing function
of distance from the midplane and its value near the disk surface
may be adequate to provide an active layer where MHD turbulence
can lead to a significant flux of angular momentum transfer 
\citep{Gammie-1996}.  The thickness of the active layer is determined 
by the penetration depth for the stellar X-rays and UV photons 
\citep{Glassgold-1997} where $\kappa_{uv} \Delta \Sigma_d \sim 1$.
The column density of dust in the active layer $\Delta \Sigma_d
\propto \kappa_{uv}^{-1} \propto 10^{-Z_d}$. If the metallicity is 
constant throughout the disk's vertical structure, the associated
fraction of gas in the active layer $\Delta \Sigma_g \sim 10^{-Z_d}
\Delta \Sigma_d \propto 10^{-2Z_d}$. Thus, the extent of a ``dead
zone'' increases with $Z_d$.

Numerical simulations \citep{Turner-2007} indicate the ionization 
fraction may indeed be suppressed in the dead zone.  Nevetheless, MHD waves 
excited in the active layer continue to induce a modest flux of angular 
momentum transfer near the mid-plane \citep{Fleming-2003, Turner-2008,
Kretke-2012}.  Reduction of the active layer may also enhance the effect 
of Ohmic dissipation and ambipolar diffusion which may further suppress MRI 
in the disk \citep{Bai-2013}. Taking these uncertainties into account, 
we consider the possibility that the effective $\alpha_\nu$ may be a decreasing
function of $Z_d$ and explore the implication on the threshold condition 
for core formation.  

In an attempt to construct a quantitative disk-structure model for this 
effect, we introduced a prescription \citep{Kretke-2012} in which 
\begin{equation}
\alpha_\nu \propto \Delta \Sigma_d \propto \kappa_0^{-1} \propto 10^{-Z_d}.
\label{eq:alphaz}
\end{equation}
Substituting this prescription into Equation (\ref{eq:criticalmdot2})
we find ${\dot M}_{9 \ {\rm res}} \propto 10^{-Z_d}$.  In this scenario,
embryos merge into supercritical cores more readily in disks with
higher metallicity due to the threshold condition for orbit crossing
rather than the availability of a richer supply of building-block
materials.  

\subsection{Simulations of embryo-disk interaction with different $Z_d$.} 

\begin{figure*}[htbp]
\includegraphics[width=0.5\linewidth,clip=true]{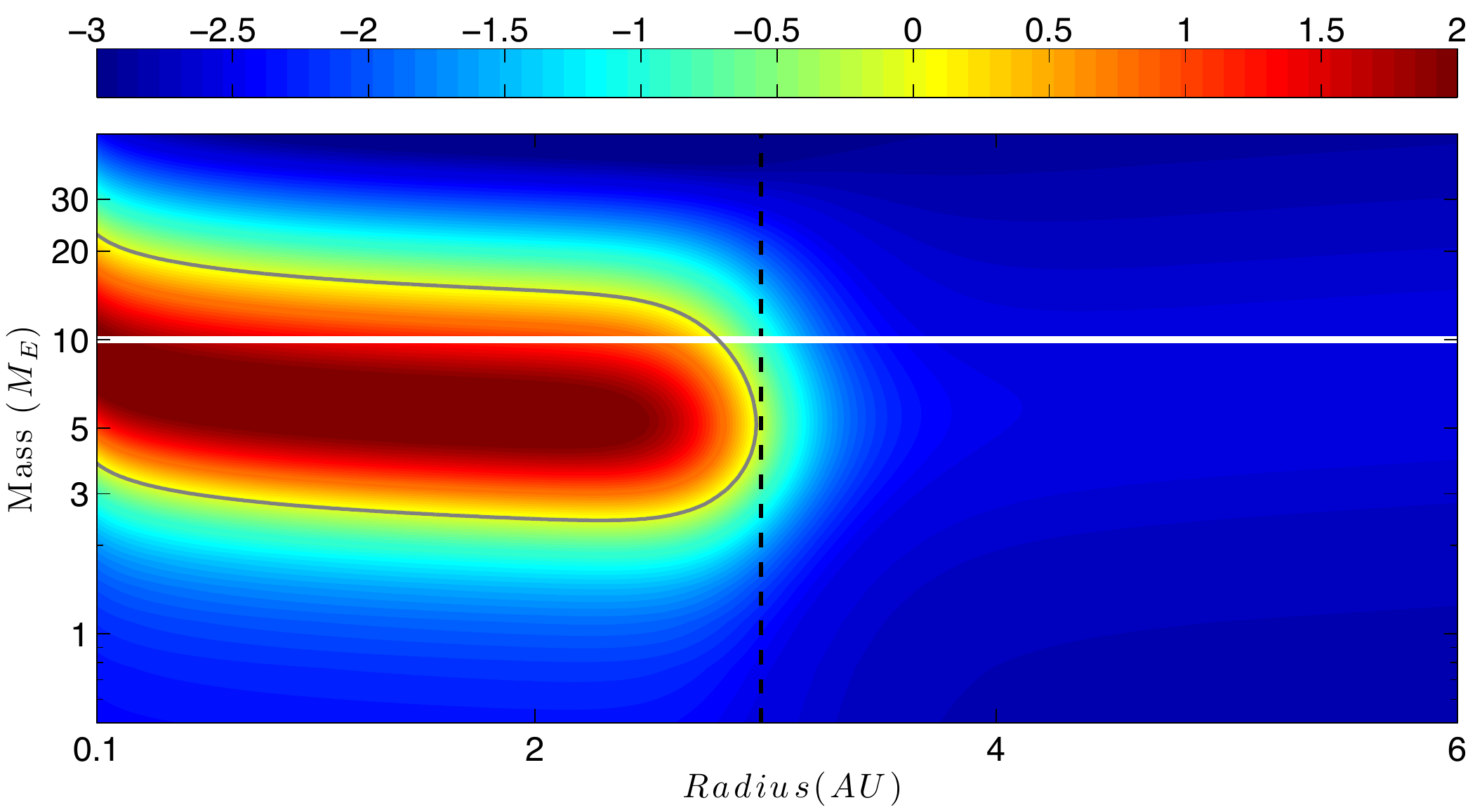}
\includegraphics[width=0.5\linewidth,clip=true]{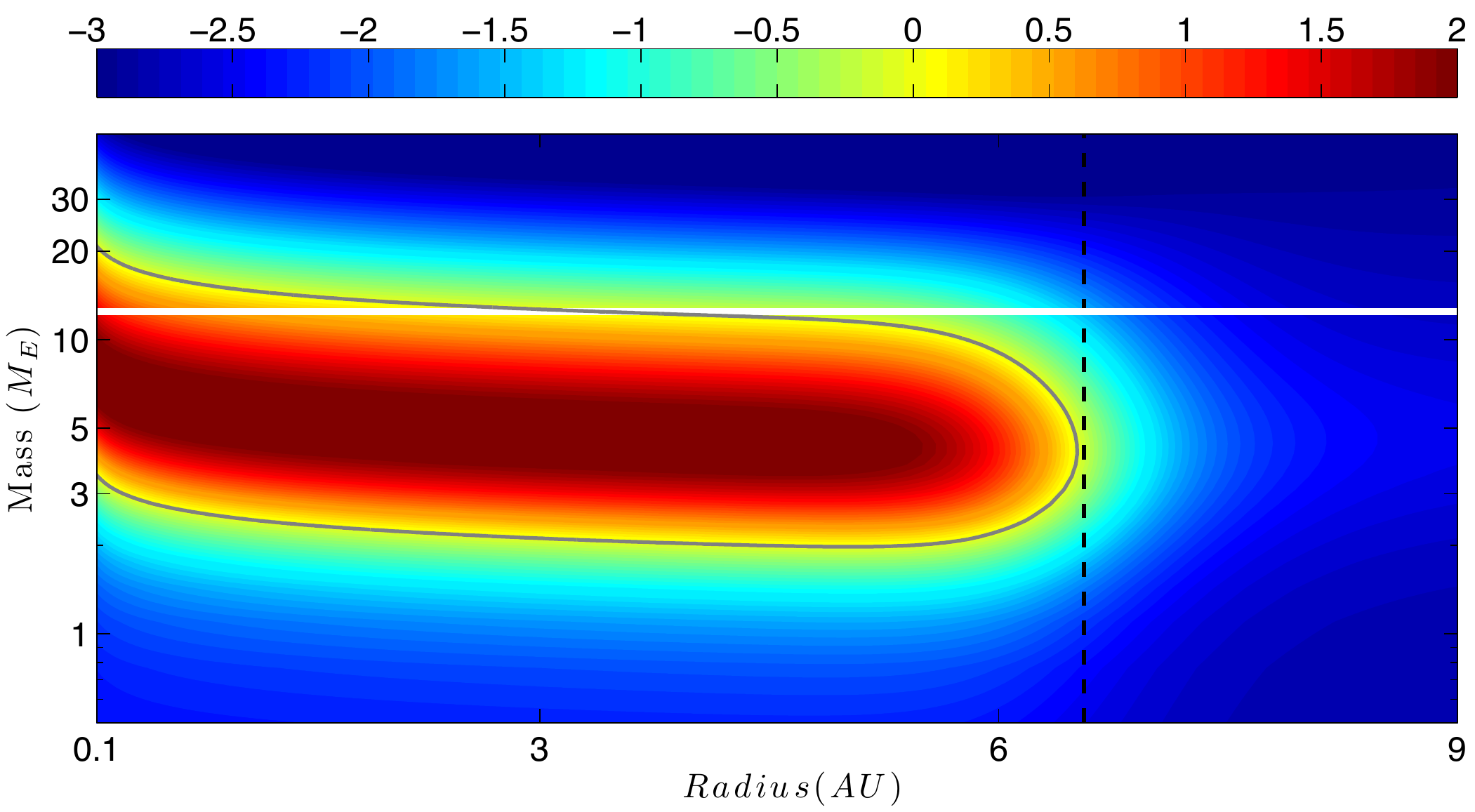}
\includegraphics[width=0.5\linewidth,clip=true]{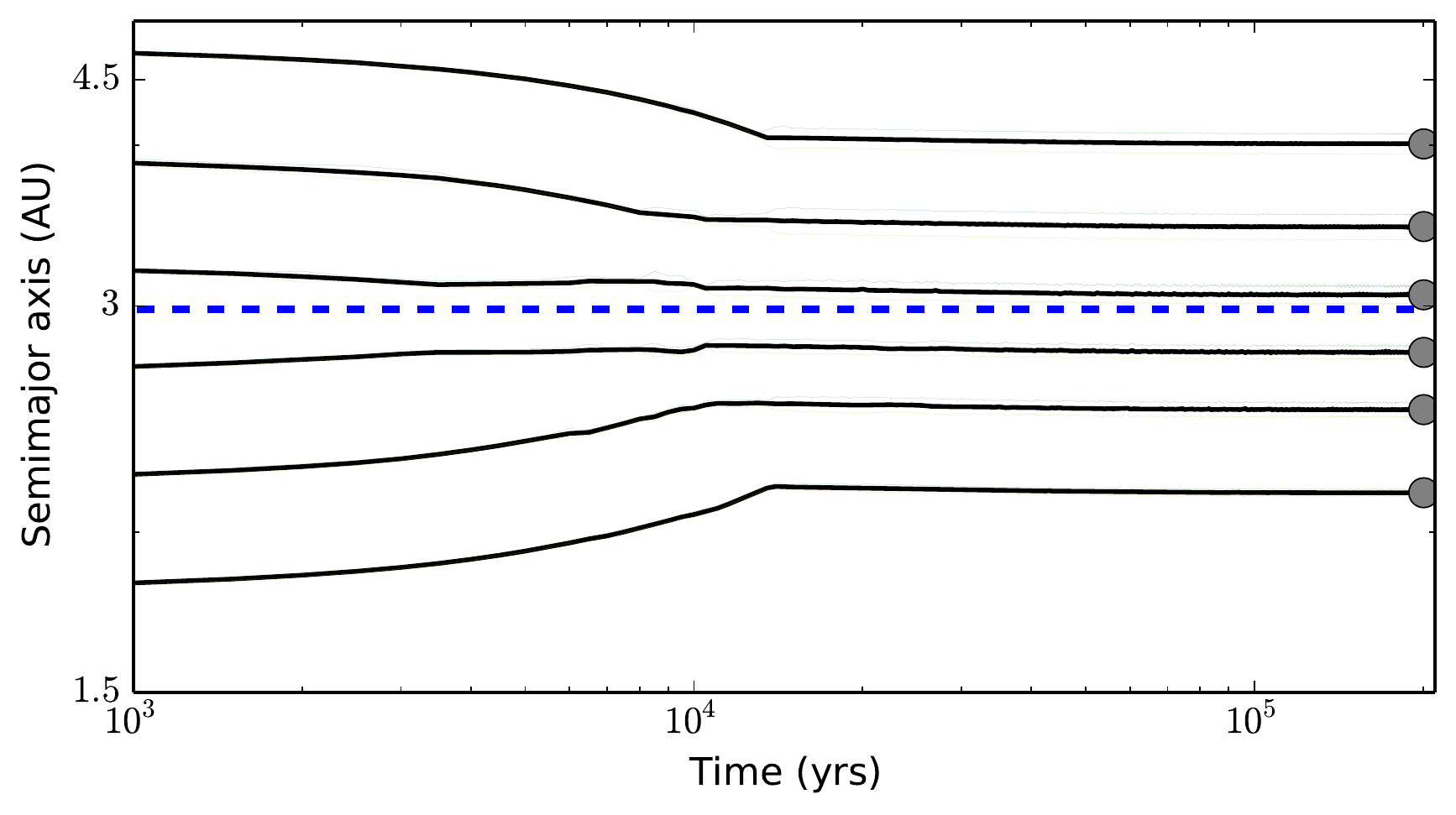}
\includegraphics[width=0.5\linewidth,clip=true]{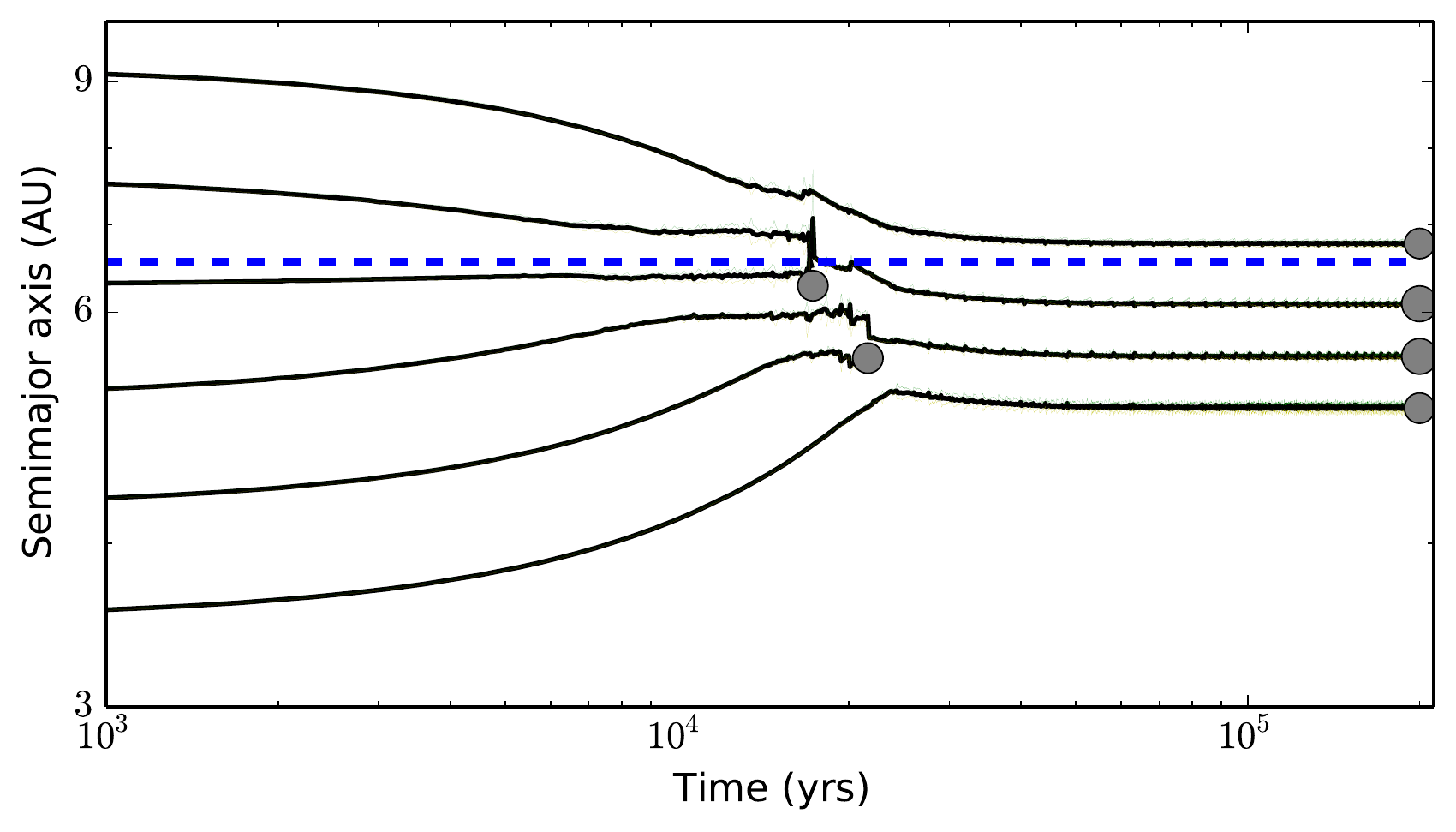}
\caption{
{\bf Top}: The type \uppercase\expandafter{\romannumeral1} 
migration coefficient ($f_a$) for a range of embryos'
mass at different locations of the disk. The black dashed line and white line represent 
the transition radius $r_{\rm trans}$ and  critical core mass ($M_{c}$).
{\bf Bottom}: orbital evolution of multiple embryos due to their
mutual perturbation and tidal interaction with their natal disks.
The black lines trace the evolution of embryos' semimajor axis and  
blue dashed line indicates the location of  $r_{\rm trap}$.  Green and 
yellow lines are embryos' apocenter and pericenter distance, respectively. 
Left and right panels have identical disk parameters  ($\rm \dot{M}
= 3\times10^{-8} \rm M_{ \odot}/\rm yr $, $\alpha_\nu =10^{-3}$ and 
$M_{\ast} =1 \ M_{\odot}$)  but slight difference in opacity ($\kappa_0= 1$ 
on the left and $\kappa_0=3$ on the right).  Both models contain six 
$5 \ M_{\oplus}$ embryos, which are initially separated in semimajor axis 
by $10 R_R$,  both interior and exterior to the  trapping radius. }
\label{bbb}
\end{figure*}

We verify the results of above analytic approximation with a series 
of numerical simulations.  For illustrative purposes, we adopt a steady
accretion rate (${\dot M}_g = 3\times 10^{-8} \ M_{\odot} \rm yr^{-1} $),
luminosity $l_\ast =1$ and mass $m_\ast = 1$  of a solar-type central
star.  In the standard model Z1, a set of fiducial opacity ($\kappa_0=1$) 
and viscosity ($\alpha_3=1$) is set to those for the solar metallicity 
($Z_d=0$).  In model Z2, we set $\kappa_0 =3$ and $\alpha_3 = 1/3$, which correspond to a metal rich disk with $Z_d=0.48$.
In both models Z1 and Z2, we neglect the effect of ice condensation by
setting $\eta_{\rm ice}=1$.  

The top two panels of Figure \ref{bbb} show the type I migration coefficient
$f_a$ in Equation (\ref{eq:adot}). For model Z1, embryos with $M_p$ in the  
range $\sim  3-16 \ M_{\oplus}$ migrate outward to $r_{\rm trans}= 3.0 \ \rm AU$.
For the more metal-rich model Z2, embryos with $M_p$ in the range of 
$\sim 2-13 \ M_{\oplus}$ migrate outward to $r_{\rm trans}=6.6 \rm AU $ (the 
red region in the right panel). A comparison between these two models 
indicates that metallicity enhancement significantly increases with the
trapping radius but slightly reduces embryos' optimum trapping mass. This 
variation is consistent with the analytic approximation $r_{\rm trans} 
\propto \kappa_0^{0.72} \propto 10^{0.72 Z_d}$ and $M_{\rm opt} 
(r_{\rm trans}) \propto \kappa_0^{-0.19} \propto 10^{-0.19 Z_d}$ from 
Equation (\ref{eq:rtrans2}) and (\ref{eq:moptdiffm2}). 
Although, with the same $M_\ast$ and ${\dot M}_g$, the effective 
temperature distributions in these two models are the same, the 
midplane temperature in the viscously heated region is elevated in models with enhanced metallicity. 
This increasing of mid-plane temperature leads to an expansion of the viscously heated region.

With the assumed $\alpha_\nu-Z_d$ prescription (Eq \ref{eq:alphaz}), 
the viscous dissipation rate actually decreases with $Z_d$ so that 
the enhanced metallicity does not significantly modify the aspect 
ratio ($H/r$) near $r_{\rm trans}$.  In Equation (\ref{eq:adot}),
the saturation for corotation resonance is at a minimum when 
$p_{\nu} \sim 1$ or $ p_{\xi} \sim 1 $ (Paper II). Minor decreases
in $\nu$ at the new $r_{\rm trans}$ also lead to slight
decline in $M_{\rm opt}(r_{\rm trans})$.   For these planets, 
\begin{equation}
{\dot m}_{9 \ {\rm res}} \simeq 60 m_\ast ^{0.07} 10^{-Z_d}
\label{eq:mdot9z}
\end{equation}
where we assume $f_{\rm res} \simeq 10$ (Papers I and II) in Equation 
\ref{eq:criticalmdot2}.

Six  $5 \ M_{\oplus}$ embryos are initially placed within and beyond
 $r_{\rm trans}$ with a separation of $10 R_{R}$. The embryos capture each other
into their MMRs in model Z1 (bottom left panel in Fig. \ref{bbb}).  This 
outcome is consistent with our estimate ${\dot m}_9 < {\dot m}_{9 \ {\rm res}}$.  
But embryos are able to marginally cross each other's orbit and undergo 
cohesive collisions in model Z2 (bottom right panel in Fig. \ref{bbb}).

In addition to the above criterion for bypassing the MMR barrier, 
the isolation mass $M_{\rm iso}$ increases (Eq. \ref{eq:misoin} 
\& \ref{eq:misoout}), their growth timescale $\tau_{\rm c, acc}$ 
 and migration time scale $\tau_I$ (Eq. \ref{eq:tauia} 
\& \ref{eq:tauib}) decrease with $Z_d$.  If, for sufficiently 
large $Z_d$, $M_{\rm iso}$ becomes much larger than $M_{\rm opt}$
so that the embryos would migrate into the central stars as their 
corotation torque becomes saturated. Such a process would deplete
$\Sigma_d$ until $M_{\rm iso}$ is reduced to $\sim M_{\rm opt}$.

The upper limit of retainable embryos in models Z1 and Z2 are larger than 
the critical core mass for the onset of efficient gas accretion $M_c$.  
The enhanced $Z_d$ also reduces the efficiency of radiation transfer 
in the envelope around these cores \citep{Ida-2004a} and increases
\begin{equation}
M_{\rm c}  \simeq  10 \kappa_0^{0.2} \ M_{\oplus} =10^{1+0.2 Z_d} \ M_\oplus 
\label{eq:mc}
\end{equation}
as indicated by the white horizontal lines in the upper panels of Figure 
\ref{bbb}.  In model Z2, the merged cores' corotation torque, though 
weakened, continues to dominate the differential Lindblad torque. 
These cores migrate outward in the viscously heated inner disk region 
and are stalled in the proximity of $r_{\rm trans}$.  Supercritical
cores initially accrete gas on the Kelvin-Helmholtz time scale 
$\tau_{\rm KH} \propto 10^{1.57 Z_d} $ (Eq \ref{eq:taukh2}). As they initiate 
efficient and runaway gas accretion, the cores evolve into gas giants 
{\it in situ}, perturb the disk structure through gap formation, 
and undergo type II migration.  

\subsection{The $\eta_J$-$Z_\ast$ correlation}
\label{sec:etajzast}

We construct a $\eta_J$-$Z_\ast$ correlation by substituting
 ${\dot M}_{a \odot}$ with $5 \times 10^{-8} \ M_{\odot} \rm yr^{-1}$ and 
${\dot M}_f$ in Equation (\ref{eq:criticalmdot}) with 
${\dot m}_{9 \ {\rm res}}$ in Equation (\ref{eq:mdot9z})
 to obtain
\begin{equation}
\begin{split}
\eta_J (M_\ast, Z_{\ast}) 
=  \frac{1}{2} \int \rm erfc \left(
{ {\rm log (1.2 m_\ast^{0.07-\eta_b} 10^{-Z_d}) \over \Delta_{{\dot M}_a}   }} 
\right) \\ 
{\rm exp} {\left[ - \left( \frac{(Z_d - Z_\ast) }{ \Delta_Z} \right)^2 \right]} d Z_d.
\end{split}
\label{eq:etaJ}
\end{equation}

\begin{figure}[ht]
\includegraphics[width=0.98\linewidth,clip=true]{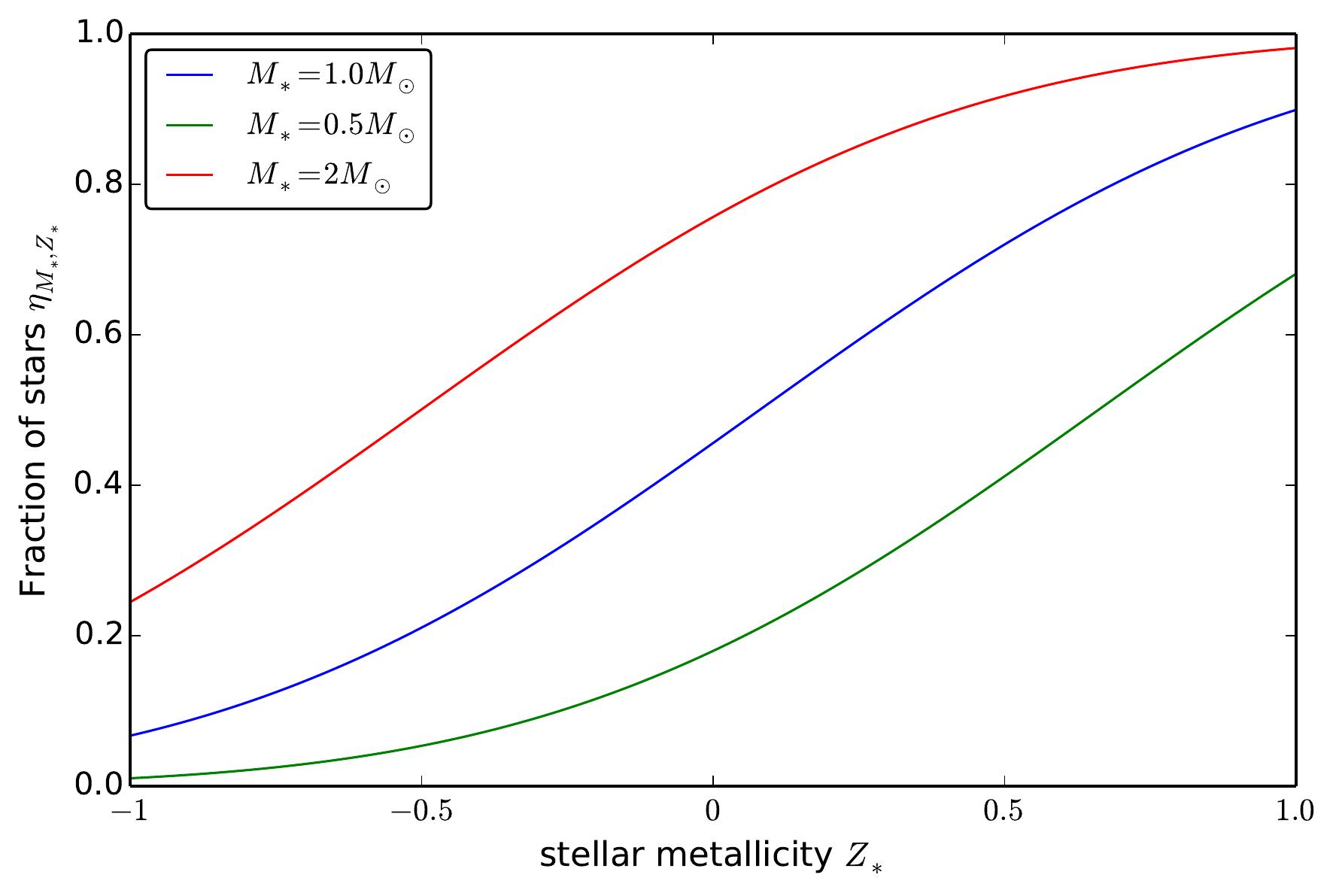}
\caption{The
$\eta_{\dot M}-Z_\ast$ correlation for different 
stellar masses. The red, green, and blue colors correspond to  $M_{\ast}  
= 2 \ M_{\odot}, 1 \ M_{\odot}, 0.5 \ M_{\odot}$, respectively.
}
\label{fig9}
\end{figure}

Numerical integration of this equation is shown in Figure \ref{fig9}.
Equation (\ref{eq:etaJ}) indicates that 

\noindent
1) the $\eta_J$-$M_\ast$ correlation is primarily due to the average disk accretion rate
${\dot M}_a$ being an increasing function of the stellar mass $M_\ast$,

\noindent
2) the $\eta_J$-$Z_\ast$ correlation is primarily due to the critical accretion rate
${\dot M}_{\rm res}$ being a decreasing function of the stellar metallicity
$Z_\ast$.  

The first conclusion is relatively robust and can be verified with future
observation.  The second conclusion is much more uncertain due to our
assumption on the  layer accretion scenario and lack of observational constraints
on correlation between the stellar and disk metallicity.  Nevertheless, 
these results provide a physical base to account for these correlations.

This theoretical $\eta_J$-$M_\ast$-$Z_\ast$ correlation generally agrees 
with the observations, though the simulated values are somewhat 
higher than the observed $\eta_J$ (also noted in \S\ref{sec:dispersion}).  
This discrepancy may be attributed in part to  (1) the
theoretical $\Sigma_g$-${\dot M}_g$ relation derived from the {\it ad hoc}
$\alpha$ prescription for viscosity, (2) uncertainty in the observationally
inferred ${\dot M}_a$-$M_\ast$ correlation, (3) orbital evolution of gas giants,
and (4) incompleteness in the observationally determined $\eta_J$ 
\citep{Clanton-2014}.

\subsection{Importance of the snow line}
\label{sec:snowline}

For simplicity, we neglected the effect of phase transition
due to ice sublimation in the above analysis.  In model Z3, 
we consider the possibility $r_{\rm trans} < r_{\rm ice}$
in the limit ${\dot M}_g < {\dot M}_{\rm ice}$ 
(see \S\ref{sec:graingrowth}).

\begin{figure*}[htbp]
\includegraphics[width=0.5\linewidth,clip=true]{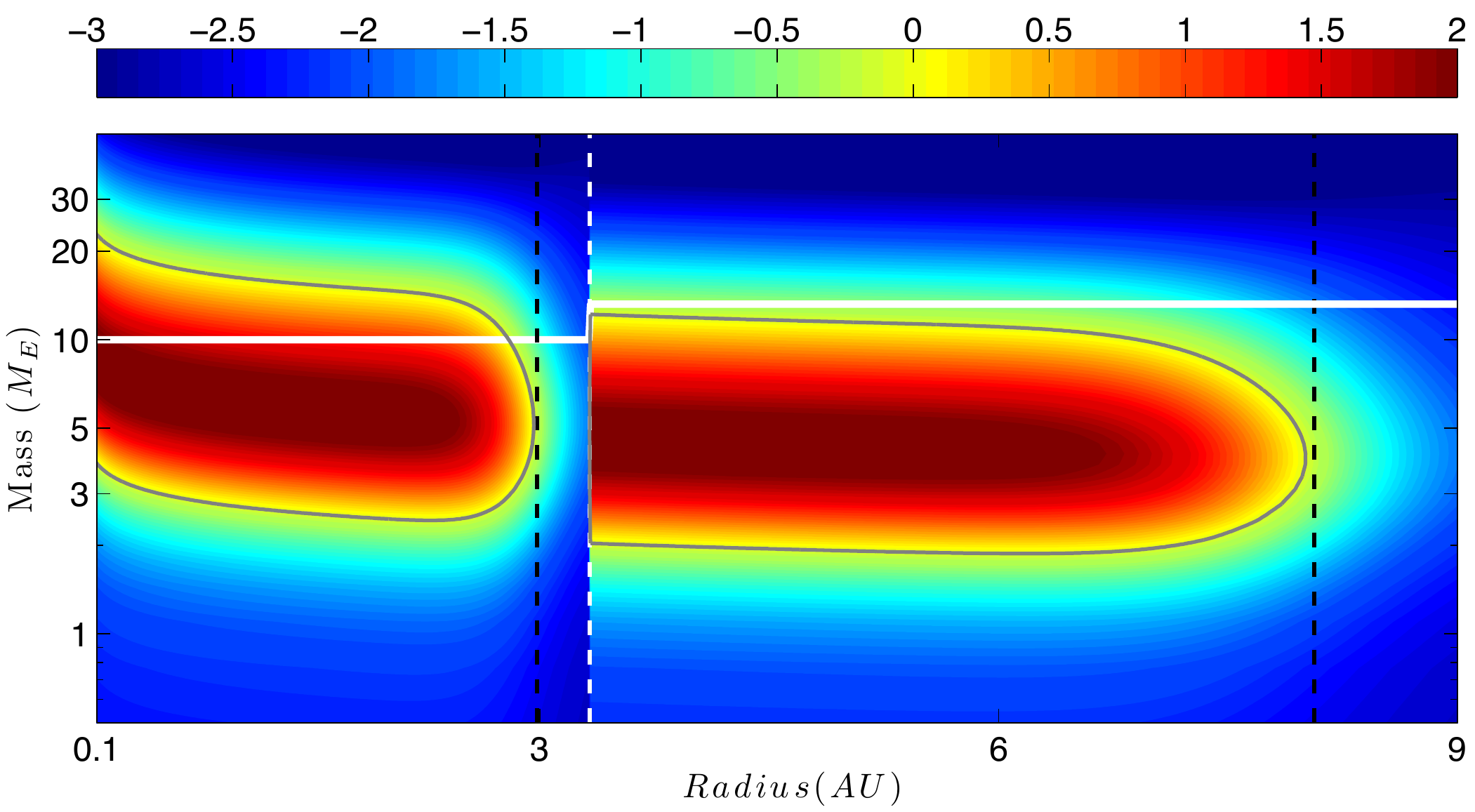}
\includegraphics[width=0.5\linewidth,clip=true]{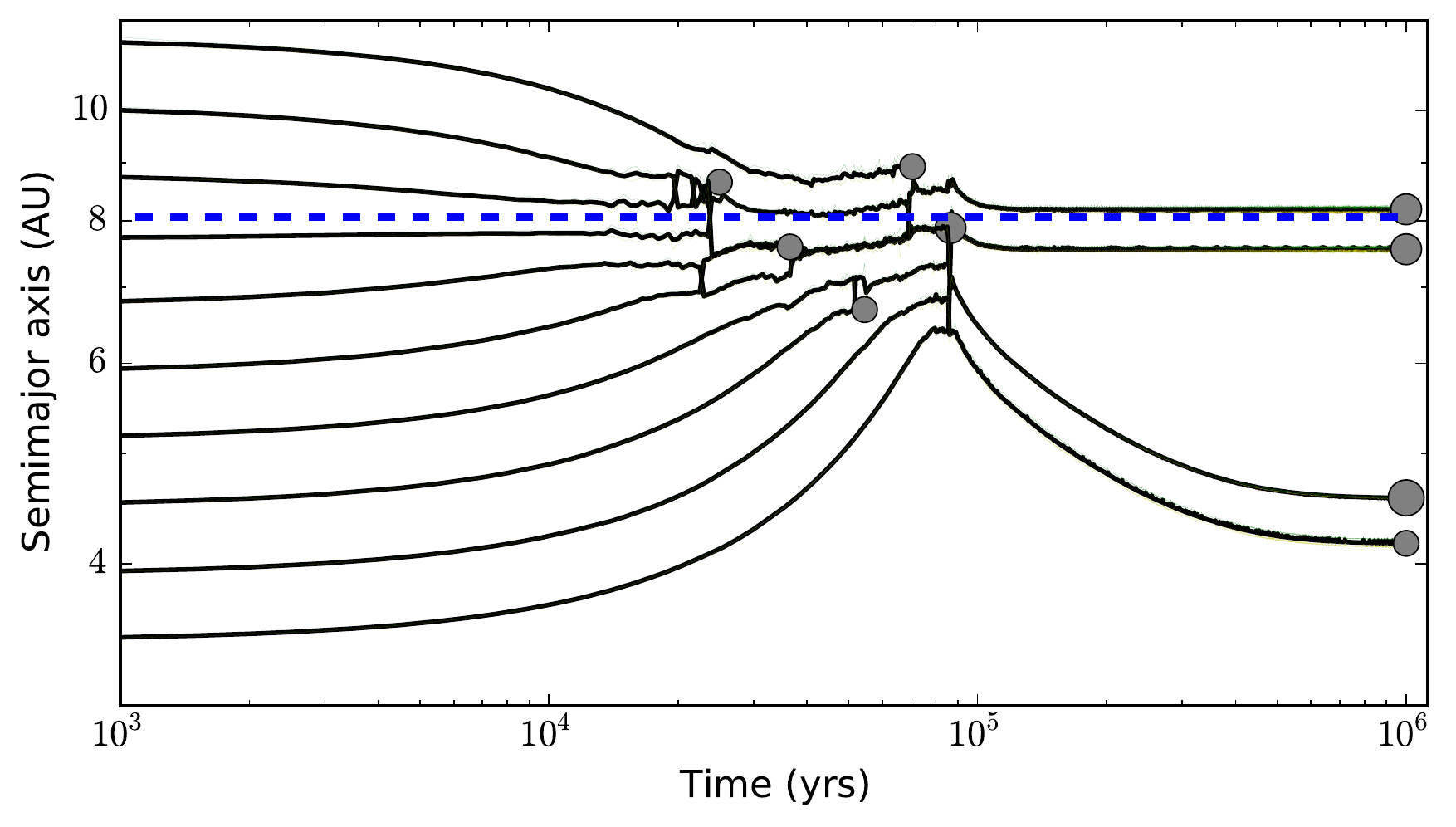}
\caption{
{\bf Left}: The type \uppercase\expandafter{\romannumeral1}
migration coefficient ($f_a$) for a range of embryos' masses at 
different locations in the disk. The two black dashed lines represent 
the transition radii $r_{\rm trans}$ associated with silicate and ice
grain opacities. The white dashed line represents the snow line.
The white solid line represents the critical core mass for the 
onset of efficient gas accretion. 
{\bf Right}: The mutual interaction between embryos and their natal 
disks in model Z3.
The black lines trace the evolution of embryos' semi-major axis and
blue dashed line indicates the location of  $r_{\rm trap}$.  Green and
yellow lines are embryos' apocenter and pericenter distance, respectively.
Disk parameters  are chosen $\rm \dot{M}= 3\times10^{-8} \rm M_{ \odot}
/\rm yr $, $\alpha_\nu =10^{-3}$, $M_{\ast} =1 \ M_{\odot}$, $\kappa_0
=1$, and $\eta_{\rm ice} = (1, 4)$  inside or outside the snow line. 
Model Z3 contains ten $3 \ M_{\oplus}$ embryos which are initially 
distributed on  
either side of the  trapping radius  with  $10R_R$ separation.
}
\label{ccc}
\end{figure*}

In this case, we set $\eta_{\rm ice}=4$ outside $r_{\rm ice}$.
All other parameters of model Z3 are identical to those of model Z1.
Figure \ref{ccc} shows that the distribution of the migration coefficient.
The white dashed line represents the snow line $r_{\rm ice}$. Since two 
different $\kappa_0$ are adopted (due to the difference in $\eta_{\rm ice}$
across the snow line), we find two different trapping locations, which are
indicated by black dashed lines.  Between these $r_{\rm trans}$, ice condensation/sublimation
 modifies the $\Sigma_g$ and $T_g$ profiles and weakens the
corotation torque.  Regardless the saturation condition, there is an
inward migration region segregated by two trapping radii. 

In order to illustrate this effect, we place ten $3 \ M_\oplus$ embryos 
across $r_{\rm ice}$ in model Z3.  Four embryos undergo convergent type  
{  \uppercase\expandafter{\romannumeral1} }  migration to the outer 
trapping radius, where they bypass their MMR barrier and merge into 
supercritical cores.  The critical core mass
$M_{c}$ which is $10 \  M_{\oplus}$ inside the snow line and $13 \ M_{\oplus}$ 
outside $r_{\rm ice}$.  

A merged embryo attained a mass of $12 \ M_{\oplus}$.  Since its mass exceeds $M_{\rm retain}$, its corotation torque is  saturated and the disk torque on it is dominated by the differential Lindblad torque.  It undergoes inward migration until it reaches the inner $r_{\rm trans}$ where its orbital evolution is stalled.  Along its migration path, it also induced the inward migration of a less massive companion embryo through their mutual MMR’s.  Two residual $6 \ M_{\oplus}$ embryos are left behind near the outer $r_{\rm trans}$.  This result highlights the possibility of forming well-separated multiple gas giant systems.


\section{Summary and Discussions}
\label{conclusion}
In this paper, we examine the cause for the $\eta_\oplus$-$\eta_J$ dichotomy,
and the origins of the $\eta_J$-$M_\ast$ and $\eta_J$-$Z_\ast$ correlations.  

Following the conventional sequential accretion scenario, we assume that
the formation of gas giants is preceded by the emergence of supercritical 
cores ($M_c>10 \ M_\odot$).  Based on the omnipresence of super-Earths 
around stars with a wide range of masses and metallicities, we assume that 
protoplanetary embryos are common. In addition, multiple-planet systems 
around low-mass and metal-deficient stars have total mass in excess of 
the critical value such the rarity of gas giants among these  
stars is not due to a lack of building-block embryos.  We suggest that  these
embryos undergo extensive convergent migration to a trapping radius likewise
 at the boundary between the viscously heated inner region and the irradiation  
heated outer region of the disk. The main deciding factor in their eventual 
fate is whether their migration is sufficiently fast to enable them to 
overcome their mutual resonant barrier.  

Using a set of steady-state disk models for protoplanetary disks around 
classical T Tauri stars, we determine the critical surface density 
distribution for the disk gas that would induce adequate torque for  
 embryos to undergo orbit crossing, close encounters
and  collisional coalescence.  Since it is difficult to measure
the surface density of both gas and dust in protostellar disks, we translate the critical condition
into critical gas accretion rate (${\dot M}_{\rm res}$).  With an {\it ad hoc} 
$\alpha$ layer structure model, we carry out analytic treatment and some 
numerical simulations to show that ${\dot M}_{\rm res}$ weakly depends on 
$M_\ast$ and decreases with $Z_d$.  Based on observational data on 
protostellar disks, we infer that the actual accretion rate increases
with the stellar mass and is independent of the stellar metallicity.
When  applied to our models, we find that embryos are
more likely to merge into supercritical cores around relatively massive
and metal-rich stars.  

Our results provide the necessary condition for the formation of gas giants.
We link this threshold condition to the cause of $\eta_J$-$M_\ast$ and 
$\eta_J$-$Z_\ast$ correlations.  In the forthcoming papers of this series, we
determine how the rapid growth of gas giants may affect the disk structure and
perturb the orbits of nearby embryos.  We also need to take into account
gas giants' type II migration and to generalize these models to 
evolving (rather than steady) disks.  During the advanced stage of their natal 
disks' evolution, both the super-Earths and gas giants may evolve into 
their asymptotic (present-day) orbital configuration as the trapping radius 
contracts with the depletion of the gas.  Finally, we will incorporate the
results of these investigations into our population synthesis models
\citep{Ida-2013} and simulate the observed $M_p$-$a_p$ distribution for 
stars with different $M_\ast$ and $Z_\ast$.


\vskip 20pt


\acknowledgements
The authors thank S. Aarseth, C. Baruteau, S. Ida, K. Kretke, H. Li,
T. Kouwenhowen, K. Schlaufman, A. Wolfgang, Y. Huang, S.
Dong, C. Ormel and C. Dominik for useful conversations and  an anonymous referee for helpful comments. This work is
supported by an UC/Lab grant and an IGPPS grant. B.Liu also
thanks T. Kouwenhowen for support by an NSFC grant. 

\clearpage


\appendix

%

\begin{table*}[htbp]
\caption{List of Notations}

\begin{tabular}{lll} \hline\hline
Variables & Meaning & Definition \\ \hline\hline
$M_{\rm c}$ & Critical core mass for initiating rapid gas accretion onto the core &          \\
$M_{\rm opt}$ &  Optimum embryo mass for unsaturated corotation torque and outward migration in viscous region & eq ~(\ref{opt}) \\
$M_{\rm retain}$ & Upper limit embyro mass for unsaturated corotation torque and outward migration in viscous region &  \\
$M_{\rm iso <} $ &  Isolation mass of the planet  for inner viscous region & eq.~(\ref{eq:misoin}) \\
$M_{\rm iso >} $ &  Planet isolation mass of the planet for outer irradiated  region & eq.~(\ref{eq:misoout}) \\
$M_{\rm p}$ & Single planet mass & \\
$M_{\rm s}$ & Total  planet mass in individual planetary system  &  \\

\hline
$M_{\ast}$ & Stellar mass  &          \\
$Z_{\ast}$ & Stellar metallicity &  \\
$\dot {M}_{\rm g}$  or $\dot {M}_{\rm d}$ & Gas accretion rate &  \\
$\dot {M}_{\rm a}$ & Average gas accretion rate & eq.~(\ref{eq:mdota}) \\
$\dot {M}_{\rm f}$ & Threshold gas accretion rate that embryos merge into retainable cores  & eq.~(\ref{eq:mdotcf}) \\
$\dot {M}_{\rm cr}$ & Normalized factor of $\dot {M}_{\rm f}$  & eq.~(\ref{eq:mdotcf}) \\
$\dot {M}_{\rm ice}$ & Threshold gas accretion rate    when $r_{\rm trans} = r_{\rm ice}$  & eq.~(\ref{mdotice})  \\
$\dot {M}_{\rm res}$ & Critical gas accretion rate for breaking mean motion resonance   &  eq.~(\ref{eq:mdotcf}) \\
$\dot {m}_{9\ {\rm res}}$ &    $\dot {M}_{\rm res} / 10^{-9} \ M_{\odot} \rm yr^{-1} $    & eq.~(\ref{eq:criticalmdot}) \\
$R_R$ &  Roche radius of the planet  &  \\
$R_p$ &   Physical radius of the planet &  \\
$k_0$ & Final mean separation in unit of Roche radius &  \\
$f_a$ & Type I migration coefficient & eq. ~(\ref{eq:adot})  \\
$\Gamma$ & Total net disk torque for a planet & eq. ~(\ref{eq:torque}) \\
$\eta_{\dot {M} (\dot {M}_f, M_\ast, Z_{\ast})}$ & Fraction of stars with $M_\ast$ and $Z_\ast$ has $\dot {M}_g$ larger than fiducial value $\dot {M}_f$ & eq.~(\ref{eq:etadotm}) \\
$\eta_a$ & Fitting power-law index for gas accretion rate and age relationship & eq.~(\ref{eq:mdota}) \\
$\eta_b$ & Fitting power-law index for gas accretion rate and stellar mass relationship & eq.~(\ref{eq:mdota}) \\
$\eta_c$ & Fitting power-law index for $\dot {M_{f}}$ and stellar mass relationship & eq.~(\ref{eq:mdotcf}) \\
$\eta_J$ & Fraction of stars containing gas giant  planets & \\
$\eta_{\oplus}$ & Fraction of stars containing super Earth planets   &  \\

\hline
$\tau_{\rm c,acc}$ & Gas accretion time scale when the core reaches critical mass & eq.~(\ref{eq:taucin}) \\
$\tau_{\rm dep}$ &  Depletion time scale of disk gas &  \\
$\tau_{\rm KH}$ &  Kelvin-Helmholtz contraction time scale of gas envelope & eq.~(\ref{eq:tau_KH}) \\
$\tau_{\rm I}$ & Type I  migration time scale of a planet & eq.~(\ref{eq:taui}) \\
$\tau_{\rm I<}$ & Type I  migration time scale of a planet in inner viscous region  & eq.~(\ref{eq:tauia}) \\
$\tau_{\rm I>}$ & Type I  migration time scale of a planet in outer irradiated  region  & eq.~(\ref{eq:tauib}) \\

\hline
$\Sigma_{\rm d}$ & Dust surface density of a disk &  \\
$\Sigma_{\rm g}$ & Gas surface density of a disk &  \\
$T_{\rm g}$ &  Disk gas temperature &  \\
$\eta_{\rm ice}$ & An enhancement factor of $\Sigma_{\rm d}$ due to ice condensation  &\\
$r_{\rm trans}$ & Trapping or transation radius separates the inner  viscous and  outer irradiated disk region    & eq.~(\ref{eq:trans1}) \\
$r_{\rm ref,ice}$ & Radius for dust destruction or ice condensation  & eq.~(\ref{eq:r_ice}) 
\\

\end{tabular}
\end{table*}

\clearpage

\bibliography{paperIII}

\begin{thebibliography}{118}
\expandafter\ifx\csname natexlab\endcsname\relax\def\natexlab#1{#1}\fi

\bibitem[{{Anders} \& {Grevesse}(1989)}]{Anders-1989}
{Anders}, E., \& {Grevesse}, N. 1989, \gca, 53, 197

\bibitem[{{Andrews} {et~al.}(2013){Andrews}, {Rosenfeld}, {Kraus}, \&
  {Wilner}}]{Andrews-2013}
{Andrews}, S.~M., {Rosenfeld}, K.~A., {Kraus}, A.~L., \& {Wilner}, D.~J. 2013,
  \apj, 771, 129

\bibitem[{{Bai} \& {Stone}(2013)}]{Bai-2013}
{Bai}, X.-N., \& {Stone}, J.~M. 2013, \apj, 769, 76

\bibitem[{{Bailli{\'e}} {et~al.}(2015){Bailli{\'e}}, {Charnoz}, \&
  {Pantin}}]{Baillie-2015}
{Bailli{\'e}}, K., {Charnoz}, S., \& {Pantin}, E. 2015, \aap, 577, A65

\bibitem[{{Baruteau} {et~al.}(2011){Baruteau}, {Cuadra}, \&
  {Lin}}]{Baruteau-2011}
{Baruteau}, C., {Cuadra}, J., \& {Lin}, D.~N.~C. 2011, \apj, 726, 28

\bibitem[{{Baruteau} {et~al.}(2014){Baruteau}, {Crida}, {Paardekooper},
  {Masset}, {Guilet}, {Bitsch}, {Nelson}, {Kley}, \&
  {Papaloizou}}]{Baruteau-2014}
{Baruteau}, C., {et~al.} 2014, Protostars and Planets VI, 667

\bibitem[{{Beckwith} {et~al.}(1990){Beckwith}, {Sargent}, {Chini}, \&
  {Guesten}}]{Beckwith-1990}
{Beckwith}, S.~V.~W., {Sargent}, A.~I., {Chini}, R.~S., \& {Guesten}, R. 1990,
  \aj, 99, 924

\bibitem[{{Bitsch} {et~al.}(2013){Bitsch}, {Crida}, {Morbidelli}, {Kley}, \&
  {Dobbs-Dixon}}]{Bitsch-2013}
{Bitsch}, B., {Crida}, A., {Morbidelli}, A., {Kley}, W., \& {Dobbs-Dixon}, I.
  2013, \aap, 549, A124

\bibitem[{{Bitsch} \& {Kley}(2010)}]{Bitsch-2010}
{Bitsch}, B., \& {Kley}, W. 2010, \aap, 523, A30

\bibitem[{{Bonfils} {et~al.}(2013){Bonfils}, {Delfosse}, {Udry}, {Forveille},
  {Mayor}, {Perrier}, {Bouchy}, {Gillon}, {Lovis}, {Pepe}, {Queloz}, {Santos},
  {S{\'e}gransan}, \& {Bertaux}}]{Bonfils-2013}
{Bonfils}, X., {et~al.} 2013, \aap, 549, A109

\bibitem[{{Buchhave} {et~al.}(2012){Buchhave}, {Latham}, {Johansen},
  {Bizzarro}, {Torres}, {Rowe}, {Batalha}, {Borucki}, {Brugamyer}, {Caldwell},
  {Bryson}, {Ciardi}, {Cochran}, {Endl}, {Esquerdo}, {Ford}, {Geary},
  {Gilliland}, {Hansen}, {Isaacson}, {Laird}, {Lucas}, {Marcy}, {Morse},
  {Robertson}, {Shporer}, {Stefanik}, {Still}, \& {Quinn}}]{Buchhave-2012}
{Buchhave}, L.~A., {et~al.} 2012, \nat, 486, 375

\bibitem[{{Buchhave} {et~al.}(2014){Buchhave}, {Bizzarro}, {Latham},
  {Sasselov}, {Cochran}, {Endl}, {Isaacson}, {Juncher}, \&
  {Marcy}}]{Buchhave-2014}
---. 2014, \nat, 509, 593

\bibitem[{{Chatterjee} \& {Tan}(2014)}]{Chatterjee-2014}
{Chatterjee}, S., \& {Tan}, J.~C. 2014, \apj, 780, 53

\bibitem[{{Chiang} \& {Youdin}(2010)}]{Chiang-2010}
{Chiang}, E., \& {Youdin}, A.~N. 2010, Annual Review of Earth and Planetary
  Sciences, 38, 493

\bibitem[{{Ciesla} \& {Cuzzi}(2006)}]{Ciesla-2006}
{Ciesla}, F.~J., \& {Cuzzi}, J.~N. 2006, \icarus, 181, 178

\bibitem[{{Clanton} \& {Gaudi}(2014)}]{Clanton-2014}
{Clanton}, C., \& {Gaudi}, B.~S. 2014, \apj, 791, 91

\bibitem[{{Coleman} \& {Nelson}(2014)}]{Coleman-2014}
{Coleman}, G.~A.~L., \& {Nelson}, R.~P. 2014, \mnras, 445, 479

\bibitem[{{Cossou} {et~al.}(2014){Cossou}, {Raymond}, {Hersant}, \&
  {Pierens}}]{Cossou-2014}
{Cossou}, C., {Raymond}, S.~N., {Hersant}, F., \& {Pierens}, A. 2014, \aap,
  569, A56

\bibitem[{{Cumming} {et~al.}(2008){Cumming}, {Butler}, {Marcy}, {Vogt},
  {Wright}, \& {Fischer}}]{Cumming-2008}
{Cumming}, A., {Butler}, R.~P., {Marcy}, G.~W., {Vogt}, S.~S., {Wright}, J.~T.,
  \& {Fischer}, D.~A. 2008, \pasp, 120, 531

\bibitem[{{Cuzzi} \& {Zahnle}(2004)}]{Cuzzi-2004}
{Cuzzi}, J.~N., \& {Zahnle}, K.~J. 2004, \apj, 614, 490

\bibitem[{{Da Rio} {et~al.}(2014){Da Rio}, {Jeffries}, {Manara}, \&
  {Robberto}}]{DaRio-2014}
{Da Rio}, N., {Jeffries}, R.~D., {Manara}, C.~F., \& {Robberto}, M. 2014,
  \mnras, 439, 3308

\bibitem[{{D'Alessio} {et~al.}(2001){D'Alessio}, {Calvet}, \&
  {Hartmann}}]{DAlessio-2001}
{D'Alessio}, P., {Calvet}, N., \& {Hartmann}, L. 2001, \apj, 553, 321

\bibitem[{{D'Antona} \& {Mazzitelli}(1994)}]{DAntona-1994}
{D'Antona}, F., \& {Mazzitelli}, I. 1994, \apjs, 90, 467

\bibitem[{{Dong} \& {Zhu}(2013)}]{Dong-2013}
{Dong}, S., \& {Zhu}, Z. 2013, \apj, 778, 53

\bibitem[{{Dressing} \& {Charbonneau}(2013)}]{Dressing-2013}
{Dressing}, C.~D., \& {Charbonneau}, D. 2013, \apj, 767, 95

\bibitem[{{Dressing} {et~al.}(2015){Dressing}, {Charbonneau}, {Dumusque},
  {Gettel}, {Pepe}, {Collier Cameron}, {Latham}, {Molinari}, {Udry}, {Affer},
  {Bonomo}, {Buchhave}, {Cosentino}, {Figueira}, {Fiorenzano}, {Harutyunyan},
  {Haywood}, {Johnson}, {Lopez-Morales}, {Lovis}, {Malavolta}, {Mayor},
  {Micela}, {Motalebi}, {Nascimbeni}, {Phillips}, {Piotto}, {Pollacco},
  {Queloz}, {Rice}, {Sasselov}, {S{\'e}gransan}, {Sozzetti}, {Szentgyorgyi}, \&
  {Watson}}]{Dressing-2015}
{Dressing}, C.~D., {et~al.} 2015, \apj, 800, 135

\bibitem[{{Endl} {et~al.}(2006){Endl}, {Cochran}, {K{\"u}rster}, {Paulson},
  {Wittenmyer}, {MacQueen}, \& {Tull}}]{Endl-2006}
{Endl}, M., {Cochran}, W.~D., {K{\"u}rster}, M., {Paulson}, D.~B.,
  {Wittenmyer}, R.~A., {MacQueen}, P.~J., \& {Tull}, R.~G. 2006, \apj, 649, 436

\bibitem[{{Ercolano} {et~al.}(2014){Ercolano}, {Mayr}, {Owen}, {Rosotti}, \&
  {Manara}}]{Ercolano-2014}
{Ercolano}, B., {Mayr}, D., {Owen}, J.~E., {Rosotti}, G., \& {Manara}, C.~F.
  2014, \mnras, 439, 256

\bibitem[{{Fischer} \& {Valenti}(2005)}]{Fischer-2005}
{Fischer}, D.~A., \& {Valenti}, J. 2005, \apj, 622, 1102

\bibitem[{{Fleming} \& {Stone}(2003)}]{Fleming-2003}
{Fleming}, T., \& {Stone}, J.~M. 2003, \apj, 585, 908

\bibitem[{{Fressin} {et~al.}(2013){Fressin}, {Torres}, {Charbonneau}, {Bryson},
  {Christiansen}, {Dressing}, {Jenkins}, {Walkowicz}, \&
  {Batalha}}]{Fressin-2013}
{Fressin}, F., {et~al.} 2013, \apj, 766, 81

\bibitem[{{Gammie}(1996)}]{Gammie-1996}
{Gammie}, C.~F. 1996, \apj, 457, 355

\bibitem[{{Garaud} \& {Lin}(2007)}]{Garaud-2007}
{Garaud}, P., \& {Lin}, D.~N.~C. 2007, \apj, 654, 606

\bibitem[{{Garcia Lopez} {et~al.}(2006){Garcia Lopez}, {Natta}, {Testi}, \&
  {Habart}}]{GarciaLopez-2006}
{Garcia Lopez}, R., {Natta}, A., {Testi}, L., \& {Habart}, E. 2006, \aap, 459,
  837

\bibitem[{{Glassgold} {et~al.}(1997){Glassgold}, {Najita}, \&
  {Igea}}]{Glassgold-1997}
{Glassgold}, A.~E., {Najita}, J., \& {Igea}, J. 1997, \apj, 480, 344

\bibitem[{{Goldreich} \& {Tremaine}(1980)}]{Goldreich-1980}
{Goldreich}, P., \& {Tremaine}, S. 1980, \apj, 241, 425

\bibitem[{{Hartmann}(1998)}]{Hartmann-1998BOOK}
{Hartmann}, L. 1998, {Accretion Processes in Star Formation}

\bibitem[{{Hartmann} {et~al.}(1998){Hartmann}, {Calvet}, {Gullbring}, \&
  {D'Alessio}}]{Hartmann-1998}
{Hartmann}, L., {Calvet}, N., {Gullbring}, E., \& {D'Alessio}, P. 1998, \apj,
  495, 385

\bibitem[{{Hasegawa} \& {Pudritz}(2010)}]{Hasegawa-2010}
{Hasegawa}, Y., \& {Pudritz}, R.~E. 2010, \apjl, 710, L167

\bibitem[{{Hayashi}(1981)}]{Hayashi-1981}
{Hayashi}, C. 1981, Progress of Theoretical Physics Supplement, 70, 35

\bibitem[{{Howard} {et~al.}(2012){Howard}, {Marcy}, {Bryson}, {Jenkins},
  {Rowe}, {Batalha}, {Borucki}, {Koch}, {Dunham}, {Gautier}, {Van Cleve},
  {Cochran}, {Latham}, {Lissauer}, {Torres}, {Brown}, {Gilliland}, {Buchhave},
  {Caldwell}, {Christensen-Dalsgaard}, {Ciardi}, {Fressin}, {Haas}, {Howell},
  {Kjeldsen}, {Seager}, {Rogers}, {Sasselov}, {Steffen}, {Basri},
  {Charbonneau}, {Christiansen}, {Clarke}, {Dupree}, {Fabrycky}, {Fischer},
  {Ford}, {Fortney}, {Tarter}, {Girouard}, {Holman}, {Johnson}, {Klaus},
  {Machalek}, {Moorhead}, {Morehead}, {Ragozzine}, {Tenenbaum}, {Twicken},
  {Quinn}, {Isaacson}, {Shporer}, {Lucas}, {Walkowicz}, {Welsh}, {Boss},
  {Devore}, {Gould}, {Smith}, {Morris}, {Prsa}, {Morton}, {Still}, {Thompson},
  {Mullally}, {Endl}, \& {MacQueen}}]{Howard-2012}
{Howard}, A.~W., {et~al.} 2012, \apjs, 201, 15

\bibitem[{{Huber} {et~al.}(2014){Huber}, {Silva Aguirre}, {Matthews},
  {Pinsonneault}, {Gaidos}, {Garc{\'{\i}}a}, {Hekker}, {Mathur}, {Mosser},
  {Torres}, {Bastien}, {Basu}, {Bedding}, {Chaplin}, {Demory}, {Fleming},
  {Guo}, {Mann}, {Rowe}, {Serenelli}, {Smith}, \& {Stello}}]{Huber-2014}
{Huber}, D., {et~al.} 2014, \apjs, 211, 2

\bibitem[{{Hubickyj} {et~al.}(2005){Hubickyj}, {Bodenheimer}, \&
  {Lissauer}}]{Hubickyj-2005}
{Hubickyj}, O., {Bodenheimer}, P., \& {Lissauer}, J.~J. 2005, \icarus, 179, 415

\bibitem[{{Ida} \& {Lin}(2004{\natexlab{a}})}]{Ida-2004a}
{Ida}, S., \& {Lin}, D.~N.~C. 2004{\natexlab{a}}, \apj, 604, 388

\bibitem[{{Ida} \& {Lin}(2004{\natexlab{b}})}]{Ida-2004b}
---. 2004{\natexlab{b}}, \apj, 616, 567

\bibitem[{{Ida} \& {Lin}(2005)}]{Ida-2005}
---. 2005, \apj, 626, 1045

\bibitem[{{Ida} \& {Lin}(2008)}]{Ida-2008}
---. 2008, \apj, 673, 487

\bibitem[{{Ida} {et~al.}(2013){Ida}, {Lin}, \& {Nagasawa}}]{Ida-2013}
{Ida}, S., {Lin}, D.~N.~C., \& {Nagasawa}, M. 2013, \apj, 775, 42

\bibitem[{{Ikoma} {et~al.}(2000){Ikoma}, {Nakazawa}, \& {Emori}}]{Ikoma-2000}
{Ikoma}, M., {Nakazawa}, K., \& {Emori}, H. 2000, \apj, 537, 1013

\bibitem[{{Johnson} {et~al.}(2010){Johnson}, {Aller}, {Howard}, \&
  {Crepp}}]{Johnson-2010}
{Johnson}, J.~A., {Aller}, K.~M., {Howard}, A.~W., \& {Crepp}, J.~R. 2010,
  \pasp, 122, 905

\bibitem[{{Johnson} {et~al.}(2007){Johnson}, {Butler}, {Marcy}, {Fischer},
  {Vogt}, {Wright}, \& {Peek}}]{Johnson-2007}
{Johnson}, J.~A., {Butler}, R.~P., {Marcy}, G.~W., {Fischer}, D.~A., {Vogt},
  S.~S., {Wright}, J.~T., \& {Peek}, K.~M.~G. 2007, \apj, 670, 833

\bibitem[{{Jones} {et~al.}(2016){Jones}, {Jenkins}, {Brahm}, {Wittenmyer},
  {Olivares}, {Melo}, {Rojo}, {Jord{\'a}n}, {Drass}, {Butler}, \&
  {Wang}}]{Jones-2016}
{Jones}, M.~I., {et~al.} 2016, ArXiv e-prints

\bibitem[{{Kim} {et~al.}(1994){Kim}, {Martin}, \& {Hendry}}]{Kim-1994}
{Kim}, S.-H., {Martin}, P.~G., \& {Hendry}, P.~D. 1994, \apj, 422, 164

\bibitem[{{Kley} \& {Nelson}(2012)}]{Kley-2012}
{Kley}, W., \& {Nelson}, R.~P. 2012, \araa, 50, 211

\bibitem[{{Kretke} \& {Lin}(2007)}]{Kretke-2007}
{Kretke}, K.~A., \& {Lin}, D.~N.~C. 2007, \apjl, 664, L55

\bibitem[{{Kretke} \& {Lin}(2012)}]{Kretke-2012}
---. 2012, \apj, 755, 74

\bibitem[{{Lambrechts} \& {Johansen}(2012)}]{Lambrechts-2012}
{Lambrechts}, M., \& {Johansen}, A. 2012, \aap, 544, A32

\bibitem[{{Laughlin} {et~al.}(2004{\natexlab{a}}){Laughlin}, {Bodenheimer}, \&
  {Adams}}]{Laughlin-2004a}
{Laughlin}, G., {Bodenheimer}, P., \& {Adams}, F.~C. 2004{\natexlab{a}}, \apjl,
  612, L73

\bibitem[{{Laughlin} {et~al.}(2004{\natexlab{b}}){Laughlin}, {Steinacker}, \&
  {Adams}}]{Laughlin-2004b}
{Laughlin}, G., {Steinacker}, A., \& {Adams}, F.~C. 2004{\natexlab{b}}, \apj,
  608, 489

\bibitem[{{Lee} \& {Peale}(2002)}]{Lee-2002}
{Lee}, M.~H., \& {Peale}, S.~J. 2002, \apj, 567, 596

\bibitem[{{Li} {et~al.}(2016){Li}, {Lin}, {Zhang}，Y., \&
  {Dong}，B}]{Li-2016}
{Li}, R., {Lin}, D., {Zhang}，Y., \& {Dong}，B. 2016, \apj \, in prep.

\bibitem[{{Lin} {et~al.}(1996){Lin}, {Bodenheimer}, \& {Richardson}}]{Lin-1996}
{Lin}, D.~N.~C., {Bodenheimer}, P., \& {Richardson}, D.~C. 1996, \nat, 380, 606

\bibitem[{{Lin} \& {Papaloizou}(1986)}]{Lin-1986}
{Lin}, D.~N.~C., \& {Papaloizou}, J. 1986, \apj, 309, 846

\bibitem[{{Lin} \& {Papaloizou}(1993)}]{Lin-1993}
{Lin}, D.~N.~C., \& {Papaloizou}, J.~C.~B. 1993, in Protostars and Planets III,
  ed. E.~H. {Levy} \& J.~I. {Lunine}, 749--835

\bibitem[{{Lissauer} {et~al.}(2011){Lissauer}, {Ragozzine}, {Fabrycky},
  {Steffen}, {Ford}, {Jenkins}, {Shporer}, {Holman}, {Rowe}, {Quintana},
  {Batalha}, {Borucki}, {Bryson}, {Caldwell}, {Carter}, {Ciardi}, {Dunham},
  {Fortney}, {Gautier}, {Howell}, {Koch}, {Latham}, {Marcy}, {Morehead}, \&
  {Sasselov}}]{Lissauer-2011}
{Lissauer}, J.~J., {et~al.} 2011, \apjs, 197, 8

\bibitem[{{Liu} {et~al.}(2015){Liu}, {Zhang}, {Lin}, \& {Aarseth}}]{Liu-2015}
{Liu}, B., {Zhang}, X., {Lin}, D.~N.~C., \& {Aarseth}, S.~J. 2015, \apj, 798,
  62

\bibitem[{{Lopez} \& {Fortney}(2014)}]{Lopez-2014}
{Lopez}, E.~D., \& {Fortney}, J.~J. 2014, \apj, 792, 1

\bibitem[{{Lyra} {et~al.}(2010){Lyra}, {Paardekooper}, \& {Mac
  Low}}]{Lyra-2010}
{Lyra}, W., {Paardekooper}, S.-J., \& {Mac Low}, M.-M. 2010, \apjl, 715, L68

\bibitem[{{Manara} {et~al.}(2012){Manara}, {Robberto}, {Da Rio}, {Lodato},
  {Hillenbrand}, {Stassun}, \& {Soderblom}}]{Manara-2012}
{Manara}, C.~F., {Robberto}, M., {Da Rio}, N., {Lodato}, G., {Hillenbrand},
  L.~A., {Stassun}, K.~G., \& {Soderblom}, D.~R. 2012, \apj, 755, 154

\bibitem[{{Marcy} {et~al.}(2008){Marcy}, {Butler}, {Vogt}, {Fischer}, {Wright},
  {Johnson}, {Tinney}, {Jones}, {Carter}, {Bailey}, {O'Toole}, \&
  {Upadhyay}}]{Marcy-2008}
{Marcy}, G.~W., {et~al.} 2008, Physica Scripta Volume T, 130, 014001

\bibitem[{{Marcy} {et~al.}(2014){Marcy}, {Isaacson}, {Howard}, {Rowe},
  {Jenkins}, {Bryson}, {Latham}, {Howell}, {Gautier}, {Batalha}, {Rogers},
  {Ciardi}, {Fischer}, {Gilliland}, {Kjeldsen}, {Christensen-Dalsgaard},
  {Huber}, {Chaplin}, {Basu}, {Buchhave}, {Quinn}, {Borucki}, {Koch}, {Hunter},
  {Caldwell}, {Van Cleve}, {Kolbl}, {Weiss}, {Petigura}, {Seager}, {Morton},
  {Johnson}, {Ballard}, {Burke}, {Cochran}, {Endl}, {MacQueen}, {Everett},
  {Lissauer}, {Ford}, {Torres}, {Fressin}, {Brown}, {Steffen}, {Charbonneau},
  {Basri}, {Sasselov}, {Winn}, {Sanchis-Ojeda}, {Christiansen}, {Adams},
  {Henze}, {Dupree}, {Fabrycky}, {Fortney}, {Tarter}, {Holman}, {Tenenbaum},
  {Shporer}, {Lucas}, {Welsh}, {Orosz}, {Bedding}, {Campante}, {Davies},
  {Elsworth}, {Handberg}, {Hekker}, {Karoff}, {Kawaler}, {Lund}, {Lundkvist},
  {Metcalfe}, {Miglio}, {Silva Aguirre}, {Stello}, {White}, {Boss}, {Devore},
  {Gould}, {Prsa}, {Agol}, {Barclay}, {Coughlin}, {Brugamyer}, {Mullally},
  {Quintana}, {Still}, {Thompson}, {Morrison}, {Twicken}, {D{\'e}sert},
  {Carter}, {Crepp}, {H{\'e}brard}, {Santerne}, {Moutou}, {Sobeck}, {Hudgins},
  {Haas}, {Robertson}, {Lillo-Box}, \& {Barrado}}]{Marcy-2014}
---. 2014, \apjs, 210, 20

\bibitem[{{Masset} {et~al.}(2006){Masset}, {Morbidelli}, {Crida}, \&
  {Ferreira}}]{Masset-2006}
{Masset}, F.~S., {Morbidelli}, A., {Crida}, A., \& {Ferreira}, J. 2006, \apj,
  642, 478

\bibitem[{{Mordasini} {et~al.}(2012){Mordasini}, {Alibert}, {Benz}, {Klahr}, \&
  {Henning}}]{Mordasini-2012}
{Mordasini}, C., {Alibert}, Y., {Benz}, W., {Klahr}, H., \& {Henning}, T. 2012,
  \aap, 541, A97

\bibitem[{{Mortier} {et~al.}(2013){Mortier}, {Santos}, {Sousa}, {Adibekyan},
  {Delgado Mena}, {Tsantaki}, {Israelian}, \& {Mayor}}]{Mortier-2013}
{Mortier}, A., {Santos}, N.~C., {Sousa}, S.~G., {Adibekyan}, V.~Z., {Delgado
  Mena}, E., {Tsantaki}, M., {Israelian}, G., \& {Mayor}, M. 2013, \aap, 557,
  A70

\bibitem[{{Movshovitz} {et~al.}(2010){Movshovitz}, {Bodenheimer}, {Podolak}, \&
  {Lissauer}}]{Movshovitz-2010}
{Movshovitz}, N., {Bodenheimer}, P., {Podolak}, M., \& {Lissauer}, J.~J. 2010,
  \icarus, 209, 616

\bibitem[{{Mulders} {et~al.}(2015){Mulders}, {Pascucci}, \&
  {Apai}}]{Mulders-2015}
{Mulders}, G.~D., {Pascucci}, I., \& {Apai}, D. 2015, \apj, 798, 112

\bibitem[{{Murray} \& {Dermott}(1999)}]{Murray-1999}
{Murray}, C.~D., \& {Dermott}, S.~F. 1999, {Solar system dynamics}

\bibitem[{{Natta} {et~al.}(2006){Natta}, {Testi}, \& {Randich}}]{Natta-2006}
{Natta}, A., {Testi}, L., \& {Randich}, S. 2006, \aap, 452, 245

\bibitem[{{Nelson}(2005)}]{Nelson-2005}
{Nelson}, R.~P. 2005, \aap, 443, 1067

\bibitem[{{Ogihara} \& {Kobayashi}(2013)}]{Ogihara-2013}
{Ogihara}, M., \& {Kobayashi}, H. 2013, \apj, 775, 34

\bibitem[{{Ormel} \& {Klahr}(2010)}]{Ormel-2010}
{Ormel}, C.~W., \& {Klahr}, H.~H. 2010, \aap, 520, A43

\bibitem[{{Ormel} \& {Kobayashi}(2012)}]{Ormel-2012}
{Ormel}, C.~W., \& {Kobayashi}, H. 2012, \apj, 747, 115

\bibitem[{{Paardekooper} {et~al.}(2010){Paardekooper}, {Baruteau}, {Crida}, \&
  {Kley}}]{Paardekooper-2010}
{Paardekooper}, S.-J., {Baruteau}, C., {Crida}, A., \& {Kley}, W. 2010, \mnras,
  401, 1950

\bibitem[{{Paardekooper} {et~al.}(2011){Paardekooper}, {Baruteau}, \&
  {Kley}}]{Paardekooper-2011}
{Paardekooper}, S.-J., {Baruteau}, C., \& {Kley}, W. 2011, \mnras, 410, 293

\bibitem[{{Papaloizou} \& {Szuszkiewicz}(2005)}]{Papaloizou-2005}
{Papaloizou}, J.~C.~B., \& {Szuszkiewicz}, E. 2005, \mnras, 363, 153

\bibitem[{{Pierens} {et~al.}(2013){Pierens}, {Cossou}, \&
  {Raymond}}]{Pierens-2013}
{Pierens}, A., {Cossou}, C., \& {Raymond}, S.~N. 2013, \aap, 558, A105

\bibitem[{{Pierens} \& {Nelson}(2008)}]{Pierens-2008}
{Pierens}, A., \& {Nelson}, R.~P. 2008, \aap, 482, 333

\bibitem[{{Pollack} {et~al.}(1996){Pollack}, {Hubickyj}, {Bodenheimer},
  {Lissauer}, {Podolak}, \& {Greenzweig}}]{Pollack-1996}
{Pollack}, J.~B., {Hubickyj}, O., {Bodenheimer}, P., {Lissauer}, J.~J.,
  {Podolak}, M., \& {Greenzweig}, Y. 1996, \icarus, 124, 62

\bibitem[{{Ram{\'{\i}}rez} {et~al.}(2011){Ram{\'{\i}}rez}, {Mel{\'e}ndez},
  {Cornejo}, {Roederer}, \& {Fish}}]{Ramirez-2011}
{Ram{\'{\i}}rez}, I., {Mel{\'e}ndez}, J., {Cornejo}, D., {Roederer}, I.~U., \&
  {Fish}, J.~R. 2011, \apj, 740, 76

\bibitem[{{Ram{\'{\i}}rez} {et~al.}(2015){Ram{\'{\i}}rez}, {Khanal}, {Aleo},
  {Sobotka}, {Liu}, {Casagrande}, {Mel{\'e}ndez}, {Yong}, {Lambert}, \&
  {Asplund}}]{Ramirez-2015}
{Ram{\'{\i}}rez}, I., {et~al.} 2015, \apj, 808, 13

\bibitem[{{Rogers}(2015)}]{Rogers-2015}
{Rogers}, L.~A. 2015, \apj, 801, 41

\bibitem[{{Ros} \& {Johansen}(2013)}]{Ros-2013}
{Ros}, K., \& {Johansen}, A. 2013, \aap, 552, A137

\bibitem[{{Ruden} \& {Lin}(1986)}]{Ruden-1986}
{Ruden}, S.~P., \& {Lin}, D.~N.~C. 1986, \apj, 308, 883

\bibitem[{{Sano} {et~al.}(2000){Sano}, {Miyama}, {Umebayashi}, \&
  {Nakano}}]{Sano-2000}
{Sano}, T., {Miyama}, S.~M., {Umebayashi}, T., \& {Nakano}, T. 2000, \apj, 543,
  486

\bibitem[{{Santos} {et~al.}(2004){Santos}, {Israelian}, \&
  {Mayor}}]{Santos-2004}
{Santos}, N.~C., {Israelian}, G., \& {Mayor}, M. 2004, \aap, 415, 1153

\bibitem[{{Schlaufman}(2015)}]{Schlaufman-2015}
{Schlaufman}, K.~C. 2015, \apjl, 799, L26

\bibitem[{{Schlaufman} \& {Laughlin}(2011)}]{Schlaufman-2011}
{Schlaufman}, K.~C., \& {Laughlin}, G. 2011, \apj, 738, 177

\bibitem[{{Shakura} \& {Sunyaev}(1973)}]{Shakura-1973}
{Shakura}, N.~I., \& {Sunyaev}, R.~A. 1973, \aap, 24, 337

\bibitem[{{Shen} {et~al.}(2005){Shen}, {Jones}, {Lin}, {Liu}, \&
  {Li}}]{Shen-2005}
{Shen}, Z.-X., {Jones}, B., {Lin}, D.~N.~C., {Liu}, X.-W., \& {Li}, S.-L. 2005,
  \apj, 635, 608

\bibitem[{{Sousa} {et~al.}(2011){Sousa}, {Santos}, {Israelian}, {Mayor}, \&
  {Udry}}]{Sousa-2011}
{Sousa}, S.~G., {Santos}, N.~C., {Israelian}, G., {Mayor}, M., \& {Udry}, S.
  2011, \aap, 533, A141

\bibitem[{{Sousa} {et~al.}(2008){Sousa}, {Santos}, {Mayor}, {Udry},
  {Casagrande}, {Israelian}, {Pepe}, {Queloz}, \& {Monteiro}}]{Sousa-2008}
{Sousa}, S.~G., {et~al.} 2008, \aap, 487, 373

\bibitem[{{Supulver} \& {Lin}(2000)}]{Supulver-2000}
{Supulver}, K.~D., \& {Lin}, D.~N.~C. 2000, \icarus, 146, 525

\bibitem[{{Tanaka} {et~al.}(2002){Tanaka}, {Takeuchi}, \& {Ward}}]{Tanaka-2002}
{Tanaka}, H., {Takeuchi}, T., \& {Ward}, W.~R. 2002, \apj, 565, 1257

\bibitem[{{Terquem}(2003)}]{Terquem-2003}
{Terquem}, C.~E.~J.~M.~L.~J. 2003, \mnras, 341, 1157

\bibitem[{{Turner} \& {Sano}(2008)}]{Turner-2008}
{Turner}, N.~J., \& {Sano}, T. 2008, \apjl, 679, L131

\bibitem[{{Turner} {et~al.}(2007){Turner}, {Sano}, \&
  {Dziourkevitch}}]{Turner-2007}
{Turner}, N.~J., {Sano}, T., \& {Dziourkevitch}, N. 2007, \apj, 659, 729

\bibitem[{{Wang} \& {Fischer}(2015)}]{Wang-2015}
{Wang}, J., \& {Fischer}, D.~A. 2015, \aj, 149, 14

\bibitem[{{Ward}(1997)}]{Ward-1997}
{Ward}, W.~R. 1997, \icarus, 126, 261

\bibitem[{{Weiss} \& {Marcy}(2014)}]{Weiss-2014}
{Weiss}, L.~M., \& {Marcy}, G.~W. 2014, \apjl, 783, L6

\bibitem[{{Whipple}(1972)}]{Whipple-1972}
{Whipple}, F.~L. 1972, in From Plasma to Planet, ed. A.~{Elvius}, 211

\bibitem[{{Wilden} {et~al.}(2002){Wilden}, {Jones}, {Lin}, \&
  {Soderblom}}]{Wilden-2002}
{Wilden}, B.~S., {Jones}, B.~F., {Lin}, D.~N.~C., \& {Soderblom}, D.~R. 2002,
  \aj, 124, 2799

\bibitem[{{Winn} \& {Fabrycky}(2015)}]{Winn-2014}
{Winn}, J.~N., \& {Fabrycky}, D.~C. 2015, \araa, 53, 409

\bibitem[{{Wolfgang} {et~al.}(2015){Wolfgang}, {Rogers}, \&
  {Ford}}]{Wolfgang-2015}
{Wolfgang}, A., {Rogers}, L.~A., \& {Ford}, E.~B. 2015, ArXiv e-prints

\bibitem[{{Wu} \& {Lithwick}(2013)}]{Wu-2013}
{Wu}, Y., \& {Lithwick}, Y. 2013, \apj, 772, 74

\bibitem[{{Zeng} \& {Sasselov}(2013)}]{Zeng-2013}
{Zeng}, L., \& {Sasselov}, D. 2013, \pasp, 125, 227

\bibitem[{{Zeng} {et~al.}(2016){Zeng}, {Sasselov}, \& {Jacobsen}}]{Zeng-2016}
{Zeng}, L., {Sasselov}, D.~D., \& {Jacobsen}, S.~B. 2016, \apj, 819, 127

\bibitem[{{Zhang} {et~al.}(2014){Zhang}, {Liu}, {Lin}, \& {Li}}]{Zhang-2014}
{Zhang}, X., {Liu}, B., {Lin}, D.~N.~C., \& {Li}, H. 2014, \apj, 797, 20

\bibitem[{{Zhou} {et~al.}(2007){Zhou}, {Lin}, \& {Sun}}]{Zhou-2007}
{Zhou}, J.-L., {Lin}, D.~N.~C., \& {Sun}, Y.-S. 2007, \apj, 666, 423

\end{thebibliography}
\end{document}